\documentclass[aip,jcp,reprint]{revtex4-1}
\usepackage{lineno,hyperref}
\usepackage{float}
\usepackage{chemformula}
\usepackage{graphicx}
\usepackage{xcolor}
\usepackage{bm}
\usepackage{amsmath,amssymb}
\DeclareUnicodeCharacter{2009}{\,} 
\usepackage[version=3]{mhchem} 

\begin{document}
	\author{Amin Bakhshandeh}
	\email{amin.bakhshandeh@ufrgs.br}  
\affiliation{Instituto de F\'isica, Universidade Federal do Rio Grande do Sul, 91501-970, Porto Alegre, RS, Brazil}
\affiliation{Departamento de F\'isico-Qu\'imica, Instituto de Qu\'imica, Universidade Federal do Rio Grande do Sul, 91501-970, Porto Alegre, RS, Brazil}

\author{Maximiliano Segala}
\email{maximiliano.segala@ufrgs.br}
\affiliation{Departamento de F\'isico-Qu\'imica, Instituto de Qu\'imica, Universidade Federal do Rio Grande do Sul, 91501-970, Porto Alegre, RS, Brazil}

\author{Thiago Colla}
\email{colla@ufop.edu.br}
\affiliation{Instituto de F\'isica, Universidade Federal de Ouro Preto, 35400-000, Ouro Preto, MG, Brazil}

 \title { Equilibrium conformations and surface charge
 	regulation of spherical polymer brushes in
 	stretched regimes}
	
	\begin{abstract}
		In the present work, we study the equilibrium conformations of linear polyelectrolytes tethered onto a spherical, oppositely charged core in equilibrium with an ionic reservoir of fixed concentration. Particular focus is placed on the situation of stretched chains, where the monomer concentration is known to display an inverse square-law decay far away from the spherical surface, which is then further extrapolated all the way down to the grafting core. While the equilibrium distributions of mobile ions are computed in the framework of a classical Density Functional Theory (cDFT) that incorporates both their size and electrostatic correlations within the grafted polyelectrolyte, the equilibrium configuration of the latter is described by its averaged radius of gyration, which is taken as a variational parameter that guarantees mechanical equilibrium across the polymer-solvent interface. The averaged particle size is then analyzed over a wide range of polymerization degrees, ionic concentrations and functionality of the polymer backbones. Two distinct regimes can be identified: at high ionic strengths, swelling of the grafted polymers is dominated by ionic entropic contribution as well as  polymer size effects, whereas at low ionic concentrations a balance between electrostatic and entropic effects is the main driven mechanism for particle stretching.   Using Monte Carlo simulations, we then proceed to investigate the effects of charge regulation when the brush core is further decorated with active functional sites randomly distributed over its surface, which act as receptors onto which dissolved acidic ions can be adsorbed. 
	\end{abstract}
	\maketitle
	\section{Introduction}
	Polymer chains are the cornerstone of most soft-matter systems~\cite{pol_phys,lik06,ver08}. The great relevance of these materials in both practical applications and theoretical investigations relies, among others, on their ease of synthesis (through a number of well-established experimental protocols~\cite{hadj01,lutz16}), the richness of topological conformations~\cite{fran75,nar10,sli14,poi14,koch15,poi15}, versatility and rather sensitive response to their environmental conditions. When embedded in a solution, their induced interactions and internal conformations can be easily triggered by suitable changes in the solvent properties~\cite{lik06}. In this sense, the intra-chain, solvent-induced monomer interactions are generally classified through the so-called solvent quality conditions, based on whether these are repulsive (good solvent), attractive (poor solvent), or ideal (the so-called $\Theta$-solvent condition) effective interactions~\cite{pol_phys,int_pol_phys,flory}. While in the latter case the conformations are dictated by entropic, elastic-like interactions, the induced interactions in good (poor) solvent favor coil-like (stretched) chain conformations~\cite{lik06,pol_phys,flory,bach14}. In a number of relevant applications in biology and material science, these interactions come from hydrophobic/hydrophilic groups attached at the polymer beads embedded in an aqueous environment~\cite{kab03,pop09,ole12,zhou19,tab19}. Functional groups can be also present at the monomer surface, in which case the dissociation of these groups in a polar medium leads to charged polymer chains (polyelectrolytes), thus releasing their counterions into the solution~\cite{kuhn98,ven17,car21}. Depending on the relevant length scales, linear macromolecules of fundamental relevance in biological applications such as RNA and DNA can be conveniently coarse-grained into flexible polyelectrolytes~\cite{zhi08,jiang09,meng11}, thereby justifying their use as prototype macromolecules in the theoretical description of many biological applications~\cite{bach14}. 
	
	In soft-matter applications, polymer chains are able to be assembled into a large number of different macroscopic configurations~\cite{cap12,cap13,yan13,ber13,han15}. The soft nature of their interaction ensures that mesoscopic particles comprised (to some extend) of polymer chains can coexist in packing fractions which largely exceed the close-packing configuration imposed by excluded volume interactions of their hard-core counterparts~\cite{lik08,moh17,gau17,col18}. This possibility entails large entropic gains (both from microscopic polymer conformations and macroscopic configurations), which results in a much wider variety of macroscopic self-assembly configurations for these soft-like particles~\cite{lik06}. Generally, polymer chains can be synthesized as free linear chains, entangled or strongly branched chains (as the case, for example, of polymer melts~\cite{doi86}, ring polymers~\cite{ber13}, microgels~\cite{kar19} or dendrimers~\cite{bos99,bal04}) or grafted chains (polymer brushes, star polymers). In the latter case, the chains are often grafted onto a macroscopic surface, thereby modifying the way the surface interacts with other particles in its environment~\cite{xu10}. By a proper understanding of the physical mechanisms that control such induced interactions, an advantage can be taken on this process by anchoring polymer chains of adjustable properties at particle surfaces, in such a way as to prompt desirable interactions between these surfaces. Indeed, the earliest applications of polymer chains in material science, back in the 1950s, were mostly concerned with the grafting of polymers onto colloidal surfaces in order to avoid their irreversible flocculation (driven by dispersion and/or gravitational forces), thus inducing a short-range, soft repulsion among them~\cite{brittain2007,vincent1974effect,van1951adsorption,mackor1951theoretical,mackor1952statistics,clayfield1966theoretical}. Most recently, applications of grafting polymer chains onto nanosurfaces involve modification of their interfacial properties~\cite{azza12,wang20}, foiling and biofoling~\cite{kri08,gro09,xu10}, as well as the control of their self-assembly into macroscopic structures with target properties~\cite{lik01}. Another promising application of grafting polyelectrolytes relies on their ability to trapping/releasing surrounding nanoparticles depending on the medium conditions, a key property in drug-delivery mechanisms~\cite{ram01,gen07,car21}. 
	
	Polymer brushes are a special class of polymer systems in which one end of the linear polymers are attached to either flat or curved substrates ~\cite{milner1991,wijmans1993,azza12}. There are basically two experimental routes in which this grafting process may take place: one in which the whole chain is attached at the surface \textit{via} chemical bonds (the so-called grafting-to process) and a second one in which free monomers are first attracted to specific surface sites, followed by a polymerization process in which the remaining free monomers are progressively linked together to these initially formed monolayers (\textit{e. g., via} highly directional, short-ranged interactions), until the fully assembled chains are found in equilibrium with the free, dissociated monomers (the so-called grafting-from process~\cite{ballauff2006,zhang2005cylindrical}).  Electrostatic interactions are usually of great relevance for driving (and further stabilizing) these processes~\cite{levin2002electrostatic,bakhshandeh2018theoretical,fisher1993criticality,levin1996criticality,bakhshandeh2011weak,bakhshandeh2019isothermal,bakhshandeh2019adsorption}. 
	They are one of the main driven forces that controls the so-called charge-regulation mechanisms, responsible for the binding/dissociation of ions into/from surface functional groups in an aqueous environment containing acidic/basic dissociated ions~\cite{lund2005charge,lund2013charge,trefalt2016charge,da2009polyelectrolyte,D0CP03633A,bk2019,derek2019,pod2019,bakhshandeh2020interaction}. 
	
	Such a Charge Regulation (CR) mechanism is a rather natural phenomenon whenever organic or inorganic functional groups are present on the surface of nanoparticles, which become dissociated in virtue of the high permittivity of the aqueous solvent. The amount of charge induced on a suspended surface bearing active functional groups is generally unknown and depends on its chemical equilibrium with the surrounding polar medium, which in turn is dictated by the free energy gain of this ionic association/dissociation process. One traditional approach for modeling the CR process in bulk solutions is the Ninham and Parsegian~\cite{NINHAM1971405} (NP) theory of ionic dissociation. However, since this theoretical framework relies on the use of bulk association constants, its extension to the (highly inhomogeneous) environment in the vicinity of curved surfaces is not straightforward~\cite{bk2019}. Quite recently, a charge regulation approach has been proposed which overcomes this drawback of NP theory by allowing for the calculation of surface association constants in the case in which functional groups are randomly arranged into lattice sites over the nanoparticle's surface~\cite{bk2019,D0CP03633A}. In this work, we shall apply this theory to compute the degree of dissociation of a spherical core comprised of functional sites which can be occupied by either acidic ionic groups or the free ends of suspended polyelectrolytes.

	When it comes to coarse-graining strategies to effectively resolve a system of grated polymers into larger length-scales, one fundamental question that arises is how to determine the equilibrium conformations of polymers chains, as this mesoscopic property depends crucially on microscopic solvent-induced intra-chain interactions, as well as on the inter-chain contributions~\cite{pol_phys,pel00,jus02,jus04,jian07,jusu09,col14,per21}. In the case of polymers attached to a common spherical core, particle conformation can be properly described through its overall radius of gyration~\cite{jus02,ver08,hansen}. This quantity is rather sensitive to a number of control parameters, such as the degree of dissociation, grafting density, the number of polymers per chain (degree of polymerization), bead size, polymer/surface interactions, among others~\cite{flory,gennes,jian07,jusu09,xu10}. In this context, scaling analysis has proved to be a powerful tool to provide estimations (in many cases with remarkable accuracy) about the particle size behavior across different parameter regimes, based on insightful and physically sound arguments~\cite{gennes,ver08,loc16}. In the case of polyelectrolytes grafted on flat and spherical surfaces, the overall mechanisms that control such scaling behavior have been rationalized by Pincus some thirty years ago~\cite{pin91}. The basic idea is that swelling of polymer chains is driven by counterion entropic forces, as these counterions condense into the polymer beads in order to achieve local electroneutrality. On the other hand, the ability of counterions to effectively screen the polymer charges on a local level depends on the underlying ionic screening length, which in turn is dictated by the overall ionic strengths. Under poor screening conditions, the stretching of the chain is dictated by a balance between entropic and electrolyte forces~\cite{col14}. Excluded volume interactions between the chain beads (self-avoiding effects) also contribute to this stretching mechanism~\cite{pin76,jus04,xu10}, which is counter-balanced by entropic (elastic-like) contributions from a Gaussian chain that prevents configurations of strong elongations.  
	
	This simple scenario is significantly changed whenever size and electrostatic correlations come into play~\cite{pin91,jian07}. Although the overall physical picture of a balance between entropic and electrostatic contributions still remains, the presence of charge and size correlations may significantly influence the fine-tuning between these contributions~\cite{levin2002electrostatic}. On one hand, size correlations generally lead to entropic loss, especially in regimes of large packing fractions close to a surface with a high grafting density. On the other hand, electrostatic correlations will try to prevent local charge neutrality and the screening of electrostatic forces, since this usually requires the packing of equally charged ions in close contact at the vicinity of an oppositely charged surface. Taking these contributions into account requires the employment of theoretical approaches that go beyond the mean-field, traditional Poisson-Boltzmann (PB) level~\cite{wu07}. In the present contribution, we shall apply a classical Density Functional Theory (cDFT) that accurately incorporates both size and electrostatic correlations into the ionic equilibrium distribution in the case of monovalent ions around polymer chains attached to an oppositely charged core. The ionic profiles will be computed under the constrain of a fixed polymer configuration with an inverse-square law decay typical of stretched chains attached to a spherical core. We then combine the cDFT approach with an equilibrium mechanical condition to compute the equilibrium radius of gyration of the star-like polymer brush.

	The aim of this work is twofold. First, we apply the aforementioned cDFT approach to investigate the equilibrium size and ionic profiles around ionic, star-like brushes under different ionic strengths and degree of polymerization. In this case, we consider that the polyelectrolytes are attached to an oppositely charged core whose surface charge is known \textit{a priori}. Next, we use a sticky sphere model to simulate the charge regulation in spherical polymer brushes and investigate the effect of connected polymers on the induced surface charge.
	The simulations are then tested against a recently developed CR theory~\cite{bk2019,D0CP03633A} to predict the density profile of polymer brush in acidic solution in a mean-field approximation.
	The remainder of this paper is organized as follows: In section II, we present the model system of an ionic star-like brush. Section III deals with the simulation details, followed by Sections IV and V in which the cDFT approach and charge regulation theory, respectively, are briefly outlined. Results are shown and discussed in Section VI. Finally, concluding remarks are outlined in Section VII, along with perspectives for future investigations.
	
	\section{Model System}
\begin{figure}[H]
		\centering
		\includegraphics[width=9cm,height=4cm]{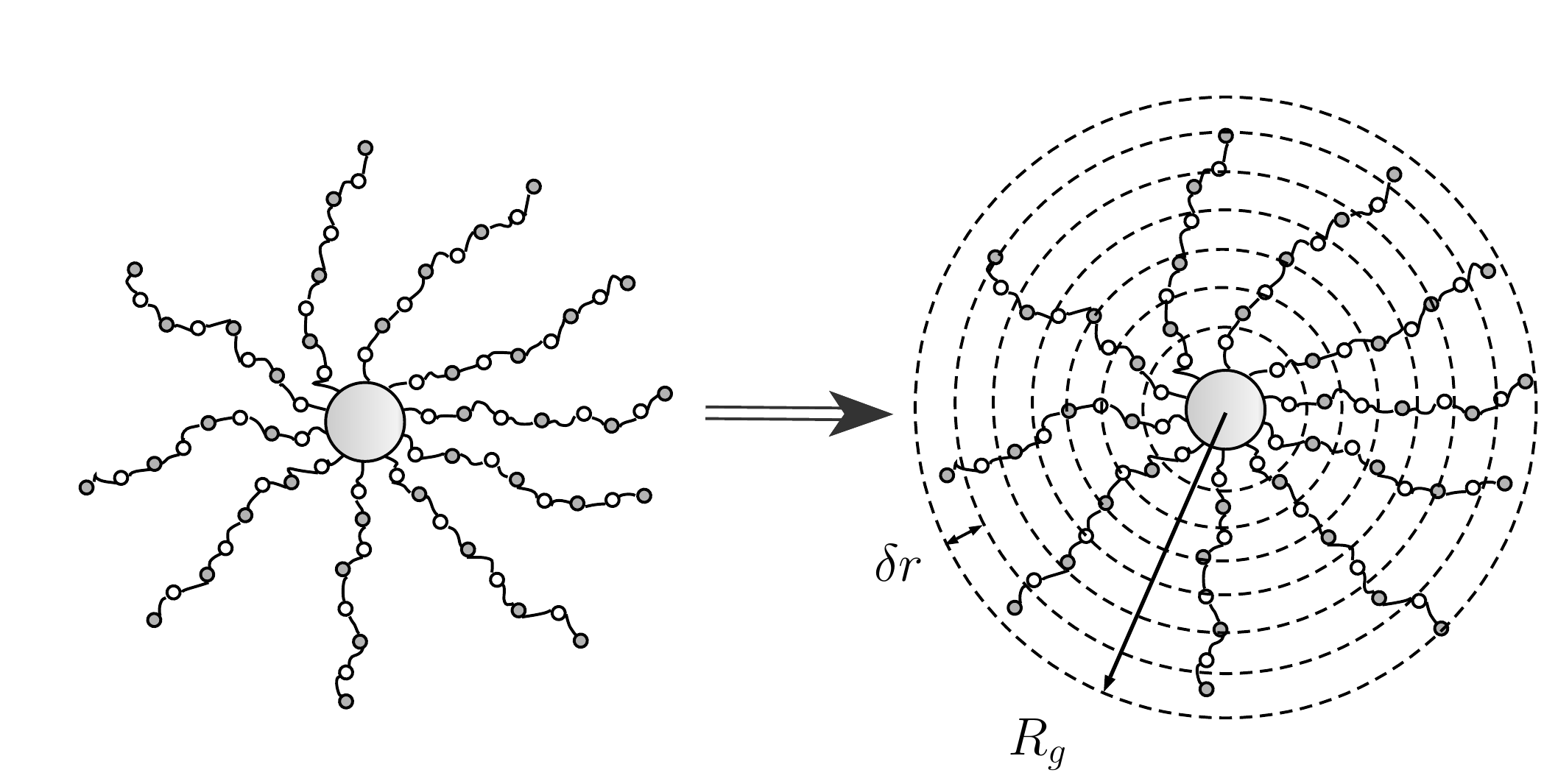}
		\caption{Sketch of our model system of a star-like polymer brush in a stretched configuration. The averaged monomer distribution in Eq. (\ref{poly_dist}) is obtained by splitting the system into a large number of concentric spherical shell of same thickness, each bearing an equal (averaged) occupation number of beads.} 
		\label{fig:fig1}
	\end{figure}

	The system under consideration consists of $f$ polymer chains grafted onto a spherical core of fixed radius $a$, bearing an effective surface charge $\sigma q $, where $q$ denotes the elementary charge. All monomers that comprise the polymer chains are represented as equally sized spheres of radius $r_b$, and possess dissociation groups on their surfaces, thus becoming ionized when immersed in an aqueous solvent.  We consider that a fraction $\alpha$ of such monomers will dissociate their functional groups, acquiring a net charge of $-q$ each, and releasing their monovalent counterions to the environment. It follows from simple symmetry considerations that the monomer distributions will be isotropic whenever the grafting density (or, equivalently, the functionality $f$) is large enough. Moreover, when the polyelectrolytes attached to the main core are in a stretched configuration, the corresponding lateral monomer fluctuations are negligible, so that the distance between neighboring monomers in a given chain is roughly the same all over the chain length. Considering this specific set-up, the distribution $\rho_b(r)$ of coarse-grained monomer beads along the chain can be obtained by chopping up these chains into spherical, concentric shells of the same thickness $\delta r$ (see Fig. \ref{fig:fig1}). It is clear that, in such a coarse-graining approach, each individual shell encloses the same averaged number of polymer beads. Therefore, the number of polymer beads $\delta N_b$ lying between a distance $r$ and $r+\delta r$ from the core center is always the same, regardless of $r$, such that $\delta N_b/\delta r=C$, with $C$ being a constant. This is a rather reasonable assumption for shells lying in the outermost layers. We notice however that this assumption may breakdown at the inner layers close to the central core, since the occupation number of a given shell is very susceptible to small orientational monomer fluctuations due to the curvature effects. We here neglect these effects by extending the constant occupation numbers all over the space \footnote{In a planar case, this so-called ``box model'' corresponds to uniform density profiles for the polymer beads.}. It then follows from the radial symmetry at large grafting densities that $\delta N_b\approx\rho_b(r)\delta V\approx 4\pi r^2\delta r\rho_b(r)$, where $\delta V$ is an element of volume at a distance $r$ from the central core~\cite{Vinicius}. As a consequence, the overall density of polymer beads should scale as $\rho_b(r)\approx C/(4\pi r^2)$. If the averaged number of monomers per chain (degree of polymerization) is $M$, the total number of polymer beads is $N_b=Mf$. Furthermore, it is reasonable to assume that the polymer density in such a coarse-graining approach will vanish beyond a given cut-off distance $R_g$ from the core center, which defines the averaged radius of gyration $R_g$ of the grafted polymer, such that $\rho_b(r)\approx C/(4\pi r^2)\Theta(R_g-r)$ (where $\Theta(x)$ equals unity if $x>0$ and zero otherwise is the unit step function). Finally, the constant $C$ can be evaluated from the normalization condition $4\pi r^2\int_{a+r_p}^{R_g}\rho_b(r)dr=Mf$, resulting in the following expression for the distribution $\rho_b(r)$ of polymer beads:
	\begin{equation}
	\rho_b(r)=\dfrac{Mf}{4\pi r^2\left(R_g-(a+r_b)\right)}.
	\label{poly_dist}
	\end{equation}
	Although the above general form of chain monomer distribution will be fixed throughout this work, its explicit dependence on the radius of gyration $R_g$ will be explored to find an optimal size of the stretched chains which accounts for the mechanical equilibrium across the brush-reservoir interface. \textcolor{black}{We also note that the model assumes a uniform monomer dissociation across the polymer networks through a constant dissociation fraction $\alpha$. However, if a charge regulation process was to be considered for the monomer dissociation, the degree of dissociation should depend on the local environment, in such a way that monomers in the vicinity of the core should have a weaker dissociation when compared to those at the external layers. By not taken these effects into account, the polymer distribution in Eq. (\ref{poly_dist}) is to be interpreted as a coarse-graining description of polyelectrolytes in stretched configurations (\textit{i. e.}, chains with large persistent lengths and/or at good solvent conditions). It is however important to notice that the proposed model should be able to properly capture some key mesoscopic mechanisms such as the dependence of the brush net charge and its averaged size on external conditions. These are typical input quantities in coarse-graining approaches designed to model these systems away from overlap regions, where most of the molecular details regarding the polymer static distributions can be safely neglected.}   
	
	The polymer brush is in equilibrium with an ionic reservoir of bulk concentration $c_s$. Therefore, in addition to cations and anions resulting from polyelectrolyte and core surface dissociation, respectively, the system also contains mobile ions which are free to move across the system-reservoir interface. Throughout this work, ionic species are modeled as monovalent hard-spheres of \textcolor{black}{radii} $r_{ion}=0.2$ nm, whereas the solvent is considered as a uniform background of dielectric constant $\varepsilon=80$. We will first assume that the core surface charge $\sigma q$ is a known constant. Later on, we shall apply a recently developed Monte Carlo simulation which properly accounts for the charge regulation mechanism~\cite{bk2019,D0CP03633A} to account for an equilibrium charge surface resulting from a given pH and number of functional groups attached to the grafting core. In that case, the grafting core is further embedded with surface acidic functional groups, modeled as randomly distributed surface patches that interact with dissolved hydronium acids of known concentration through a sticky-like potential that mimic particle binding~\cite{bk2019,D0CP03633A}.

	\section{Simulation details}

\begin{figure}[H]
		\centering
		\includegraphics[width=0.9\linewidth]{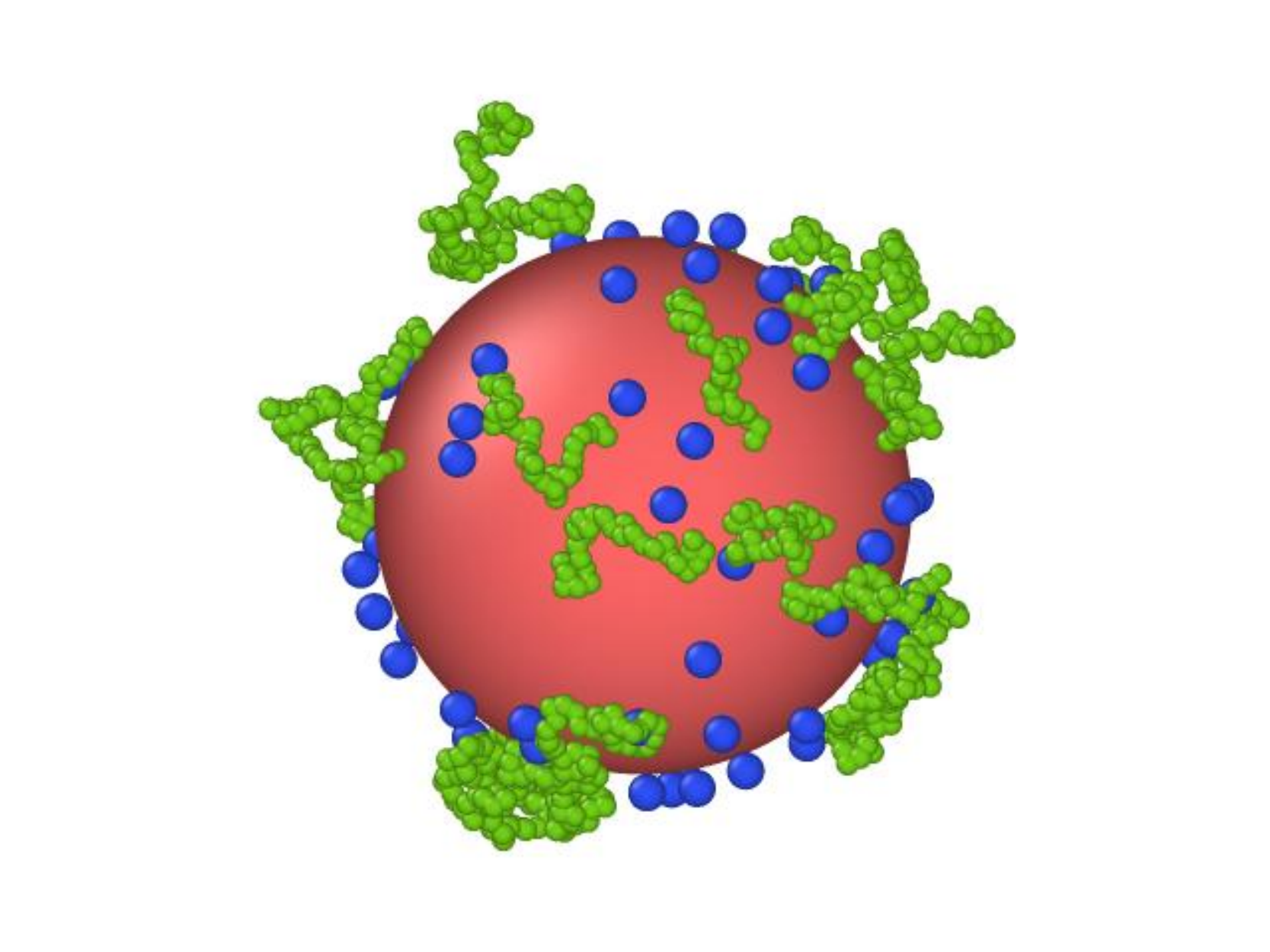}
		\caption{Spherical polymer brush. The hard core sphere is fixed on the center of the cell, having polyelectrolytes connected on its surface, along with sticky sphere sites.}
		\label{fig:fig2}
	\end{figure}

	Our simulation system consists of a spherical Wigner-Seitz (WS), at the center of which the spherical polymer brush is located. This star-shaped polymer is modeled as a hard core spherical particle with a total of $f$ grafted polyelectrolytes (functionality), each composed of $M$ spherical monomers with radius equal $r_b=0.2$,~nm and charge $-q$, having one of its ends bonded to the surface. Apart from the grafted polymers, the surface also comprises $N_{site}$  sticky sphere sites, randomly distributed over its surface, as depicted in Fig~\ref{fig:fig2}. 
	The monomers are connected in sequence \textit{via} a \textcolor{black}{harmonic spring potential} $U_b(R)$ which models the molecular bonds between neighboring monomers,~\cite{bakhshandeh2019adsorption,bakhshandeh2019isothermal,akinchina2004diblock,wernersson2013spreading,angelescu2014monte,Vinicius} 
	\begin{equation}\label{eq1}
	U_b(R)=\frac{A}{2}(r-r_0)^2,
	\end{equation}
	where $A=0.97k_BT$ ($T$ being the temperature and $k_B$ the Boltzmann constant), $r$ is the distance between monomers and $r_0$ represents the equilibrium distance between adjacent monomers, which we here take to be $r_0=0.5$~nm. 
	The active sites, randomly placed at the core surface, are  modeled  as spheres  of   diameters $0.4$~nm. These sites should account for the following   reaction:
	\begin{equation}\label{r1}    
	\ch{HA} +\ch{H_2 O}   \rightleftarrows \ch{H_3 O+}  + \ch{A-}.
	\end{equation} 
	\textcolor{black}{Since the process of charge regulation takes place only on central core, the polymer charges remain unaffected during CR. The effects of CR on the grafted polymers are thus indirect, as the changing the core surface charge will affect the  distribution of ions across the polymer networks.} To model the above reaction  between \ch{H+} and \ch{A}, an square well interaction potential is considered as follows~\cite{derek2019,bk2019}:
	\begin{equation}  
	u(r) =
	\begin{cases}
	\infty, & r<d,\\
	-\epsilon, &          d<r< d+\delta_r,\\
	0, &        r> d+\delta_r,
	\end{cases} 
	\label{Eq6}  
	\end{equation}  
	where  $d$ is the particle diameter and $\epsilon$ is the interaction strength. With this form for $u(r)$, it can be shown that in the Baxter limit (\textit{i. e.}, when the depth of the potential well potential goes to infinity, while its width goes to zero), the Boltzmann factor becomes~\cite{derek2019,bk2019}:
	\begin{equation} 
	e^{-\beta u(r)} =1+l_g \delta(r-d),
	\end{equation}
	where $l_g$ is the so-called sticky length, given by~\cite{bk2019,D0CP03633A}
	\begin{equation} 
	l_g=\delta_r(\mathrm{e}^\epsilon -1).
	\end{equation}
	As a result, the total interaction potential can be written as: 
	\begin{equation} 
	\begin{split}
	U =  \sum_{i=1}^{N}\sum_{j=i+1}^{N}  \frac{  q ^2}{\varepsilon |\bold{r}_i-\bold{r}_{j}|}+\sum{}^{'} u(r) + \sum_{adj} U_b(|\bold{r}_i-\bold{r}_j|),
	\end{split}
	\end{equation}
	where  the first term represents the direct Coulombic interactions between both free and bound monomers, the second term stands for the interaction between the surface functional sites and free hydronium ions, while the last term is the bounding interaction between neighboring monomers. The prime  on the second summation indicates that the sum is constrained to unbound sites only, since  we are considering monovalent functional groups so that each site can react just with one hydronium ion.  In our simulations, we consider a hard sphere core of radius $a=8$~nm. 
	The solvent is considered as a uniform dielectric constant $\varepsilon = 80$, such that the Bjerrum length, defined as $\lambda_B=q^2/\epsilon k_B T$, is $\lambda_B= 0.72$~nm. The simulations were performed using a canonical Monte Carlo algorithm~\cite{frenkel2001understanding} in combination with a typical Metropolis algorithm~\cite{metropolis1953equation}. Simulation runs of $5 \times 10^6$ MC steps for equilibration followed by $10^6$ steps for production have been considered.

	\section{Density Functional Theory}
	
	When size and charge correlation effects among mobile ions and the grafted polyelectrolytes become relevant, the traditional mean-field approximation becomes inaccurate, thus requiring the use of more elaborated approaches. While the employment of large-scale computer simulations to describe these systems over a wide range of system parameters may demand quite a bit of computational resources, alternative theoretical frameworks such as classical DFT and Self-Consistent field theories (SCFT) have emerged as promising candidates to perform systematic investigations spanning a large region of the system parameters that characterize polymer systems~\cite{xu10}. The cDFT has yet the advantage of allowing for natural incorporation of different species coexisting with the polymeric chains ~\cite{li06,li06_2,wu07,jian07,zhi08,xu10}, as well as interactions between different constituents and different levels of approximation for the various particle correlations. Here, we shall employ a simplified version of the cDFT in which the ionic equilibrium is achieved under the constrain of fixed configuration of polymer chains given by Eq. (\ref{poly_dist}). 
	
	The mobile ions as well as the bound monomers are subject to the influence of an external field that comprises their interactions with the charged core. The equilibrium ionic profiles are the ones that satisfy the Euler-Lagrange condition~\cite{hansen},
	\begin{equation}
	\dfrac{\delta \Omega[\{\rho_i(\bm{r})\}]}{\delta \rho_{i}(\bm{r})}=0,
	\label{el}
	\end{equation}
	where $\Omega$ represents the grand-potential density-functional, which reproduces the grand-potential at thermodynamic equilibrium, $\Omega_0$, when evaluated at the equilibrium profiles that satisfy (\ref{el}). Strictly speaking, the Euler-Lagrange condition (\ref{el}) is to be enforced upon the whole set of equilibrium profiles, including both free monomers (mobile ions) and bound ones (polymer beads). However, we propose here an alternative, two-step route to equilibrium, as follows. First, the Euler-Lagrange condition, Eq. (\ref{el}), is employed on the subset of free monomer distributions alone, while the distribution of bound monomers comprising the polyelectrolytes retains its functional form given by Eq. (\ref{poly_dist}).  The equilibrium condition is then achieved by considering $R_g$ as a variational parameter that ensures a global minimum of $\Omega$ in this particular functional subspace spanned by the family of parametric functions in Eq. (\ref{poly_dist}). As we shall see, this whole process amounts to the physical requirements of osmotic and mechanical equilibrium across the polymer-solvent interface under the constrain of stretched-like polymer conformations described by Eq. (\ref{poly_dist}).
	
	It is convenient to express $\Omega$ in terms of an intrinsic free energy $\mathcal{F}[\{\rho_i(\bm{r})\}]$, defined such that
	\begin{equation}
	\Omega[\{\rho_i(\bm{r})\}]=\mathcal{F}[\{\rho_i(\bm{r})\}]+\sum_i\int\left(\phi_i(\bm{r})-\mu_i\right)d\bm{r},
	\label{Fint}
	\end{equation}
	where $\phi_i(\bm{r})$ is the external potential which acts upon the $i$th component, and $\mu_i$ is the corresponding chemical potential at the particle reservoir. In terms of this quantity, the equilibrium condition can be re-expressed as~\cite{hansen,xu10}
	\begin{equation}
	\dfrac{ \delta\mathcal{F}[\{\rho_i(\bm{r})\}]}{\delta \rho_i(\bm{r})}=\mu_i-\phi_i(\bm{r}).
	\label{el2}
	\end{equation}
	The rhs of the above expression can be interpreted as an effective chemical potential which incorporates the effects from the external potentials. The intrinsic free energy can be split into ideal gas (entropic) and excess (internal interactions) contributions, 
	$\delta \mathcal{F}[\{\rho_i(\bm{r})\}]=\mathcal{F}_{id}[\{\rho_i(\bm{r})\}]+\mathcal{F}_{ex}[\{\rho_i(\bm{r})\}]$. While the latter functional is generally unknown, the ideal gas functional has an exact form given by:
	\begin{equation}
	\mathcal{F}_{id}[\{\rho_i(\bm{r})\}]=\sum_i\int\rho_i(\bm{r})\left[\ln(\Lambda^3\rho_i(\bm{r}))-1\right]d\bm{r},
	\label{Fid}
	\end{equation}
	where the summation is carried out over the ionic components (free monomers) alone. Notice that the ideal gas contributions are naturally decoupled from one another. Likewise, the excess contributions can be further split into electrostatic and hard-core (\textit{i. e.}, excluded volume) interactions, $\mathcal{F}_{ex}=\mathcal{F}_{elec}+\mathcal{F}_{hc}$. Apart from ion-ion interactions, these contributions also contain ion-polymer as well as polymer-polymer interactions. The hard-core functional can be accurately evaluated in the framework of the Fundamental Measure Theory (FMT)~\cite{ros89,ros02,roth02}. In this approach, the quantity $\mathcal{F}_{hc}$ is written as a local functional of a suitable set of weighted densities $n_{\alpha}{(\bm{r})}$ as
	\begin{equation}
	\mathcal{F}_{hc}[\rho_i(\bm{r})]=\int \Phi(\{n_{\alpha}(\bm{r})\})d\bm{r},
	\label{Fhc}
	\end{equation}
	where the free-energy density $\Phi(\{n_{\alpha}(\bm{r})\})$ is a known function of the weighted densities, chosen such as to recover some well-known bulk properties of a hard-sphere system in the limit of homogeneous profiles, in addition to the proper dimensional crossovers in strong confinements~\cite{roth02,hansen}.  Following the original FMT formulation, the weighted densities  are either scalar or vector quantities, defined in terms of corresponding weighted densities $\omega^{(\alpha)}_i(\bm{r}-\bm{r}')$ and $\bm{\omega}^{(\alpha)}_i(\bm{r}-\bm{r}')$ as
	\begin{eqnarray}
	n_{\alpha}(\bm{r})=\sum_i\int \rho_{i}(\bm{r}')\omega^{(\alpha)}_i(\bm{r}-\bm{r}')d\bm{r}, \label{n_esc}\\
	\bm{n}_{\alpha}(\bm{r})=\sum_i\int \rho_{i}(\bm{r}')\bm{\omega}^{(\alpha)}_i(\bm{r}-\bm{r}')d\bm{r}, \label{n_vec}
	\end{eqnarray}
	where the summations are performed over all the different species (free and bound ions, in the present situation). The set of weighted densities are explicitly given by $\omega^{(2)}_i(r)=\delta(r-a_i)$, $\omega^{(3)}_i(r)=\Theta(a_i-r)$, $\omega^{(1)}_i(r)=\dfrac{\omega_i^{(0)}(r)}{4\pi a_i}$, $\omega^{(0)}_i(r)=\dfrac{\omega_i^{(0)}(r)}{4\pi a_i^2}$, $\bm{\omega}^{(2)}(\bm{r})=-\nabla \omega^{(3)}_i(r)=\delta(r-a_i)\hat{\bm{e}}_r$, and $\bm{\omega}_i^{(1)}=\dfrac{\bm{\omega}_i^{(2)}(\bm{r})}{4\pi a_i}$. In these expressions, $a_i$ stands for the radius of $i$th monomer component, $\Theta(r)$ is equals $1$ if $r>0$ and $0$ otherwise, while $\hat{\bm{e}}_r\equiv\bm{r}/r$ is the unitary vector pointing in the radial direction. Here we adopt the so-called White-Bear (WB) approximation in which the free-energy density is taken to be~\cite{yu02,roth02,roth10}
	\begin{eqnarray}
	\Phi(\{n_i\}) & =\dfrac{\left(n_3+(1-n_3)^2\ln(1-n_3)\right)\left(n_2^3-3n_2|\bm{n}_2|^2\right)}{36\pi n_3^2(1-n_3)^2}\nonumber\\
	&+\dfrac{n_1n_2-\bm{n}_1\cdot\bm{n}_2}{1-n_3}-n_0\ln(1-n_3),
	\label{Phi}
	\end{eqnarray}
	such as to reproduce the Mansoori-Carnahan-Starling-Leland (MCSL) equation of state~\cite{man71} for a hard-sphere mixture in the bulk limit. 
	
	The electrostatic part $\mathcal{F}_{el}$ of the excess free energy can be conveniently written as a superposition of mean-field and residual (\textit{i. e.}, over mean-field) contributions~\cite{wu07}, $\mathcal{F}_{el}=\mathcal{F}_{mf}+\mathcal{F}_{res}$. The mean-field contribution amounts to neglect all electrostatic pair correlations beyond ideal gas, and has its explicit form given by
	
	\begin{equation}
	\mathcal{F}_{mf}[\rho_i(\bm{r})]=\dfrac{1}{2}\int \varrho(\bm{r})\psi(\bm{r})d\bm{r},
	\label{Fmf}
	\end{equation}
	where $\varrho(\bm{r})$ represents the local charge density at point $\bm{r}$, while $\psi(\bm{r})$ is the averaged, mean field potential at that point. The charge density  $\varrho(\bm{r})$  can be written in terms of the underlying ionic equilibrium distributions as
	\begin{equation}
	\varrho(\bm{r})=\sum_i z_iq\rho_i(\bm{r})+\sigma q\delta(r-a-r_c)-\alpha q\rho_b(r),
	\label{rho_tot}
	\end{equation}
	where $z_i=\pm 1$ for the case of monovalent counterions, $\sigma$ is the (for now fixed) surface charge density on the grafting core. The last term above represents the charge distribution of the polyelectrolytes, with $\alpha$ being the fraction monomers with dissociated charges. The averaged electrostatic potential in (\ref{Fmf}) can be written as
	\begin{equation}
	\psi(\bm{r})=\int\dfrac{\varrho(\bm{r}')}{\varepsilon|\bm{r}-\bm{r}'|}d\bm{r}'.
	\label{psii}
	\end{equation}
	Finally, the residual free energy is obtained using a second-order Taylor functional expansion around the corresponding homogeneous fluid:
	\begin{eqnarray}
	&\mathcal{F}_{res}[\rho_i(\bm{r})]  =   \displaystyle{\mathcal{F}_{res}(\rho_i^0)+\sum_i\mu_i^{0}\int\delta\rho_i(\bm{r})d\bm{r}}\nonumber\\
	&-\displaystyle{\dfrac{k_BT}{2}\sum_j\int c^{res}_{ij}(|\bm{r}-\bm{r}'|)\delta\rho_i(\bm{r})\delta\rho_j(\bm{r}')d\bm{r}d\bm{r}'},
	\label{Fres}
	\end{eqnarray}
	where $\mu_i^{0}=\dfrac{\delta F_{res}}{\delta\rho_i(\bm{r})}\biggr\arrowvert_{\rho_i(\bm{r})=\rho_i^{0}}$ is the homogeneous electrostatic contribution to the excess chemical potential in bulk, and  $c^{res}_{ij}(|\bm{r}-\bm{r}'|)\equiv \dfrac{-\delta\beta\mathcal{F}_{el}}{\delta\rho_i(\bm{r})\delta\rho_j(\bm{r}')}\biggr\arrowvert_{\rho_i(\bm{r}),\rho_j(\bm{r}')=\rho_i^{0},\rho_j^0}$ is the direct residual pair correlation function in a uniform fluid of concentrations $\rho_i^0$. The corresponding residual chemical potential is
	\begin{equation}
	\mu^{res}_i(\bm{r})\equiv\dfrac{\delta\mathcal{F}_{res}}{\delta\rho_i(\bm{r})}=\mu_i^{0}-k_BT\sum_j\int c^{res}_{ij}(|\bm{r}-\bm{r}'|)\delta\rho_j(\bm{r}')d\bm{r}'.
	\label{psi}
	\end{equation}
	The bulk chemical potentials and residual direct correlations are here computed in the framework of the Mean-Spherical-Approximation (MSA), which allows for analytic expressions for these quantities~\cite{blum75,li04}. It is important to notice that mean field and size contributions must be removed away from chemical potentials and pair correlations calculated in the bulk fluid, thus avoiding double counting of these contributions. In a similar way, the hard-sphere contribution to the excess chemical potential can be obtained from the functional derivative of Eq. (\ref{Fhc}), yielding:
	\begin{equation}
	\mu^{hs}_i(\bm{r})=\sum_{\alpha}\int f_{\alpha}(\bm{r}')\omega_i^{(\alpha)}(\bm{r}'-\bm{r})d\bm{r}',
	\label{mu_hs}
	\end{equation}
	where $f_{\alpha}(\bm{r}')\equiv \dfrac{\partial\Phi(n_{\alpha})}{\partial n_\alpha}$ is a local function obtained from Eq. (\ref{Phi}). In the case of vector densities, a contraction over the three vector components is implicit above.
	
	Application of the Euler Lagrange (\ref{el2}) equation for the free-energy functionals defined above leads to the following expression for the ionic equilibrium distributions:
	\begin{equation}
	\rho_i({r})=\rho_i^{0}\exp\left(-\beta qz_i\delta\psi({r})-\beta\delta\mu^{hc}_i(r)-\beta \delta\mu^{res}_i(r)\right),
	\label{rhoi}
	\end{equation}
	where $\delta\mu(r)$ and $\delta\psi(r)$ denote the difference between local and bulk chemical potentials and averaged potential, respectively. In the above expression, usage has been made of the fact that the ionic distributions are radially symmetric. The equilibrium ionic profiles in Eq. (\ref{rhoi}) can be numerically evaluated by solving a set of integral equations resulting from the combination of the above expression with relations (\ref{psi})   and (\ref{mu_hs}). Once this is accomplished, an equilibrium grand-potential $\Omega'_0[\rho_i(\bm{r})]$ can be computed by evaluating Eq. (\ref{Fint}) at the target equilibrium distributions that solve Eq. (\ref{rhoi}). However, as discussed above, this quantity does not represent the global equilibrium condition in the proposed approach. Instead, it is the quantity  that satisfies the Euler-Lagrange condition, Eq. (\ref{el}), subject to the constrain of fixed polymer-chain distributions described by Eq. (\ref{poly_dist}) with a given polymer size $R_g$. A global equilibrium condition can thus be achieved by further requiring the minimization of the grand-potential with respect to $R_g$, viewed in this context as a variational parameter that controls the equilibrium polymer conformations. It is important to notice that the grand-potential functional described above properly accounts for polymer-ion and polymer-polymer (\textit{i. e.}, intra- and inter- chain) electrostatic and size interactions. However, it does not account for the ideal-gas contribution of the polymer chains, which in turns depends on their averaged conformation. We must then add such a contribution to $\Omega'_0$ before imposing the global equilibrium condition. Modeling the polymer chains as freely-jointed, gaussian-like  chains, entropic contributions are described by the following elastic-like free-energy
	\begin{equation}
	\beta\mathcal{F}_{pol}=\dfrac{3f(R_g-a)^2}{2M\sigma_b^2},
	\label{Fpol}
	\end{equation}
	where $\sigma_b=2r_b$ is the bead diameter (corresponding to an elastic constant of $k=3f/M\sigma_B^2$), and $R_g-a$ is the average end-to-end polymer distance. Taking this contribution into account, the global equilibrium requirement can finally be translated into the following condition:
	\begin{equation}
	\dfrac{\partial\beta\Omega'_0(R_g)}{\partial R_g}+\dfrac{3f(R_g-a)}{M\sigma_b^2}=0.
	\label{el_g}
	\end{equation}
	This final condition enforced upon the equilibrium grand-potential allows one to compute the equilibrium profiles of both free monomers (mobile ions) and bound monomers (polyelectrolytes) under the constrain of stretched-like chains conformations that behave like (\ref{poly_dist}). It is important to stress that the quantity  $\Omega'_0(R_g)$ defined above properly takes care of electrostatic and size correlations between polymers and free ions. The optimal polymer size $R_g$ that satisfies Eq. (\ref{el_g}) should then naturally incorporate such non-trivial effects, being accurate enough in the wide-range domains of stretch-like polymer conformations. 
	
	\section{Charge regulation approach}

	In practical situations, the  surface charge on the core particle after dissociation takes place depends on the chemical reaction described in Eq.~(\ref{r1}). Accounting for such reaction equilibrium requires the incorporation of a charge regulation mechanism which describe binding between free ions and functional groups. Recently, a theory has been proposed to properly describe this phenomena in the case of charged surfaces bearing active sites. Here we shall briefly outline the basic aspects of the theory, referring the reader to Refs. \cite{bk2019} and \cite{D0CP03633A} for further details.

	The effective charge of the core particle depends on the equilibrium constant of the reaction shown in Eq.~(\ref{r1}), defined as the ratio between the activity coefficients of bound and free components once the chemical equilibrium has been established. This quantity is usually calculated based on reactions that take place in the bulk, which strongly  contracts with a highly inhomogeneous surface environment. As recently shown in Ref.~\cite{bk2019}, an affiliated surface constant $K_S$ can be defined, being related to its bulk counterpart $K_B$ by the following expression~\cite{bk2019}:
	\begin{equation}
	K_{S} = \frac{K_{B}}{2}\mathrm{e}^{-\lambda/d -\beta \mu},
	\label{ks}
	\end{equation}
	where $d$ is diameter of ion and $\mu$ is the work performed on a free hydronium ion which comes from infinity up to close contact with a given surface site. Considering a one-component plasma (OCP) in two dimensions, this quantity can be expressed as~\cite{bk2019}: 
	\begin{equation}
	\mu=\frac{3 m\lambda_B \sqrt{N_{site} } }{4 a }.
	\label{mu_qq}
	\end{equation}
	where $N_{site}$ is the number of active sites on the surface, and $m=1.106$ is the Madulung constant~\cite{levin2002electrostatic}. If there were no grafted polymers on the surface, $N_{site}$ would be the number of sites representing active functional groups, but now since we there also polymer attached on the surface, we should add the effect of these species to $\mu$. To this end, we consider the replacement replace $N\rightarrow N_{site} + f $ above. The reason for this change is that, as a hydronium reaches the sites, it will feel the interaction from the sites and also from the $f$ monomer monolayer grafted onto the surface. The bulk constant $K_B$ can be written as
	\begin{equation}
	K_{B} = 4\pi d^2 l_g \mathrm{e}^{b},
	\label{ka}
	\end{equation}
	where $b=\lambda_B/d$. Once the surface equilibrium constant is computed from Eqs. (\ref{ka}), (\ref{mu_qq}) and (\ref{ks}), the effective surface on the core can be obtained using the following expression~\cite{bk2019,D0CP03633A}
	\begin{equation}
	\sigma_{eff} = \dfrac{N_{site}}{4 \pi\left(a+r_{a}\right)^2}\left(\dfrac{K_{S}  c_a \mathrm{e}^{-\beta \varphi_0}}{\left(1+ K_{S} c_a \mathrm{e}^{-\beta \varphi_0}\right)}-1\right),
	\label{sig_eff}
	\end{equation}
	where $r_a$ and $c_a$ denote the radius and bulk concentration of hydronium ions, respectively, and $\varphi_0$ is the potential of mean force evaluated at the ion-core contact distance. In terms of the latter quantity, the local ionic profiles can be written as~\cite{bk2019}:
	
	\begin{eqnarray}
	&&\rho_{\ch{H^+}}(r) = c_a~\mathrm{e}^{-\beta \left( u(r) \varphi_{H}\right)} \label{Eq13}\\ 
	&& \rho_+(r)= c_s~\mathrm{e}^{-\beta \varphi_{+}(r)}\\
	&&\rho_-(r) = (c_a+c_s)~\mathrm{e}^{ \beta \varphi_-(r)},	
	\label{Eq14}
	\label{Eq15}
	\end{eqnarray}  
	where $\varphi_i(r)$ denotes the potential of mean force of the $i-$th  component at point $r$. As one can see, the part of sticky reaction has been included in Eq.~\ref{Eq13} through the potential $u(r)$ between dissolved hydronium ions and the sticky surface sites, defined in Eq. (\ref{Eq6}). Apart from Eq. (\ref{sig_eff}), the above ionic distributions are also linked to the surface charge through the satisfy Poisson equation: 
	\begin{equation}
	\nabla^2\psi(r)=-\dfrac{4\pi q}{\varepsilon}\left(\sigma_{eff}\delta(r-a-r_{ion})+\rho_{\ch{H+}}+\rho_{+}(r)-\rho_{-}(r)-\alpha \rho_b(r)\right).
	\label{poisson}
	\end{equation}
	In the mean-field approach, the potential of mean force is identified with the averaged potential, $\varphi_i(r)\rightarrow z_iq\psi(r)$, and the above equation recovers the Poisson-Boltzmann equation for the ionic distributions, which has to be solved in a self-consistent fashion such that the potential at close contact satisfies Eq. (\ref{sig_eff}). When correlations are non-negligible, the potential of mean force is related to the excess chemical potential, \textit{i. e.}, with the free excess energy cost for inserting a particle at a given position. In this case, the potential of mean force for each ionic specie in Eqs. (\ref{Eq13}), (\ref{Eq14}) and (\ref{Eq15}) can be directly associated with the Boltzmann factors appearing in Eq.(\ref{rhoi}). This procedure allows one to compute equilibrium properties of star-like polyelectrolytes with a fixed number of functional sites on its core, instead of assuming a fixed charge which is insensitive to ionic and polymer distributions.

	\section{Results and discussion}

	We shall now apply the theoretical approaches described above to analyze some key properties of grafted polyelectrolytes, namely the equilibrium ionic structure around the grafted polymers, the averaged size of the star polymer under different conditions, as well as the charge regulation process that dictates the equilibrium effective charge assign to the grafting core. \textcolor{black}{The analysis herein are restricted to the case of equally sized ions. In the SI material, results are presented for the case of an asymmetric electrolyte comprised of bigger cations.}

	\subsection{Ionic structure}
	
\begin{figure}[H]
		\centering
		\includegraphics[width=5.5cm,height=4cm]{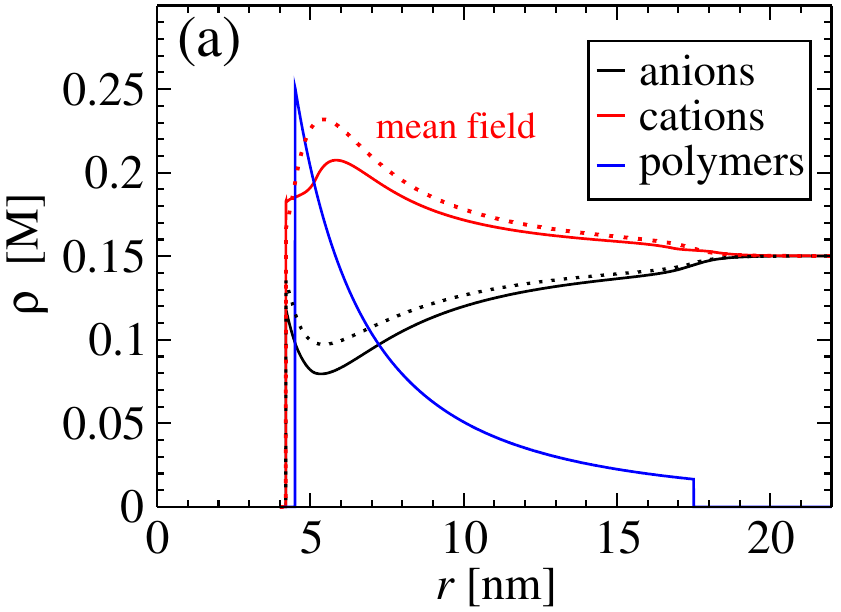}
		\includegraphics[width=5.5cm,height=4cm]{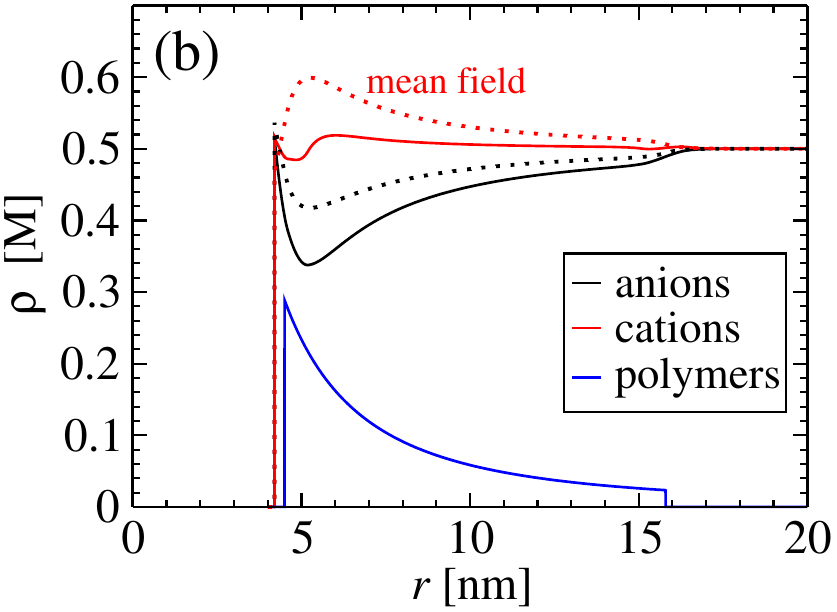}
		\includegraphics[width=5.5cm,height=4cm]{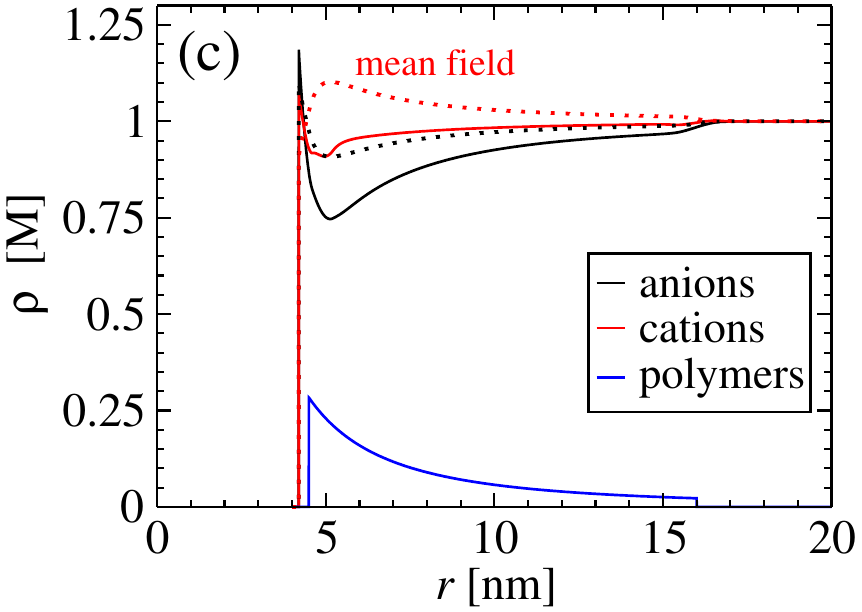}
		\caption{Ionic distributions around a star polymer with core size $a=4$~nm, comprising $f=10$ grafted polymers with a total of $M=50$ monomers each. All polymers are fully ionized, $\alpha =1$. The core surface charge is fixed at $\sigma = 0.01$~C/m$^2$, while the ionic reservoir concentrations are $c_s=0.15$~M (a),  $c_s=0.5$~M (b) and  $c_s=0.5$~M (c). While full lines are results from the cDFT approach, dotted lines refer to mean-field predictions. The fixed polymer distribution (blue lines) is also shown for the sake of completeness.}
		\label{fig:fig3}
	\end{figure}

	We start by applying the proposed cDFT  approach to investigate the ionic distributions in the vicinity of the grafted polyelectrolyte chains. As stated in the Introduction, the ionic arrangement around the charged monomers will be dictated by the best compromise between local neutrality -- which favors counterion condensation at both polymer beads and surface core -- and entropic contributions, which prevents the emergence of highly inhomogeneuos structures necessary to achieve electroneutrality at a local level. The balance between these competing effects is strongly influenced by the amount of salt in the external reservoir. The higher the reservoir ionic concentration, the weaker the entropic penalty for achieving local neutrality, since osmotic flux  allows for a large amount of ions to enter the polyelectrolyte region in order to neutralize its charge, yet avoiding the building-up strong local inhomogeneities. Furthermore, the presence of charge and size correlations will play an important role on this fine tuning between energetic gains and entropic penalties. While size correlations tend to underweight configurations of high local packing fractions, the presence of electrostatic correlations inhibits the emergence of high charge modulations. These general features can be observed in Fig.~\ref{fig:fig3}, which shows equilibrium ionic profiles around a total of $f=10$ fully ionized ($\alpha=1$) polymer chains made up of $M=50$ monomers at different ionic concentrations $c_s$. Unless otherwise specified, the discussions below are restricted to star polymers with a fixed core size of $a=4$~nm bearing a total of $f=10$ attached polymers, whose monomers have radii $r_b=0.4$~nm.  
	
	From Fig. \ref{fig:fig3} we can see that, as the salt concentration increases, the distributions become less homogeneous with respect to their bulk values, indicating an entropy gain at larger ionic concentrations. Alternatively, this large inhomogeneity can also be assigned to a poor screening of electrostatic interactions at low ionic strengths, which increases the range of electrostatic interactions.  Another clear effect of salt addition is the enhancement of ionic correlations, which are stronger at the vicinity of the core surface. Indeed, the deviations from mean-field theory (which amounts to take $\mu^{hc}=\mu^{res}=0$ in Eq. (\ref{rhoi})), represented as dotted lines in Fig.~\ref{fig:fig3}, become more pronounced as the ion concentration grows larger. While the mean-field approach always predict anion condensation and a cation depletion from the (positively charged) core, followed by a region of maximum/minimum concentration due to mutual repulsion, profiles from cDFT show a far more complex structure, clearly brought about by ionic correlations. In this case, a strong attraction of anions to the charged core leads to a contaminant second layer of cations, which are attracted to this region due to ionic correlations. This is followed by a small minimum in both profiles, most likely due to size correlations. Indeed, such layering-like profiles are a well-known feature resulting from packing effects in the vicinity of interfaces~\cite{xu10,pat14,col16}. 
	
	\begin{figure}[H]
		\centering
		\includegraphics[width=5.5cm,height=4cm]{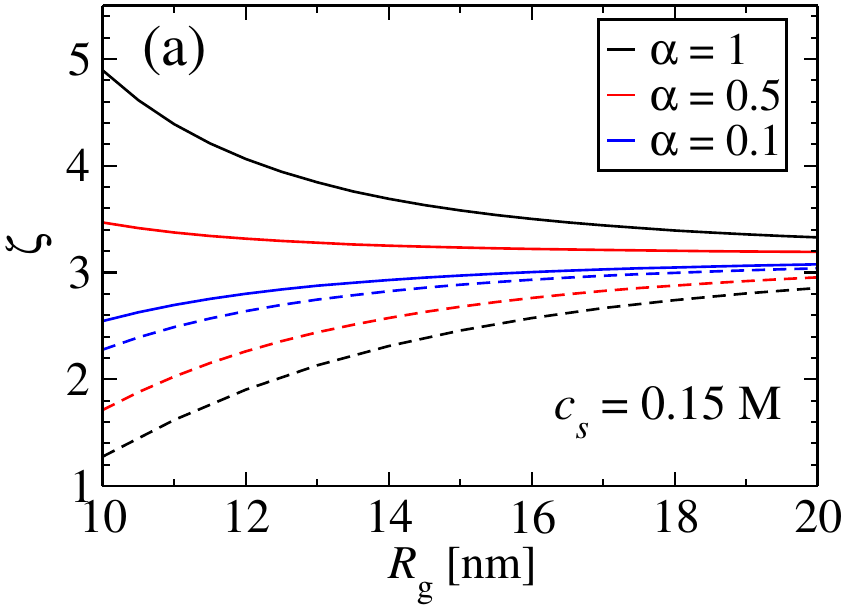}
		\includegraphics[width=5.5cm,height=4cm]{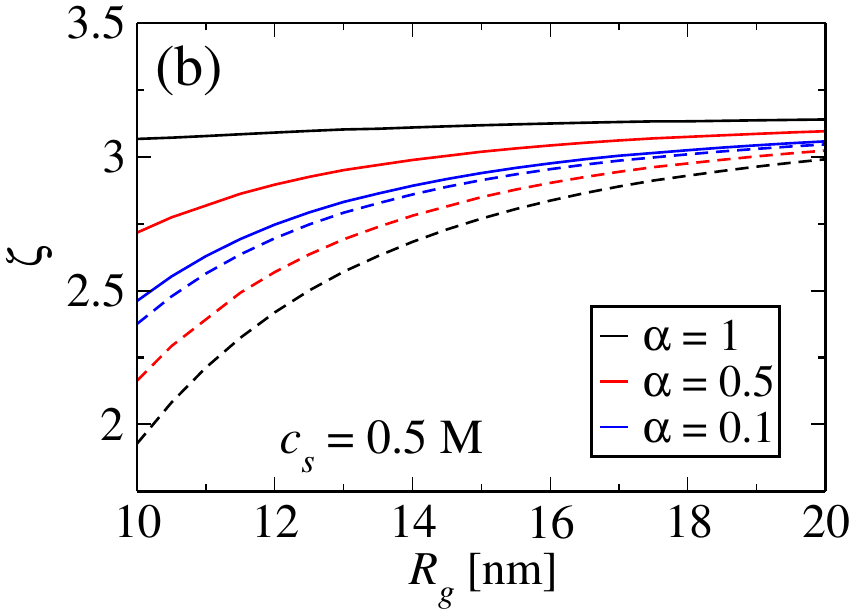}
		\caption{Ionic adsorptions $\zeta$ as a function of the radius of gyration $R_g$ for different dissociation fractions $\alpha$. The degree of polymerization is $M=50$, whereas the salt concentrations are $c_s=0.15$~M (a) and  $c_s=0.5$~M (b). Solid lines represent cation adsorption, while dashed-lines stand for anion adsorptions.}
		\label{fig:fig4}
	\end{figure}
	
	Further insights on the interplay between electrostatic and entropic contributions can be gained by looking at the so-called adsorption coefficients, $\zeta_i$, which measure the ratio between the averaged concentration $\bar{c}_i$ of the $i$th component adsorbed onto a confining region and that of the bulk reservoir, far away from the interface. In the present situation, the adsorption coefficients $\zeta_{\pm}=\bar{c}_{\pm}/c_s$ quantifies the excess of cations (anions) adsorbed onto the star polymer with respect to their reservoir concentrations. Since the polymer chains and the surface core are oppositely charged, the adsorption coefficients are always larger than unity for both cations and anions. The behavior of these quantities as functions of the star-polymer size are shown in Fig.~\ref{fig:fig4}, for star polymers with $M=50$ monomers per polymers grafted onto a core of surface charge $\sigma=0.01$~C/m$^2$, at different fractions of monomer dissociation and salt concentrations. Some very nice features can be highlighted from this figure. We first notice that, as expected, the adsorption of cations always increase as the fraction of dissociation $\alpha$ becomes higher, since an increasingly large amount of cations will be adsorbed from the reservoir, in such a way as to neutralize the polymer charge locally. For the same reason, the opposite trend is observed for the anion adsorptions. These anions adsorptions are always increasing functions of the polymer size. This is mainly due to entropic reasons: as the star polymer size grows, the ions have more room to move along this region, which favors they diffusion into the polymer chains. Concomitantly, the polymer concentration decreases locally, decreasing the energy cost of anion adsorption.  In the case of cations (solid lines), the situation is not so straightforward, and two different regimes can be identified.  At large ionic concentrations, the adsorption is an increasing function of $R_g$, suggesting that electroneutrality can be already achieved at small polymer sizes, in such a way that further cation adsorption is driven mostly by entropic effects. On the other hand, at low ionic concentrations the adsorption into highly charged polymer chains decreases as the polymer size increases, indicating an opposite flux in which counterions are spelled out from the polymer chains back into the ionic reservoir as the polymer size grows larger. This effect results from the balance between entropic contributions and charge neutrality at low salt concentrations and high polymer charge. The large cation adsorption at small system sizes $R_g$ implies that the reservoir has to provide the system with a very large amount of counterions in an attempt to neutralize the polymer charge. Such a large adsorption results in a very large entropic penalty (large $\zeta$), which is released as $R_g$ increases, with the reservoir ``taking back'' its counterions attached at the polymer chains. 
	
\begin{figure}[h!]
		\centering
		\includegraphics[width=5.5cm,height=4cm]{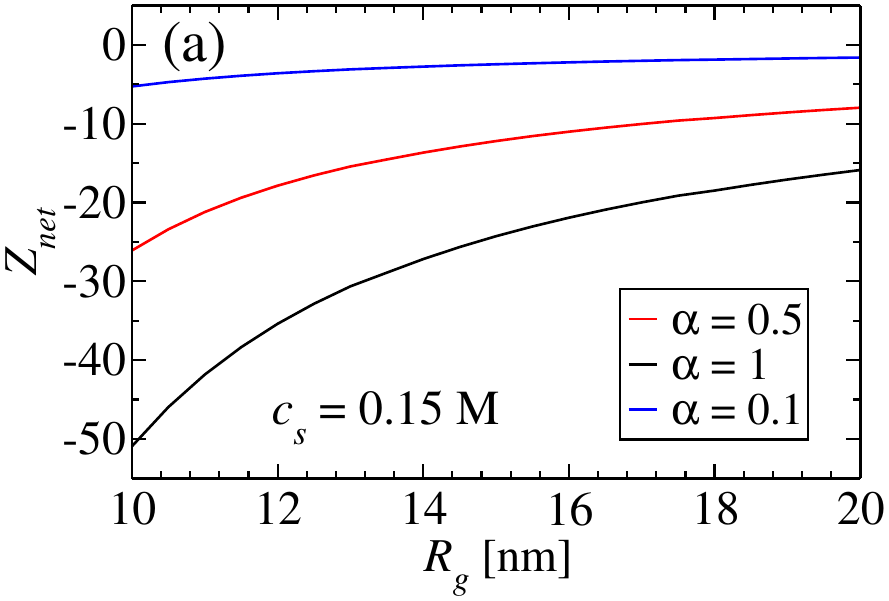}
		\includegraphics[width=5.5cm,height=4cm]{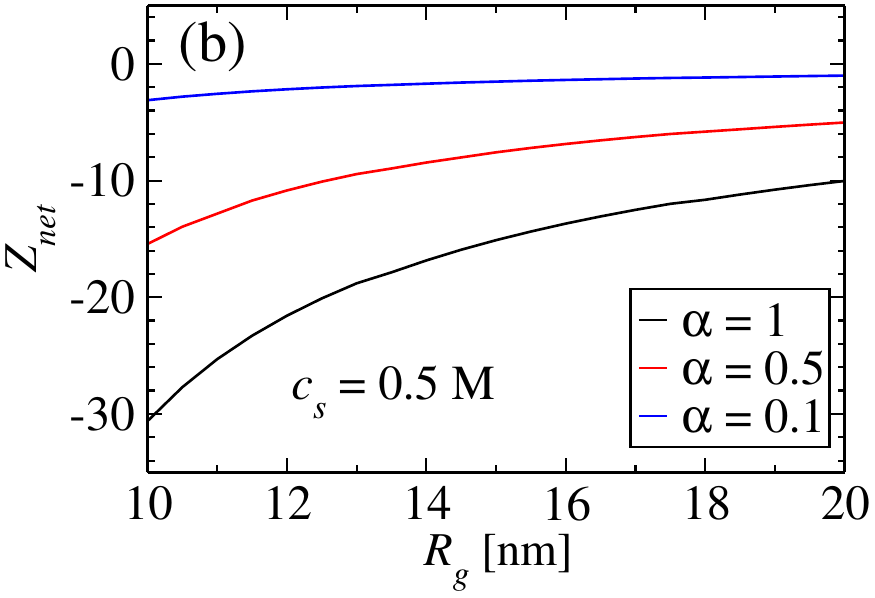}
		\caption{Net charge $Z_{net}$ of the star polymer, comprising the fixed polymer and surface charge, in addition to the charge of mobile ions adsorbed into this region. All grafted polymers are made of $M=50$ monomers, and the grafting core has a surface charge of $\sigma =0.01$~C/m$^2$. The ionic reservoir concentrations are $c_s=0.15$~M (a) and $c_s=0.5$~M (b).}
		\label{fig:fig5}
	\end{figure}

	As a final measurement of the ionic equilibrium distributions across the polymer-reservoir interface, we show in Fig.~\ref{fig:fig5} the net charge of the star polymer -- defined as the overall charge lying within a distance $R_g$ from the core center -- as a function of $R_g$. This quantity always decreases (in absolute value) as the chains become larger, reflecting the reduction in energy cost for stretched chain configurations. This mechanism is also indicated in Fig.~\ref{fig:fig4}, where the adsorption coefficients tend to a constant at large system sizes. While at small dissociation fractions the net charge is roughly independent of ionic concentrations, at larger polymer charges the salt concentration plays an important role in determining the degree of ionic condensation that partially neutralize the polymer charges. Despite the strong counterion adsorption in the case of small salt concentrations and strongly charged polymers, the overall charge of the star polymer is still very high in this regime, particularly at smaller radius of gyrations (see black curves in Figs.~\ref{fig:fig4}a and \ref{fig:fig5}a). This points to a poor screening of polymer charges, along with a strong entropic penalty in this regime. These effects should clearly result in a larger stretching force that pushes the polymer chains toward regions of larger radius of gyrations. \textcolor{black}{We also notice that the mechanisms responsible for ionic adsorption can be enhanced \textit{via} size correlation effects by considering ions with different sizes. The size asymmetry can be tuned to control the amount of condensed counterions in the polymer networks, as well as the resulting effective charge assigned to the brush as a while (see Supplementary Information for further details).}
	
	\subsection{Polymer conformations}
	
\begin{figure}[h!]
		\centering
		\includegraphics[width=5.5cm,height=4cm]{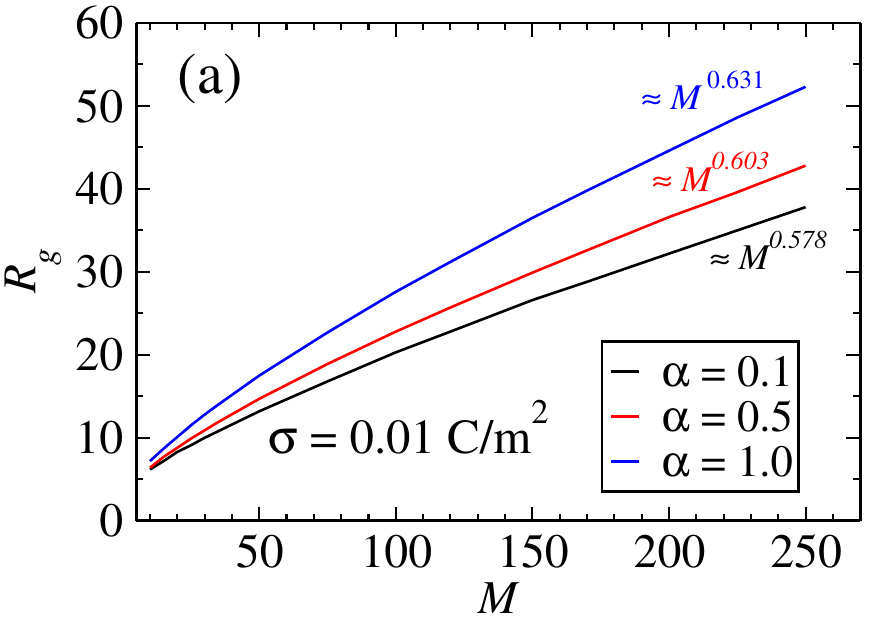}
		\includegraphics[width=5.5cm,height=4cm]{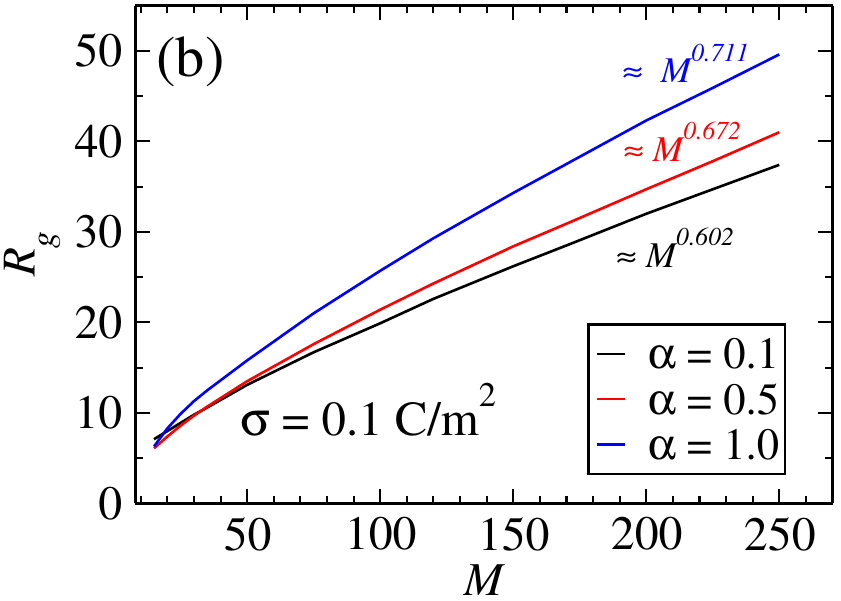}
		\caption{Averaged radius of gyration $R_g$ as a function of the polymer degree of polymerization $M$ at different degrees of dissociation $\alpha$. The core charges are set to be $\sigma=0.01$~C/m$^2$ (a) and $\sigma=0.1$~C/m$^2$ (b), while the ionic reservoir concentration is kept fixed at $c_s=0.15$~M.}
		\label{fig:fig6}
	\end{figure}

	We shall now focus on the equilibrium structure of polymer chains, which in the present approach is characterized by their averaged radius of gyration. This quantity is bound to satisfy Eq. (\ref{el_g}), which is nothing but the condition of vanishing osmotic pressure across the system-reservoir interface. Once again, it is important to notice that the equilibrium sizes $R_g$ implicitly incorporates charge and size effects in both polymer and ion interactions through the proposed cDFT formalism. Our major assumption is that the coarse-graining bead distribution keeps its functional form as given by Eq. (\ref{poly_dist}) as mobile ions relax to their equilibrium positions. Apart from the mechanisms discussed above that prescribe ionic equilibrium, in the case of mechanical stability that controls the averaged star polymer size, another contributions star to play a role: the polymer-core interactions and the chain entropy. The two contributions have a similar effect of preventing strong chain stretching. Since the polymers and the core are oppositely charged, their mutual electrostatic attraction will push the polymer beads towards the main core. This contribution is counter-balanced by ionic condensation and the electrostatic repulsion between equally charged chain monomers. On the other hand, entropic effects as described by a flexible gaussian chain, Eq. (\ref{Fpol}), favors orientational disorder among the links that connect neighboring beads, thus preventing stretching-like chains configurations. These trends can be observed in Fig. \ref{fig:fig6}, which shows the radius of gyration $R_g$ as function of the degree of polymerization $M$ at different dissociation fractions and core surface charges. Here, the salt concentration is fixed at a $c_s=0.15$~M, corresponding to physiological conditions. A rather weak dependence of $R_g$  on the core charge is observed, which can be assigned to a sufficiently strong condensation of anions very close to the core surface, effectively screening the polymer-core attractive interactions. Such a binding mechanism should be even enhanced upon increasing of salt concentration. The star polymer swelling is in this case governed by both steric and charge monomer repulsion, which becomes more pronounced by increasing the degree of polymerization $M$ and fraction of dissociation $\alpha$, respectively.  A similar growth of particle size is observed at sufficiently large $M$ in all cases, in accordance with classical scaling arguments. Fig. \ref{fig:fig6} also shows the scaling behavior of $R_g$ at large $M$. Notice that the scaling exponents are very close to the Daoud-Cotton scaling law $R_g\sim M^{3/5}$~\cite{dao82}. This scaling law can be easily understood in terms of a balance between polymer size and entropic effects. While the size contribution to the osmotic pressure is $\beta\Pi_{hs}\approx \dfrac{B_2}{2}\rho_b^2$, where $B_2=\pi\sigma_b^6/6$ is the second viral coefficient and $\rho_b=3Mf/4\pi R_g^3$ is the averaged polymer concentration, it follows from Eq. (\ref{Fpol}) that the entropic chain contribution is $\beta\Pi_{el}=-3f/(4\pi M\sigma_b^2 R_g)$. The force balance condition $\beta\Pi_{el}+\beta\Pi_{hs}=0$ thus provide the scaling behavior $R_g\sim M^{3/5}$. Since in this regime of large $R_g$ the system is almost charge-neutral, the deviations from this ideal behavior can be assign to the swelling contribution from the condensed counterions. This is why the deviations become more pronounced at the largest dissociation fraction $\alpha=1$.  
	
\begin{figure}[h!]
		\centering
		\includegraphics[width=5.5cm,height=4cm]{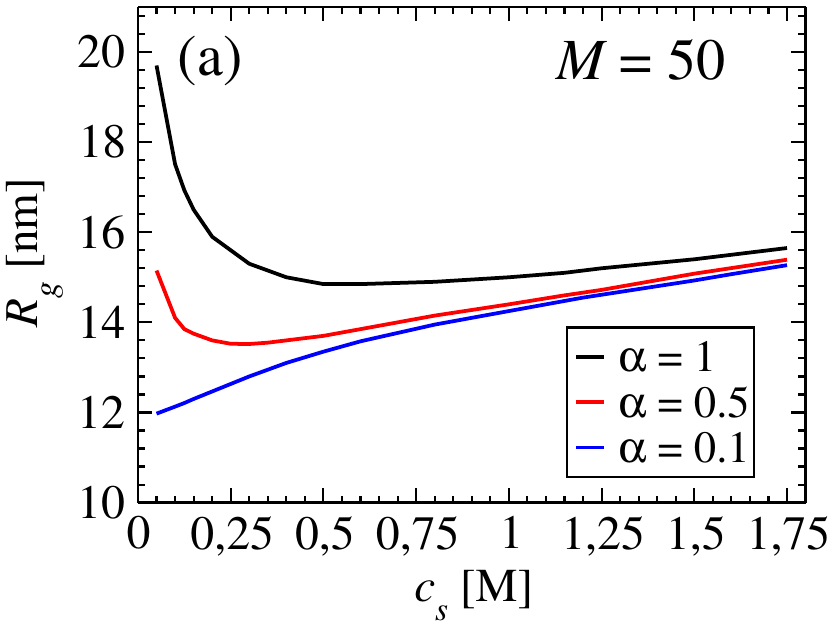}
		\includegraphics[width=5.5cm,height=4cm]{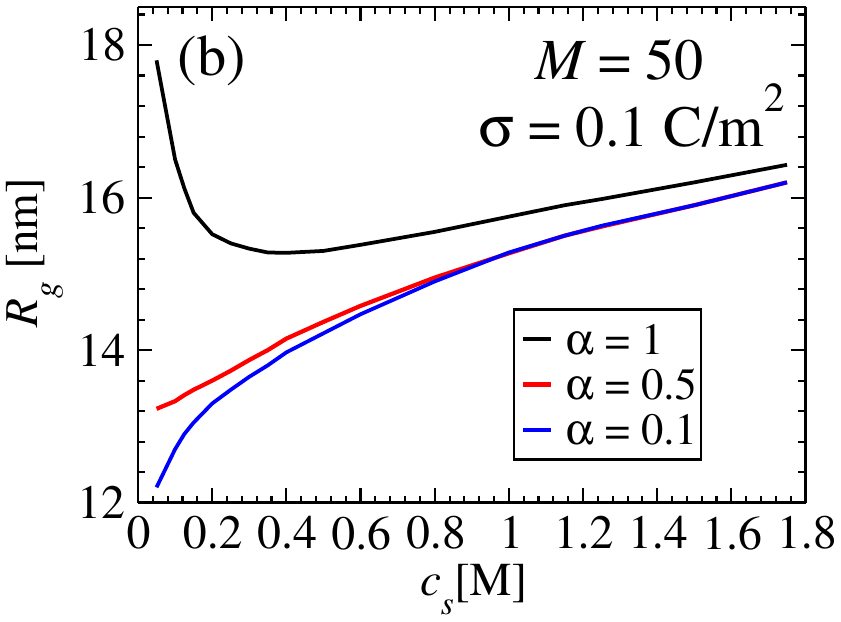}
		\caption{Averaged radius of gyration $R_g$ as a function salt concentration  for different degrees of dissociation $\alpha$ and at a fixed degree of polymerization $M=50$. The core charges are  $\sigma=0.01$~C/m$^2$ (a) and $\sigma=0.1$~C/m$^2$ (b).}
		\label{fig:fig7}
	\end{figure}

	The effects of ionic strengths on the polymer conformations are summarized in Fig.~\ref{fig:fig7}, which shows the radius of gyration as a function of salt concentrations at different degrees of polymerization and two target core surfaces of $\sigma=0.01$~C/m$^2$ (a) and $\sigma=0.1$~C/m$^2$ (b). First, we point out that the polymer sizes does not change significantly as the salt concentration is changed over the wide range from $c_s\approx  0.05$~M up to $c_s\approx  1.80$~M for a given degree of polymerization. Again, this weak dependence suggests that the main mechanisms that control polymer expansion/contraction are polymers steric and electrostatic repulsion, together with entropic effects. At very large salt concentrations, the star-polymer sizes become roughly independent on their charge, and increase linearly with the increasing of salt concentration. In this regime, electrostatic interactions are strongly screened, and particle expansion is dominated by ionic entropy, as salt particles become adsorbed at the polymer chains, thereby providing a stretching force on their beads. In the opposite regime of low salt concentrations, the competition between entropic and electrostatic interactions become an important ingredient in controlling the stable mechanical condition. This is reflected on the non-monotonic behavior of $R_g$ at small salt concentrations, which becomes more clear at small surface charges and higher degrees of dissociation (see black and red curves at Fig~\ref{fig:fig7}a). In this regime, the particle size undergoes a minimum at repetitively small salt concentrations, and then becomes to increase upon reduction of ionic reservoir concentrations. This behavior is to be contrasted with the one at the smallest dissociation degree, $\alpha=0.1$, in which $R_g$ grows monotonically at all ionic strengths. This interesting behavior at small $c_s$ is closely related to the increase in counterion adsorption and net charges reported in Figs. \ref{fig:fig4} and \ref{fig:fig5} at small $R_g$ as the salt concentration decreases. These effects lead to a stretching force that attempts to reduce energy and entropic costs resulting, respectively, from large net charges and strong adsorptions. Since these stretching mechanisms should be enhanced by further decreasing the salt concentration, the polymer chains tend to expand, rather then contract, upon decreasing of the salt concentrations in this regime.  The threshold between electrostatic/entropic balance at low ionic strengths and entropic dominance at large salt concentrations is reflected in the minimum of $R_g$, which is pushed towards larger salt concentrations at stronger dissociations and smaller surface charges. Finally, we remark that the core charge plays a relevant role on the mechanical equilibrium mechanism only at small salt concentrations and relatively large dissociation fractions $\alpha$. 
	
	\subsection{Charge regulation}
	
	Thus far we have considered polymers grafted onto a core of fixed charge. We will now apply the simulation and theoretical techniques presented above to address the charge regulation process when the core surface charge is not know \textit{a priori}.  Instead, this quantity should result from a chemical equilibrium balance which is controlled by solution pH and the number of functional groups attached to the core surface. In our simulations, we consider the colloidal particles with radius $0.8$~nm and $60$ active sites randomly distributed over its surface, in addition to grafted polymers with $M=10$ monomers each.  The colloidal particle is placed inside the cell with $c_a=20$~mM strong acid (considered to be \ch{HCl}) and the sticky length of site has been throughout fixed ay $l_g=50$~nm. Notice that in this case both the grafting core and the polymer chains are negatively charged.  In the theoretical approach, we considered a rescaled spring constant through the replacement  $\sigma_b\rightarrow r_0$ in Eq. (\ref{Fpol}), where $r_0$ is the equilibrium distance between neighboring monomers across the chain, defined in Eq. (\ref{eq1}). This is necessary because Eq. (\ref{Fpol}) is based on a tangentially-connected chain with fixed link length $\sigma_b$, whereas in our simulations the distance between neighboring monomers are able to undergo fluctuations through the bound potential Eq. (\ref{eq1}), with an average monomer-to-monomer distance of $\approx r_0$.
	
	\textcolor{black}{It is important to notice that simulations have been performed in conditions that mimic tethered chains in $\Theta-$solvents. Therefore, their equilibrium configurations will strongly deviate from the stretched chain distribution described by Eq. (1). Since the ionic concentrations (and the resulting charge regulation mechanism) are relatively insensitive to the molecular details of polymer distributions in the limit of small chains, we focus our analysis in this situation, pointing out that in the case of long chains the assumptions underlying Eq. (\ref{poly_dist}) lead to strong deviations from simulation results. In that case, the theory has to be modified to describe chain conformations beyond the stretching-chain limit.}
	
	\begin{figure}[H]
		\centering
		\includegraphics[width=0.7\linewidth]{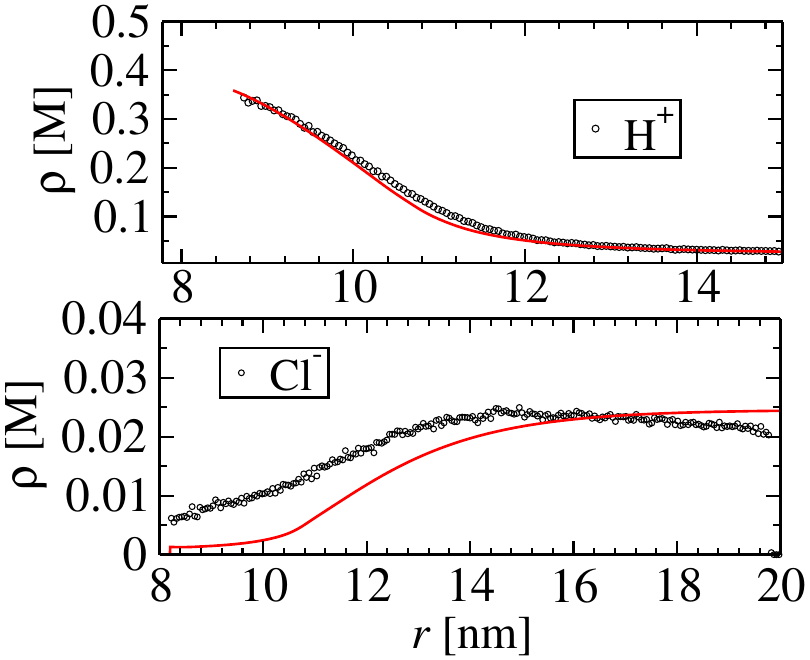}
		\caption{Comparison between simulation result and theory. The concentration of acid inside the cell is $c_a=20mM$. The circles are simulation results, while solid lines are predictions from the \textcolor{black} cDFT. The value of $l_g$ is fixed at $50$~nm, the number of acidic sites is $N_{site}=60$ and that of grafted polyelectrolytes if $f=60$. }
		\label{fig:fig8}
	\end{figure}
\begin{figure}[H]
		\centering
		\includegraphics[width=0.7\linewidth]{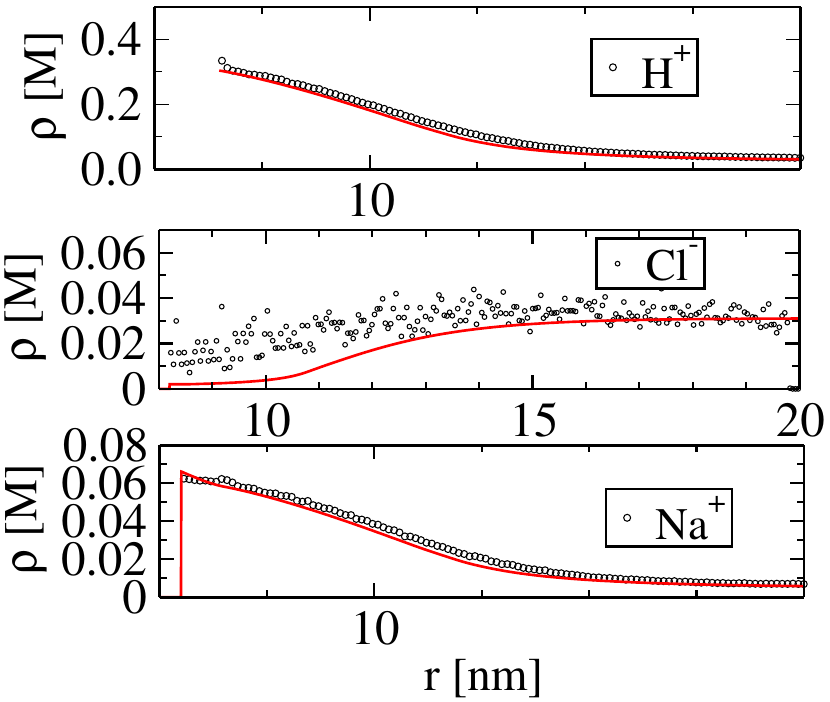}
		\caption{Density profile of \ch{H+} (top panel), \ch{Cl-} (middle panel) and \ch{Na+} (bottom panel). The system contains added salt of bulk concentration $c_s=10$~mM, and the core particle has $N_{site}=60$ acidic sites and $f=60$ grafted polymers. Symbols are data from simulation and solid lines are the predictions from PB theory.}
		\label{fig:fig9}
	\end{figure}

	Figure \ref{fig:fig8} shows ionic density profiles of  \ch{H+} (top panel) and \ch{Cl-} (bottom panel), in the case where the concentration of acid inside the cell is $c_a=20$~mM, with no added salt ($c_s=0$), with a total of $f=60$ grafting polymers. The theory agrees very well with simulation results for the case of dissolved \ch{H+}. However, in the case of \ch{Cl-} larger deviations between theory and simulation become to appear, with the cDFT profiles underestimating the simulation predictions. The deviation of \ch{Cl-} is related to thermal fluctuations that avoids co-ions to relax to their bulk values far away from the polymer chains, requiring the use of a very large simulation cell. In order to study the effect of added salt to the system, we have consider, in Fig.~\ref{fig:fig9}, simulation and theoretical results for the ionic profiles in presence of $c_s=10$~mM  dissolved \ch{NaCl}. Again, a very good agreement is observed for the counterion profiles, although the co-ion distribution is still underestimated by the theoretical approach due to large thermal fluctuations that favors homogeneous distributions of this component all over the simulation box. 

\begin{figure}[H]
		\centering
		\includegraphics[width=0.7\linewidth]{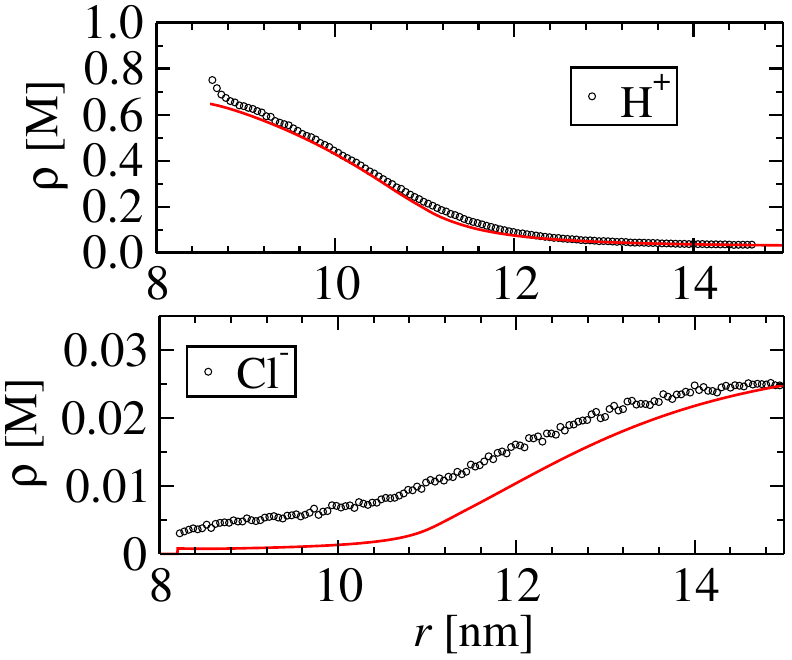}
		\caption{Density profile of ions around a spherical core comprising $N_{site}=60$ functional sites and  $f=120$ grafted polymers, each carrying a total of  $M= 10$ connected monomers. The number of active sites on the core surface is $N_{site}= 60$. Top panel shows the distribution of \ch{H+} ions, while the bottom panel displays the density profile of \ch{Cl-} ions.The solid lines are prediction of cDFT and circles are the result of MC simulation.}
		\label{fig:fig10}
	\end{figure}

\begin{figure}[H]
		\centering
		\includegraphics[width=0.7\linewidth]{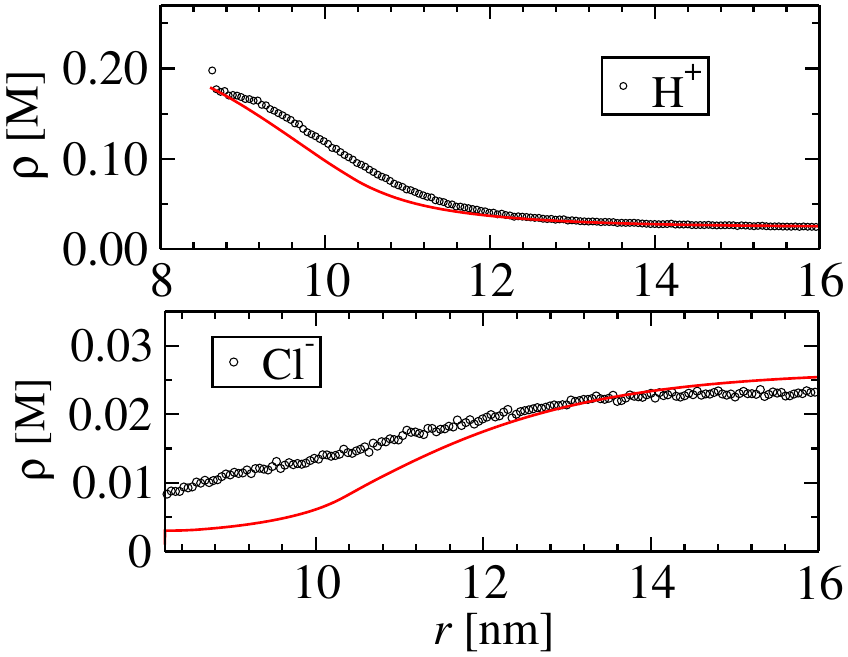}
		\caption{Density profile of ions around a spherical core. The parameters are same as in Fig~\ref{fig:fig10}, but now with a core possessing a total of $f=30$ grafted polymers.}
		\label{fig:fig11}
	\end{figure}

	It is interesting to check the effects of changing the number of grafting polymers $f$ on the ionic distribution and core effective charge. These effects are summarized in Figs.~\ref{fig:fig10} and \ref{fig:fig11}. In Fig. ~\ref{fig:fig10}, we consider $f=120$ grafted polymers with $M=10$ monomers each, grafted onto a core made of $60$ active sites polymers. Again, the strong acid concentration is $c_a=20$~mM, with $c_s=10$~mM added salt. The larger number of grafted polymer results in an exclusion volume barrier that prevents hydronium  ions to reach the active sites, thus decreasing their reactions with active sites, which leads to a large effective charge at the core surface. As a result, the counterion adsorption into the core gets larger, in spite of the strong steric effects due to the high grafting surface. In Fig.  ~\ref{fig:fig11}, density profiles are shown for the case of a small number of grafted polymers, $f=30$, attached to a core containing $N_{site}=60$ active sites. We now observe a smaller concentration of hydronium ions close to the core surface. This is clearly due to the reduction in steric repulsion between the grafted polymers and hydronium ions, which are now able to get close enough to the surface and link together with the available sites. As a result, the effective charge is decreased with respect to the case of larger densities of grafted polymers. 

	\section{Conclusions and final remarks}

	In this work, we have studied the equilibrium ionic distributions, as well as the averaged sizes of polyelectrolytes attached to a central charge core (star-like ionic polymer brushes). To this end, we have employed a cDFT approach which considers a fixed charged surface assigned to the core, followed by a recently developed Monte Carlo technique in which the core charge is not known \textit{a priori}, and is rather obtained as an output from a charge regulation process that mimics ionic association. Our cDFT approach has been worked out in a two-step minimization approach in which optimal ionic profiles are obtained via an Euler-Lagrange condition under the constrain of fixed, stretched-like chain configurations. This is followed by a mechanical equilibrium condition which allows one to compute an equilibrium radius of gyration that characterizes the family of stretched chain conformations under consideration. We then proceeded to analyze the ionic distributions around the polymer chains, along with the equilibrium size of these objects over different regions in parameter space. We have shown that, while osmotic equilibrium is mainly controlled by competing electrostatic and entropic effects, the mechanical stability depends crucially on the interplay between steric, electrostatic and elastic (ideal chain) interactions.  While at large salt concentration stretching with added salt is dictated mostly by counterion entropy, at small ionic strengths the fine balance between electrostatic and entropic interactions is shown to have a non-trivial effect on the equilibrium conformations. We have also checked that the size and electrostatic correlations incorporated into the cDFT approach are key ingredients to properly describe the fine tuning between charge and size effects, leading to ionic profiles that considerably deviate from mean field predictions. The surface regulation process that controls the core surface charge was then investigated using a previously reported Baxter sticky model for describing the chemical reaction between acidic groups and dissolved hydronium ions of given concentrations~\cite{bk2019,D0CP03633A}.  In particular, the effects of added salt and grafting density have been investigated on the underlying ionic distributions. It was shown that the grafting of  polyelectrolytes increases the adsorption of counterions close to the core surface since the sticky surface sites become increasingly inaccessible due to steric interactions with the grafted polymers.
	
	One key assumption in the approach proposed herein is that the grafted chains always retain a unfold configuration, regardless of the external conditions. While this constraint can be accurate enough in the case of $\Theta$ and good solvents, the presence of attractive effective interactions is known to induce coil-like configurations in the polymer chains, which clearly go beyond the scope of the proposed model. Besides, the assumption of fully stretched beads might be rather questionable at close distances to the grafting surface, especially when the grafting density becomes sufficiently high. This is due to strong steric effects in this region, which will induce layering-like monomer profiles that deviate from the inverse square power law underlying the assumption of fully stretched chains~\cite{xu10}. Therefore,  extensions of this work might include a larger number class of polymer conformations. This can be achieved in the context of the cDFT, where a bound, sticky-like potential is introduced between neighboring beads of the same chain, while inter-chain interactions are incorporated \textit{via} steric and electrostatic interactions~\cite{wu07,xu10,jiang18}.  The free energy of polymerization is also usually included in different levels of approximation. We also notice that the proposed model can be easily extended to address the question of macroions uptake/releasement at the polymer network, a key point related to many practical applications. \textcolor{black}{As a final remark, we note that the present model can be combined with conventional methods, like the one proposed by Szleifer and co-workers~\cite{Szleifer1,Szleifer2}, to incorporate charge regulation effects at the tethered polyelectrolytes.}

	\section{Supplementary Information}
	\section{Polymer brushes in the presence of an asymmetric electrolyte}
	
	The present material contains additional results aiming to summarize the overall effects of ion size asymmetry, a topic not discussed in the main text. Particular focus is placed on the role played by size-effects on ionic structure and their adsorption onto the polymer network. In order to access these effects, we have fixed the anion diameter at $\sigma_-=2r_-=0.4$~nm, while the cation size has been set to $\sigma_{+}=2r_{+}=0.4$~nm,  and $\sigma_{+}=2r_+=0.8$~nm. The core has radius $a=4$~nm and is positively charged, with a fixed charge surface of $\sigma=0.1$~C/m$^2$. There are $f=10$ grafted polymers, each comprised of $M=30$ monomers of radii $r_{mon}=0.5$~nm and charge $-q$ (full dissociation regime).

	\begin{figure}[h!]
		\centering
		\includegraphics[width=4.6cm,height=3.5cm]{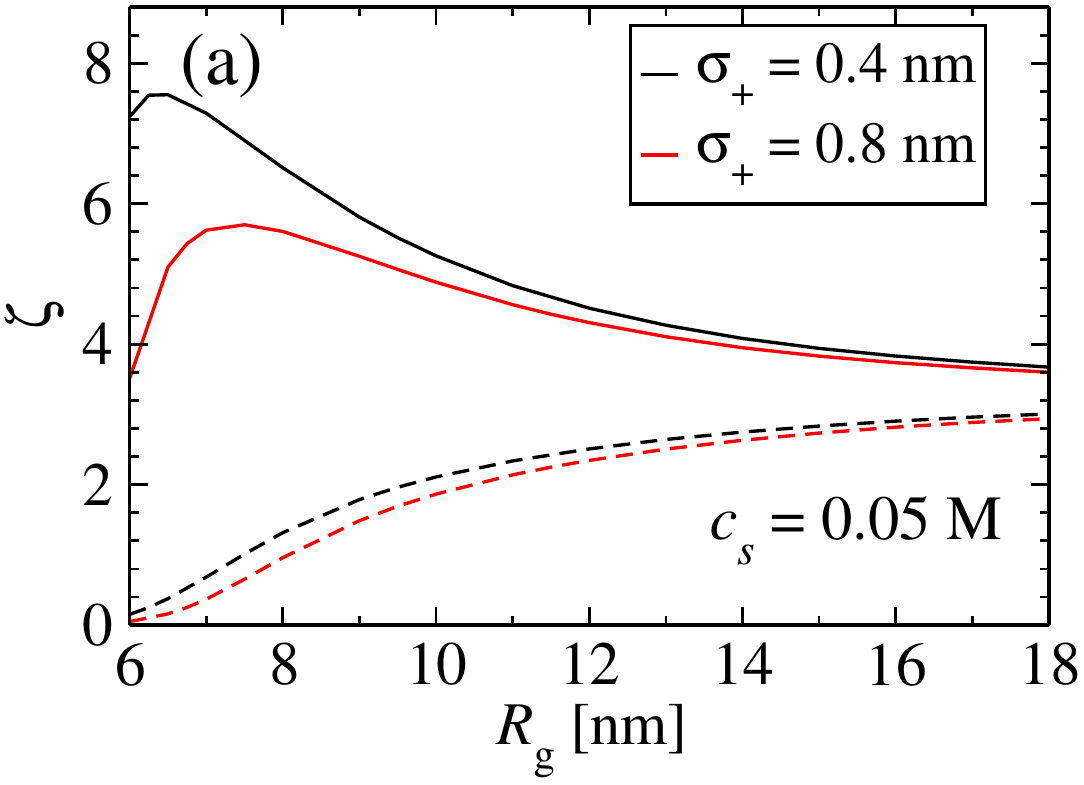}
		\includegraphics[width=4.6cm,height=3.5cm]{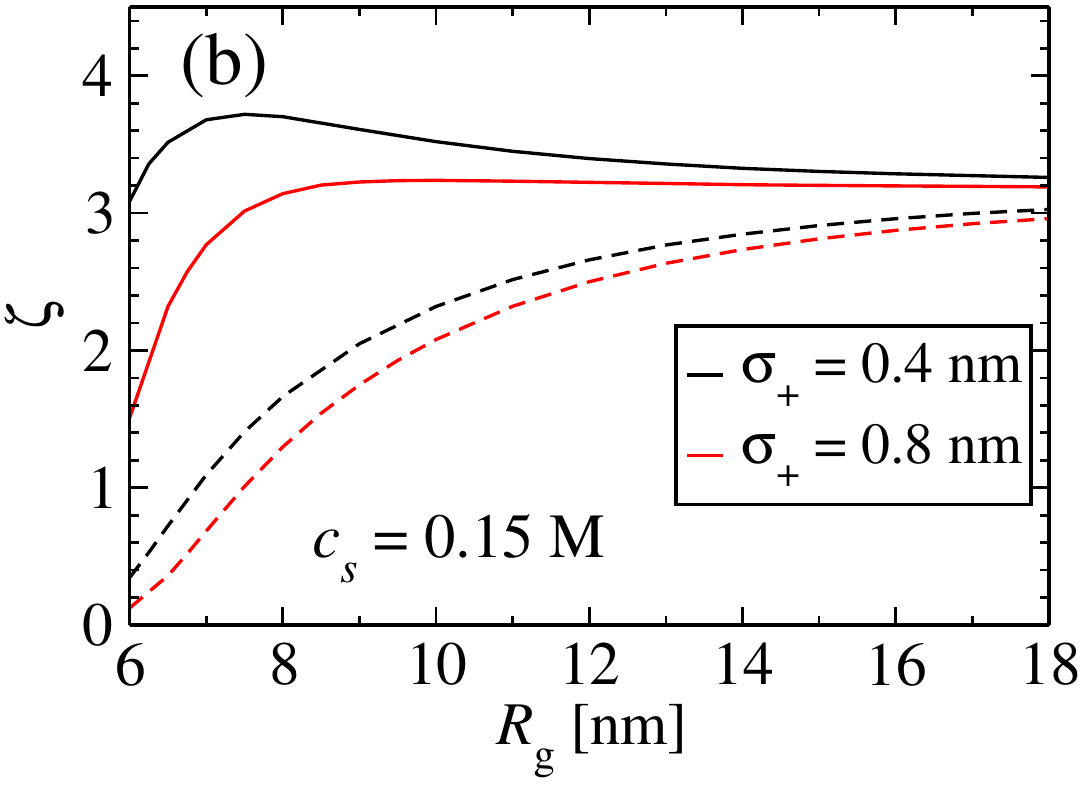}
		\includegraphics[width=4.6cm,height=3.5cm]{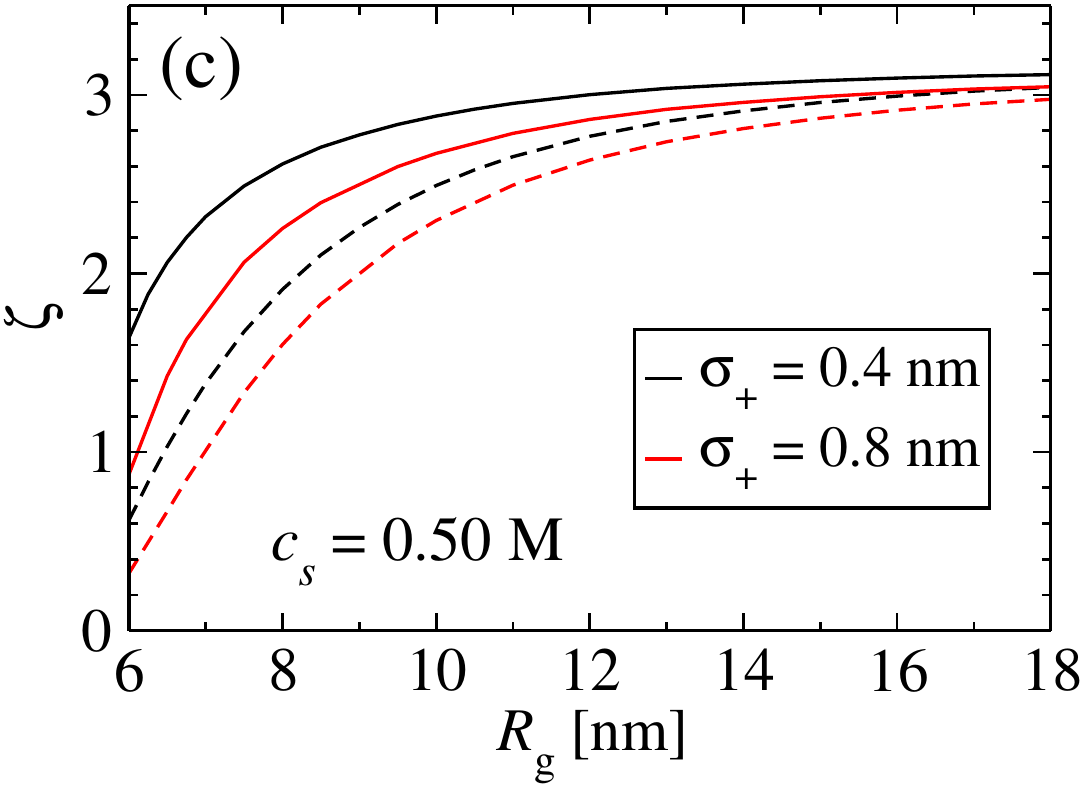}
		\caption{Adsorption coefficients as functions of brush size, at different salt concentrations and considering cations of diameter $\sigma_+= 0.4$~nm (black curves) and $\sigma_+= 0.8$~nm (red curves), keeping the anion size fixed at $\sigma_-=0.4$~nm. Full curves represent cation adsorptions, while dot curves stand for anion adsorption coefficients. The bulk ionic concentrations are $c_s=0.05$~M (a), $c_s=0.15$~M (b) and $c_s=0.5$~M (c).}
		\label{fig:SI_fig1}
	\end{figure}

	\begin{figure}[h!]
		\centering
		\includegraphics[width=4.9cm,height=3.5cm]{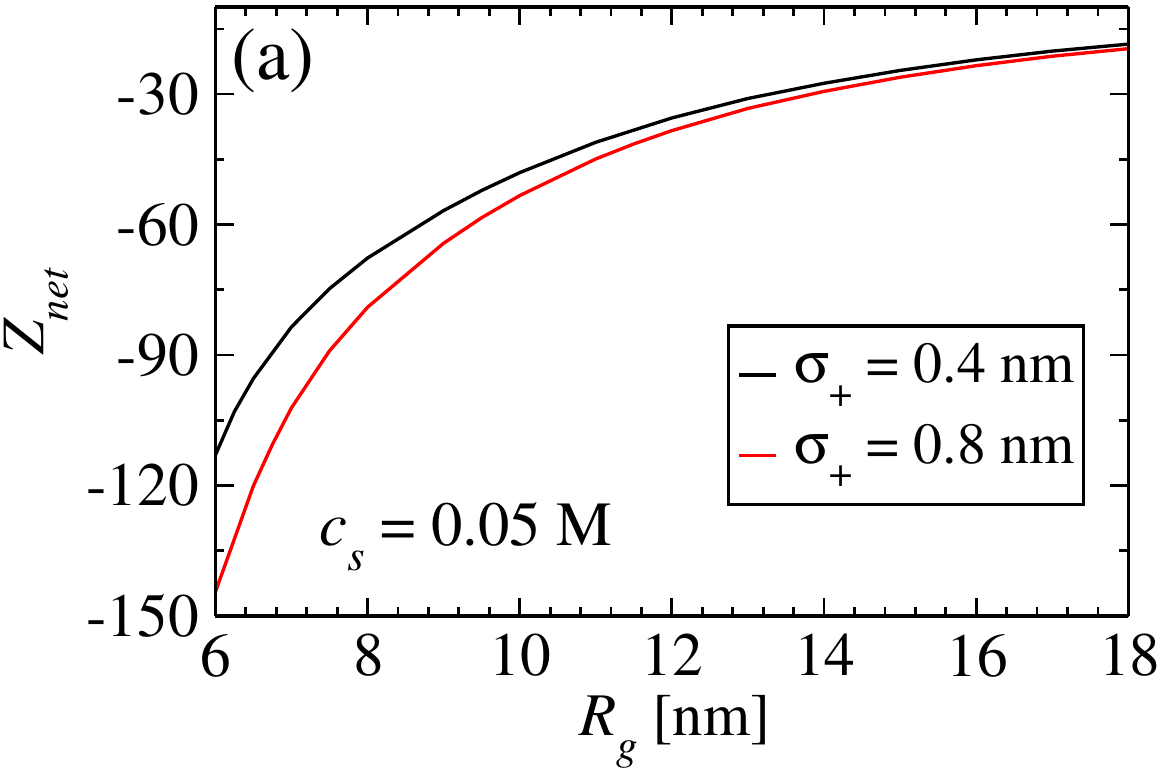}
		\includegraphics[width=4.9cm,height=3.5cm]{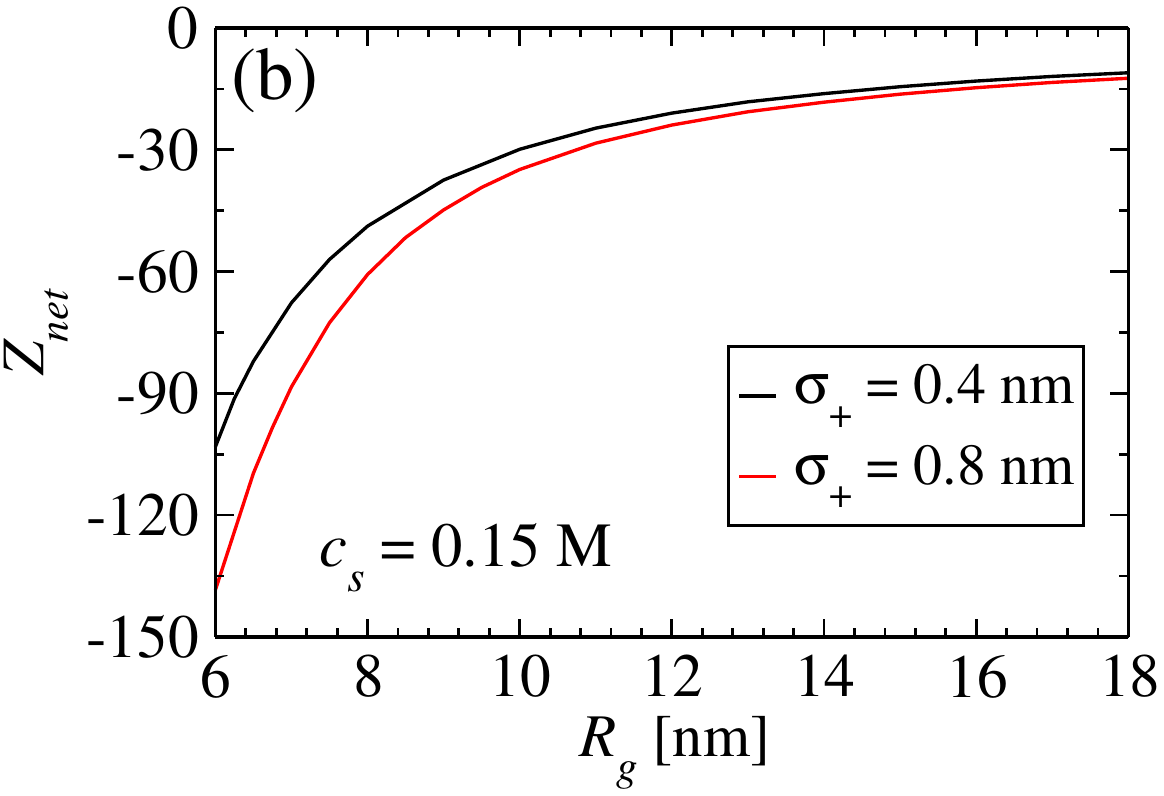}
		\includegraphics[width=4.9cm,height=3.5cm]{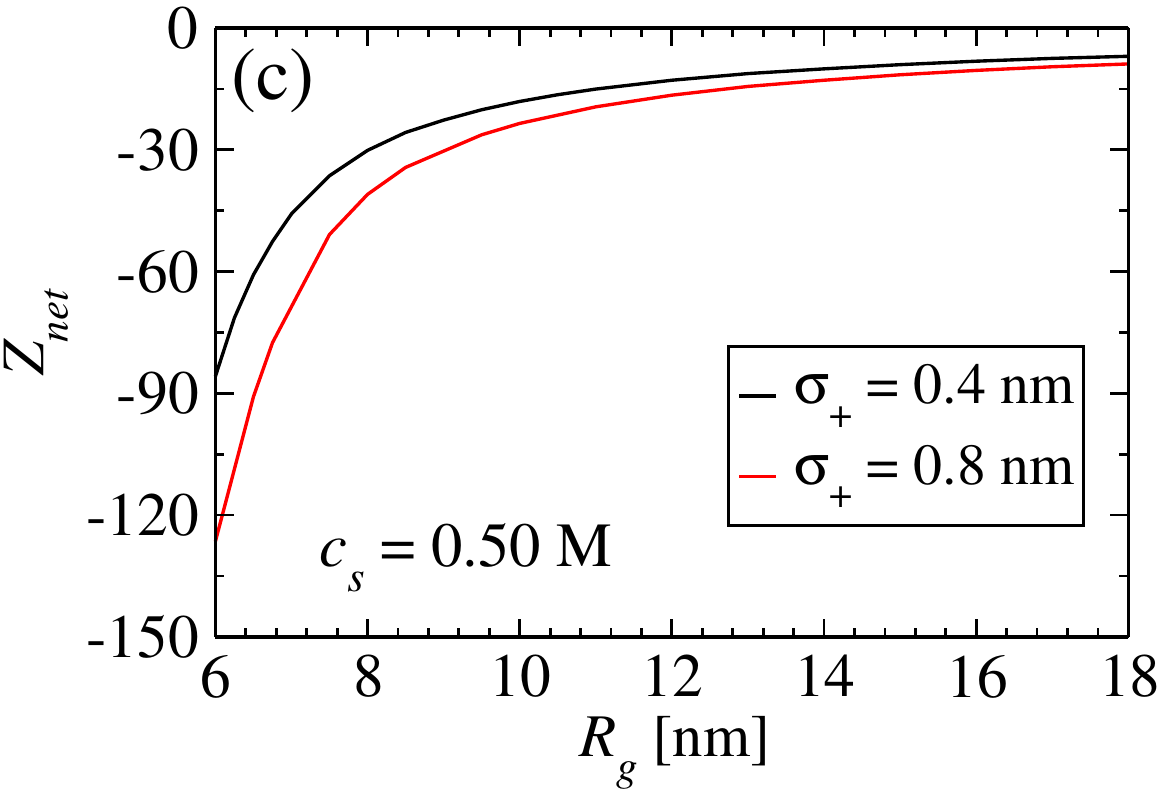}
		\caption{Brush net charge as a function of brush size, at different salt concentrations and considering cations of diameter $\sigma_+= 0.4$~nm (black curves) and $\sigma_+= 0.8$~nm (red curves), keeping the anion size fixed at $\sigma_-=0.4$~nm. The bulk ionic concentrations are $c_s=0.05$~M (a), $c_s=0.15$~M (b) and $c_s=0.5$~M (c).}
		\label{fig:SI_fig2}
	\end{figure}
	
	When a region of strong confinement is in equilibrium with an ionic reservoir, ion size and charge asymmetries are known to play a major role in determining the adsorption of different ionic components into this region, an effect generally known as ionic selectivity. In the case of crowded environments, exclusion volume effects favor a stronger adsorption of smaller components, which on the other hand might lead to a charge unbalance and local breakdown of electroneutrality. Apart from its direct influence on size correlations, the changing in ionic sizes also impacts electrostatic correlations. This can be quantified \textit{via} the so-called coupling parameter, defined as the ratio between electrostatic and thermal energies at closest ionic contact. As the ionic radii increase, this parameter becomes smaller, resulting in weaker electrostatic positional correlations. While this effect should favor a stronger adsorption of equally charged ions (e. g., at the vicinity of an oppositely charged substrates) upon increase of their size, exclude volume effects try to prevent such strong local packing. In the present situation of ions in contact with tethered polyelectrolytes, these combined effects should result in a smaller selective adsorption of cations into the brush region when the cation size increases. While the gain in electrostatic energy resulting from bringing a cation in a closest contact to a oppositely charged monomer decreases, their mutual hard-sphere repulsion becomes more pronounced, both effects leading to a smaller adsorption onto the polymer networks. These effects are summarized in Fig. \ref{fig:SI_fig1} and  \ref{fig:SI_fig2}, in which the adsorption coefficients and brush net charges, respectively, are shown as function of the brush size for different ionic concentrations. At small particle sizes, the crowded environment within the polymer networks results in a smaller particle adsorption upon increase of cation sizes. A similar behavior is observed for anions, though in a smaller level due to their reduced size. As the brush size progressively increases, the cations find more room available to move in, such that size asymmetry effects become negligible.  This results in a decrease of the brush net charge, as the amount of condensed cations is not enough to neutralize the polyelectrolyte charges. 
	
\begin{figure}[h!]
		\centering
		\includegraphics[width=4.9cm,height=3.5cm]{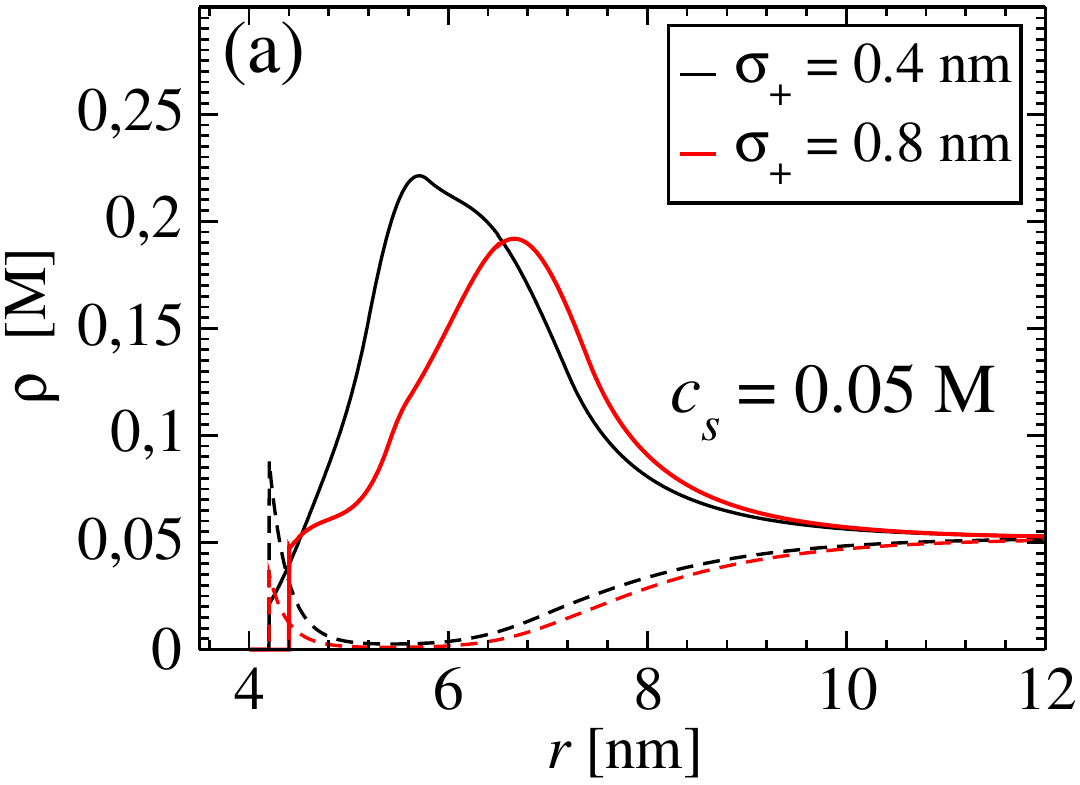}
		\includegraphics[width=4.9cm,height=3.5cm]{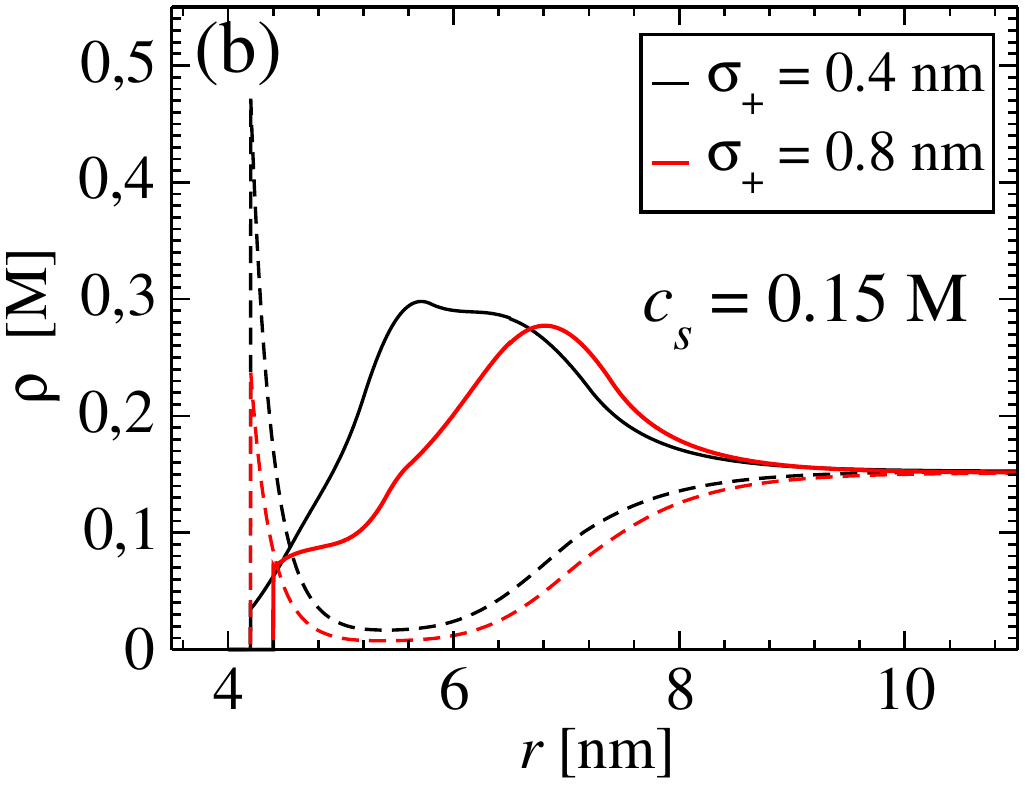}
		\includegraphics[width=4.9cm,height=3.5cm]{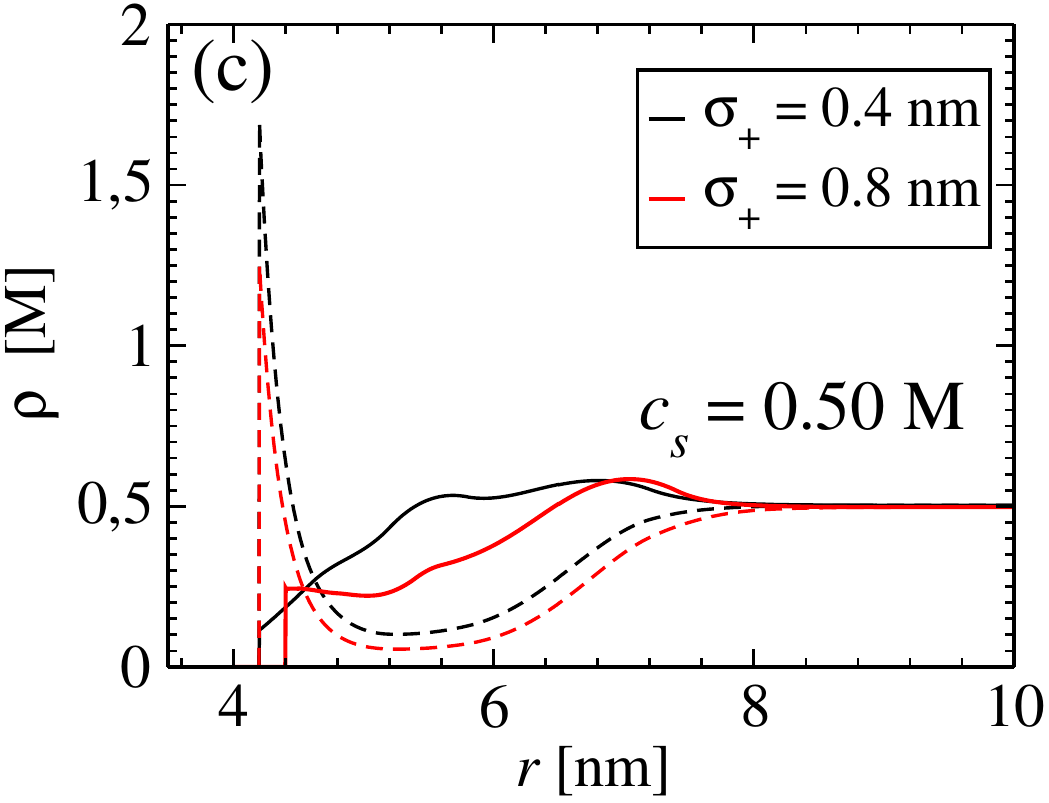}
		\caption{Ionic profiles at different salt concentrations, and considering cations of diameter $\sigma_+= 0.4$~nm (black curves) and $\sigma_+= 0.8$~nm (red curves), keeping the anion size fixed at $\sigma_-=0.4$~nm. Full curves represent cation profiles, while dot curves are anion local densities. The bulk ionic concentrations are $c_s=0.05$~M (a), $c_s=0.15$~M (b) and $c_s=0.5$~M (c).}
		\label{fig:SI_fig3}
	\end{figure}

	The effects of size asymmetries in local ionic structure can be observed in Fig. \ref{fig:SI_fig3}, in which ionic profiles for both symmetric and asymmetric electrolytes are compared at different size asymmetries, for a brush size of $R_g=6.5$~nm. We notice that cations are moved away from the core region as their size is increased. This is clearly due to steric effects, as the outer regions of the brush process smaller local packing fractions. It is also interesting to note that the increase of cation size has a major effect on counterion (i. e., anion) adsorption onto the core surface. This can be interpreted as indirect effect: as cations are spelled out from the core region, the monomers at that region are not neutralized by condensed counterions (cations), and will therefore provide a strong repulsion for anions approaching this region. Even though the net charges are not strongly affected by size asymmetry, we conclude that the local charge profiles can be strongly modified within the brush network. These effects might be particularly relevant for controlling the adsorption of suspended particles onto the polymer backbones.

\begin{thebibliography}{109}%
	\makeatletter
	\providecommand \@ifxundefined [1]{%
		\@ifx{#1\undefined}
	}%
	\providecommand \@ifnum [1]{%
		\ifnum #1\expandafter \@firstoftwo
		\else \expandafter \@secondoftwo
		\fi
	}%
	\providecommand \@ifx [1]{%
		\ifx #1\expandafter \@firstoftwo
		\else \expandafter \@secondoftwo
		\fi
	}%
	\providecommand \natexlab [1]{#1}%
	\providecommand \enquote  [1]{``#1''}%
	\providecommand \bibnamefont  [1]{#1}%
	\providecommand \bibfnamefont [1]{#1}%
	\providecommand \citenamefont [1]{#1}%
	\providecommand \href@noop [0]{\@secondoftwo}%
	\providecommand \href [0]{\begingroup \@sanitize@url \@href}%
	\providecommand \@href[1]{\@@startlink{#1}\@@href}%
	\providecommand \@@href[1]{\endgroup#1\@@endlink}%
	\providecommand \@sanitize@url [0]{\catcode `\\12\catcode `\$12\catcode
		`\&12\catcode `\#12\catcode `\^12\catcode `\_12\catcode `\%12\relax}%
	\providecommand \@@startlink[1]{}%
	\providecommand \@@endlink[0]{}%
	\providecommand \url  [0]{\begingroup\@sanitize@url \@url }%
	\providecommand \@url [1]{\endgroup\@href {#1}{\urlprefix }}%
	\providecommand \urlprefix  [0]{URL }%
	\providecommand \Eprint [0]{\href }%
	\providecommand \doibase [0]{http://dx.doi.org/}%
	\providecommand \selectlanguage [0]{\@gobble}%
	\providecommand \bibinfo  [0]{\@secondoftwo}%
	\providecommand \bibfield  [0]{\@secondoftwo}%
	\providecommand \translation [1]{[#1]}%
	\providecommand \BibitemOpen [0]{}%
	\providecommand \bibitemStop [0]{}%
	\providecommand \bibitemNoStop [0]{.\EOS\space}%
	\providecommand \EOS [0]{\spacefactor3000\relax}%
	\providecommand \BibitemShut  [1]{\csname bibitem#1\endcsname}%
	\let\auto@bib@innerbib\@empty
	\bibitem [{\citenamefont {Rubinstein}\ and\ \citenamefont
		{Ralph~H.}(2003)}]{pol_phys}%
	\BibitemOpen
	\bibfield  {author} {\bibinfo {author} {\bibfnamefont {M.}~\bibnamefont
			{Rubinstein}}\ and\ \bibinfo {author} {\bibfnamefont {C.}~\bibnamefont
			{Ralph~H.}},\ }\href@noop {} {\emph {\bibinfo {title} {Polymer Physics}}},\
	Vol.~\bibinfo {volume} {1}\ (\bibinfo  {publisher} {Oxford University
		Press},\ \bibinfo {address} {New York},\ \bibinfo {year} {2003})\BibitemShut
	{NoStop}%
	\bibitem [{\citenamefont {Likos}(2006)}]{lik06}%
	\BibitemOpen
	\bibfield  {author} {\bibinfo {author} {\bibfnamefont {C.~N.}\ \bibnamefont
			{Likos}},\ }\href@noop {} {\bibfield  {journal} {\bibinfo  {journal} {Soft
				Matter}\ }\textbf {\bibinfo {volume} {2}},\ \bibinfo {pages} {478} (\bibinfo
		{year} {2006})}\BibitemShut {NoStop}%
	\bibitem [{\citenamefont {{Lo Verso}}\ and\ \citenamefont
		{Likos}(2008)}]{ver08}%
	\BibitemOpen
	\bibfield  {author} {\bibinfo {author} {\bibfnamefont {F.}~\bibnamefont {{Lo
					Verso}}}\ and\ \bibinfo {author} {\bibfnamefont {C.~N.}\ \bibnamefont
			{Likos}},\ }\href@noop {} {\bibfield  {journal} {\bibinfo  {journal}
			{Polymer}\ }\textbf {\bibinfo {volume} {49}},\ \bibinfo {pages} {1425}
		(\bibinfo {year} {2008})}\BibitemShut {NoStop}%
	\bibitem [{\citenamefont {Hadjichristidis}\ \emph {et~al.}(2001)\citenamefont
		{Hadjichristidis}, \citenamefont {Pitsikalis}, \citenamefont {Pispas},\ and\
		\citenamefont {Iatrou}}]{hadj01}%
	\BibitemOpen
	\bibfield  {author} {\bibinfo {author} {\bibfnamefont {N.}~\bibnamefont
			{Hadjichristidis}}, \bibinfo {author} {\bibfnamefont {M.}~\bibnamefont
			{Pitsikalis}}, \bibinfo {author} {\bibfnamefont {S.}~\bibnamefont {Pispas}},
		\ and\ \bibinfo {author} {\bibfnamefont {H.}~\bibnamefont {Iatrou}},\
	}\href@noop {} {\bibfield  {journal} {\bibinfo  {journal} {Chemical Reviews}\
		}\textbf {\bibinfo {volume} {101}},\ \bibinfo {pages} {3747} (\bibinfo {year}
		{2001})}\BibitemShut {NoStop}%
	\bibitem [{\citenamefont {Lutz}\ \emph {et~al.}(2016)\citenamefont {Lutz},
		\citenamefont {Lehn}, \citenamefont {Meijer},\ and\ \citenamefont
		{Matyjaszewski}}]{lutz16}%
	\BibitemOpen
	\bibfield  {author} {\bibinfo {author} {\bibfnamefont {J.-F.}\ \bibnamefont
			{Lutz}}, \bibinfo {author} {\bibfnamefont {J.-M.}\ \bibnamefont {Lehn}},
		\bibinfo {author} {\bibfnamefont {E.~W.}\ \bibnamefont {Meijer}}, \ and\
		\bibinfo {author} {\bibfnamefont {K.}~\bibnamefont {Matyjaszewski}},\
	}\href@noop {} {\bibfield  {journal} {\bibinfo  {journal} {Nature Reviews
				Materials}\ }\textbf {\bibinfo {volume} {1}},\ \bibinfo {pages} {16024}
		(\bibinfo {year} {2016})}\BibitemShut {NoStop}%
	\bibitem [{\citenamefont {Frank-Kamenetskii}, \citenamefont {Lukashin},\ and\
		\citenamefont {Vologodskii}(1975)}]{fran75}%
	\BibitemOpen
	\bibfield  {author} {\bibinfo {author} {\bibfnamefont {M.~D.}\ \bibnamefont
			{Frank-Kamenetskii}}, \bibinfo {author} {\bibfnamefont {A.~V.}\ \bibnamefont
			{Lukashin}}, \ and\ \bibinfo {author} {\bibfnamefont {A.~V.}\ \bibnamefont
			{Vologodskii}},\ }\href@noop {} {\bibfield  {journal} {\bibinfo  {journal}
			{Nature}\ }\textbf {\bibinfo {volume} {258}},\ \bibinfo {pages} {398–402}
		(\bibinfo {year} {1975})}\BibitemShut {NoStop}%
	\bibitem [{\citenamefont {Narros}, \citenamefont {Moreno},\ and\ \citenamefont
		{Likos}(2010)}]{nar10}%
	\BibitemOpen
	\bibfield  {author} {\bibinfo {author} {\bibfnamefont {A.}~\bibnamefont
			{Narros}}, \bibinfo {author} {\bibfnamefont {A.~J.}\ \bibnamefont {Moreno}},
		\ and\ \bibinfo {author} {\bibfnamefont {C.~N.}\ \bibnamefont {Likos}},\
	}\href@noop {} {\bibfield  {journal} {\bibinfo  {journal} {Soft Matter}\
		}\textbf {\bibinfo {volume} {6}},\ \bibinfo {pages} {2435} (\bibinfo {year}
		{2010})}\BibitemShut {NoStop}%
	\bibitem [{\citenamefont {Slimani}\ \emph {et~al.}(2014)\citenamefont
		{Slimani}, \citenamefont {Bacova}, \citenamefont {Bernabei}, \citenamefont
		{Narros}, \citenamefont {Likos},\ and\ \citenamefont {Moreno}}]{sli14}%
	\BibitemOpen
	\bibfield  {author} {\bibinfo {author} {\bibfnamefont {M.~Z.}\ \bibnamefont
			{Slimani}}, \bibinfo {author} {\bibfnamefont {P.}~\bibnamefont {Bacova}},
		\bibinfo {author} {\bibfnamefont {M.}~\bibnamefont {Bernabei}}, \bibinfo
		{author} {\bibfnamefont {A.}~\bibnamefont {Narros}}, \bibinfo {author}
		{\bibfnamefont {C.~N.}\ \bibnamefont {Likos}}, \ and\ \bibinfo {author}
		{\bibfnamefont {A.~J.}\ \bibnamefont {Moreno}},\ }\href@noop {} {\bibfield
		{journal} {\bibinfo  {journal} {ACS Macro Letters}\ }\textbf {\bibinfo
			{volume} {3}},\ \bibinfo {pages} {611} (\bibinfo {year} {2014})}\BibitemShut
	{NoStop}%
	\bibitem [{\citenamefont {Poier}, \citenamefont {Likos},\ and\ \citenamefont
		{Matthews}(2014)}]{poi14}%
	\BibitemOpen
	\bibfield  {author} {\bibinfo {author} {\bibfnamefont {P.}~\bibnamefont
			{Poier}}, \bibinfo {author} {\bibfnamefont {C.~N.}\ \bibnamefont {Likos}}, \
		and\ \bibinfo {author} {\bibfnamefont {R.}~\bibnamefont {Matthews}},\
	}\href@noop {} {\bibfield  {journal} {\bibinfo  {journal} {Macromolecules}\
		}\textbf {\bibinfo {volume} {47}},\ \bibinfo {pages} {3394} (\bibinfo {year}
		{2014})}\BibitemShut {NoStop}%
	\bibitem [{\citenamefont {Koch}\ \emph {et~al.}(2015)\citenamefont {Koch},
		\citenamefont {Panagiotopoulos}, \citenamefont {Lo~Verso},\ and\
		\citenamefont {Likos}}]{koch15}%
	\BibitemOpen
	\bibfield  {author} {\bibinfo {author} {\bibfnamefont {C.}~\bibnamefont
			{Koch}}, \bibinfo {author} {\bibfnamefont {A.~Z.}\ \bibnamefont
			{Panagiotopoulos}}, \bibinfo {author} {\bibfnamefont {F.}~\bibnamefont
			{Lo~Verso}}, \ and\ \bibinfo {author} {\bibfnamefont {C.~N.}\ \bibnamefont
			{Likos}},\ }\href@noop {} {\bibfield  {journal} {\bibinfo  {journal} {Soft
				Matter}\ }\textbf {\bibinfo {volume} {11}},\ \bibinfo {pages} {3530}
		(\bibinfo {year} {2015})}\BibitemShut {NoStop}%
	\bibitem [{\citenamefont {Poier}\ \emph {et~al.}(2015)\citenamefont {Poier},
		\citenamefont {Likos}, \citenamefont {Moreno},\ and\ \citenamefont
		{Blaak}}]{poi15}%
	\BibitemOpen
	\bibfield  {author} {\bibinfo {author} {\bibfnamefont {P.}~\bibnamefont
			{Poier}}, \bibinfo {author} {\bibfnamefont {C.~N.}\ \bibnamefont {Likos}},
		\bibinfo {author} {\bibfnamefont {A.~J.}\ \bibnamefont {Moreno}}, \ and\
		\bibinfo {author} {\bibfnamefont {R.}~\bibnamefont {Blaak}},\ }\href@noop {}
	{\bibfield  {journal} {\bibinfo  {journal} {Macromolecules}\ }\textbf
		{\bibinfo {volume} {48}},\ \bibinfo {pages} {4983} (\bibinfo {year}
		{2015})}\BibitemShut {NoStop}%
	\bibitem [{\citenamefont {Doi}(1996)}]{int_pol_phys}%
	\BibitemOpen
	\bibfield  {author} {\bibinfo {author} {\bibfnamefont {M.}~\bibnamefont
			{Doi}},\ }\href@noop {} {\emph {\bibinfo {title} {Introduction to Polymer
				Physics}}}\ (\bibinfo  {publisher} {Oxford University Press},\ \bibinfo
	{address} {New York},\ \bibinfo {year} {1996})\BibitemShut {NoStop}%
	\bibitem [{\citenamefont {Flory}(1969)}]{flory}%
	\BibitemOpen
	\bibfield  {author} {\bibinfo {author} {\bibfnamefont {P.~J.}\ \bibnamefont
			{Flory}},\ }\href@noop {} {\emph {\bibinfo {title} {Statistical Mechanics of
				Chain Molecules}}}\ (\bibinfo  {publisher} {John Wiley},\ \bibinfo {address}
	{New York},\ \bibinfo {year} {1969})\BibitemShut {NoStop}%
	\bibitem [{\citenamefont {Bachmann}(2014)}]{bach14}%
	\BibitemOpen
	\bibfield  {author} {\bibinfo {author} {\bibfnamefont {M.}~\bibnamefont
			{Bachmann}},\ }\href@noop {} {\emph {\bibinfo {title} {Thermodynamics and
				Statistical Mechanics of Macromolecular Systems}}}\ (\bibinfo  {publisher}
	{Cambridge University Press},\ \bibinfo {year} {2014})\BibitemShut {NoStop}%
	\bibitem [{\citenamefont {Kabanov}\ \emph {et~al.}(2003)\citenamefont
		{Kabanov}, \citenamefont {Zezin}, \citenamefont {Kasaikin}, \citenamefont
		{Zakharova}, \citenamefont {Litmanovich},\ and\ \citenamefont
		{Ivleva}}]{kab03}%
	\BibitemOpen
	\bibfield  {author} {\bibinfo {author} {\bibfnamefont {V.~A.}\ \bibnamefont
			{Kabanov}}, \bibinfo {author} {\bibfnamefont {A.~B.}\ \bibnamefont {Zezin}},
		\bibinfo {author} {\bibfnamefont {V.~A.}\ \bibnamefont {Kasaikin}}, \bibinfo
		{author} {\bibfnamefont {J.~A.}\ \bibnamefont {Zakharova}}, \bibinfo {author}
		{\bibfnamefont {E.~A.}\ \bibnamefont {Litmanovich}}, \ and\ \bibinfo {author}
		{\bibfnamefont {E.~M.}\ \bibnamefont {Ivleva}},\ }\href@noop {} {\bibfield
		{journal} {\bibinfo  {journal} {Macromolecular Symposia}\ }\textbf {\bibinfo
			{volume} {191}},\ \bibinfo {pages} {111} (\bibinfo {year}
		{2003})}\BibitemShut {NoStop}%
	\bibitem [{\citenamefont {Popa-Nita}\ \emph {et~al.}(2009)\citenamefont
		{Popa-Nita}, \citenamefont {Rochas}, \citenamefont {David},\ and\
		\citenamefont {Domard}}]{pop09}%
	\BibitemOpen
	\bibfield  {author} {\bibinfo {author} {\bibfnamefont {S.}~\bibnamefont
			{Popa-Nita}}, \bibinfo {author} {\bibfnamefont {C.}~\bibnamefont {Rochas}},
		\bibinfo {author} {\bibfnamefont {L.}~\bibnamefont {David}}, \ and\ \bibinfo
		{author} {\bibfnamefont {A.}~\bibnamefont {Domard}},\ }\href@noop {}
	{\bibfield  {journal} {\bibinfo  {journal} {Langmuir}\ }\textbf {\bibinfo
			{volume} {25}},\ \bibinfo {pages} {6460} (\bibinfo {year}
		{2009})}\BibitemShut {NoStop}%
	\bibitem [{\citenamefont {Olea}(2012)}]{ole12}%
	\BibitemOpen
	\bibfield  {author} {\bibinfo {author} {\bibfnamefont {A.~F.}\ \bibnamefont
			{Olea}},\ }\enquote {\bibinfo {title} {Hydrophobic polyelectrolytes},}\ in\
	\href@noop {} {\emph {\bibinfo {booktitle} {Ionic Interactions in Natural and
				Synthetic Macromolecules}}}\ (\bibinfo  {publisher} {John Wiley {\&} Sons,
		Ltd},\ \bibinfo {year} {2012})\ Chap.~\bibinfo {chapter} {7}, pp.\ \bibinfo
	{pages} {211--233}\BibitemShut {NoStop}%
	\bibitem [{\citenamefont {Zhou}\ and\ \citenamefont {Zhao}(2019)}]{zhou19}%
	\BibitemOpen
	\bibfield  {author} {\bibinfo {author} {\bibfnamefont {X.}~\bibnamefont
			{Zhou}}\ and\ \bibinfo {author} {\bibfnamefont {K.}~\bibnamefont {Zhao}},\
	}\href@noop {} {\bibfield  {journal} {\bibinfo  {journal} {Journal of Polymer
				Science Part B: Polymer Physics}\ }\textbf {\bibinfo {volume} {57}},\
		\bibinfo {pages} {1716} (\bibinfo {year} {2019})}\BibitemShut {NoStop}%
	\bibitem [{\citenamefont {Tabandeh}\ and\ \citenamefont {Leon}(2019)}]{tab19}%
	\BibitemOpen
	\bibfield  {author} {\bibinfo {author} {\bibfnamefont {S.}~\bibnamefont
			{Tabandeh}}\ and\ \bibinfo {author} {\bibfnamefont {L.}~\bibnamefont
			{Leon}},\ }\href@noop {} {\bibfield  {journal} {\bibinfo  {journal}
			{Molecules}\ }\textbf {\bibinfo {volume} {24}} (\bibinfo {year}
		{2019})}\BibitemShut {NoStop}%
	\bibitem [{\citenamefont {Kuhn}, \citenamefont {Levin},\ and\ \citenamefont
		{Barbosa}(1998)}]{kuhn98}%
	\BibitemOpen
	\bibfield  {author} {\bibinfo {author} {\bibfnamefont {P.~S.}\ \bibnamefont
			{Kuhn}}, \bibinfo {author} {\bibfnamefont {Y.}~\bibnamefont {Levin}}, \ and\
		\bibinfo {author} {\bibfnamefont {M.~C.}\ \bibnamefont {Barbosa}},\
	}\href@noop {} {\bibfield  {journal} {\bibinfo  {journal} {Macromolecules}\
		}\textbf {\bibinfo {volume} {31}},\ \bibinfo {pages} {8347} (\bibinfo {year}
		{1998})}\BibitemShut {NoStop}%
	\bibitem [{\citenamefont {Meka}\ \emph {et~al.}(2017)\citenamefont {Meka},
		\citenamefont {Sing}, \citenamefont {Pichika}, \citenamefont {Nali},
		\citenamefont {Kolapalli},\ and\ \citenamefont {Kesharwani}}]{ven17}%
	\BibitemOpen
	\bibfield  {author} {\bibinfo {author} {\bibfnamefont {V.~S.}\ \bibnamefont
			{Meka}}, \bibinfo {author} {\bibfnamefont {M.~K.}\ \bibnamefont {Sing}},
		\bibinfo {author} {\bibfnamefont {M.~R.}\ \bibnamefont {Pichika}}, \bibinfo
		{author} {\bibfnamefont {S.~R.}\ \bibnamefont {Nali}}, \bibinfo {author}
		{\bibfnamefont {V.~R.}\ \bibnamefont {Kolapalli}}, \ and\ \bibinfo {author}
		{\bibfnamefont {P.}~\bibnamefont {Kesharwani}},\ }\href@noop {} {\bibfield
		{journal} {\bibinfo  {journal} {Drug Discovery Today}\ }\textbf {\bibinfo
			{volume} {22}},\ \bibinfo {pages} {1697} (\bibinfo {year}
		{2017})}\BibitemShut {NoStop}%
	\bibitem [{\citenamefont {Cazorla-Luna}\ \emph {et~al.}(2021)\citenamefont
		{Cazorla-Luna}, \citenamefont {Martín-Illana}, \citenamefont
		{Notario-Pérez}, \citenamefont {Ruiz-Caro},\ and\ \citenamefont
		{Veiga}}]{car21}%
	\BibitemOpen
	\bibfield  {author} {\bibinfo {author} {\bibfnamefont {R.}~\bibnamefont
			{Cazorla-Luna}}, \bibinfo {author} {\bibfnamefont {A.}~\bibnamefont
			{Martín-Illana}}, \bibinfo {author} {\bibfnamefont {F.}~\bibnamefont
			{Notario-Pérez}}, \bibinfo {author} {\bibfnamefont {R.}~\bibnamefont
			{Ruiz-Caro}}, \ and\ \bibinfo {author} {\bibfnamefont {M.-D.}\ \bibnamefont
			{Veiga}},\ }\href@noop {} {\bibfield  {journal} {\bibinfo  {journal}
			{Polymers}\ }\textbf {\bibinfo {volume} {13}} (\bibinfo {year}
		{2021})}\BibitemShut {NoStop}%
	\bibitem [{\citenamefont {Zhidong}, \citenamefont {Jianzhong},\ and\
		\citenamefont {Zhen-Gang}(2008)}]{zhi08}%
	\BibitemOpen
	\bibfield  {author} {\bibinfo {author} {\bibfnamefont {L.}~\bibnamefont
			{Zhidong}}, \bibinfo {author} {\bibfnamefont {W.}~\bibnamefont {Jianzhong}},
		\ and\ \bibinfo {author} {\bibfnamefont {W.}~\bibnamefont {Zhen-Gang}},\
	}\href@noop {} {\bibfield  {journal} {\bibinfo  {journal} {Biophysical
				Journal}\ }\textbf {\bibinfo {volume} {94}},\ \bibinfo {pages} {737}
		(\bibinfo {year} {2008})}\BibitemShut {NoStop}%
	\bibitem [{\citenamefont {Jiang}, \citenamefont {Wang},\ and\ \citenamefont
		{Wu}(2009)}]{jiang09}%
	\BibitemOpen
	\bibfield  {author} {\bibinfo {author} {\bibfnamefont {T.}~\bibnamefont
			{Jiang}}, \bibinfo {author} {\bibfnamefont {Z.-G.}\ \bibnamefont {Wang}}, \
		and\ \bibinfo {author} {\bibfnamefont {J.}~\bibnamefont {Wu}},\ }\href@noop
	{} {\bibfield  {journal} {\bibinfo  {journal} {Biophysical Journal}\ }\textbf
		{\bibinfo {volume} {96}},\ \bibinfo {pages} {3065} (\bibinfo {year}
		{2009})}\BibitemShut {NoStop}%
	\bibitem [{\citenamefont {Meng}\ \emph {et~al.}(2011)\citenamefont {Meng},
		\citenamefont {Hjelm}, \citenamefont {Hu},\ and\ \citenamefont
		{Wu}}]{meng11}%
	\BibitemOpen
	\bibfield  {author} {\bibinfo {author} {\bibfnamefont {D.}~\bibnamefont
			{Meng}}, \bibinfo {author} {\bibfnamefont {R.}~\bibnamefont {Hjelm}},
		\bibinfo {author} {\bibfnamefont {J.}~\bibnamefont {Hu}}, \ and\ \bibinfo
		{author} {\bibfnamefont {J.}~\bibnamefont {Wu}},\ }\href@noop {} {\bibfield
		{journal} {\bibinfo  {journal} {Biophysical Journal}\ }\textbf {\bibinfo
			{volume} {101}},\ \bibinfo {pages} {2476} (\bibinfo {year}
		{2011})}\BibitemShut {NoStop}%
	\bibitem [{\citenamefont {Capone}\ \emph {et~al.}(2012)\citenamefont {Capone},
		\citenamefont {Coluzza}, \citenamefont {LoVerso}, \citenamefont {Likos},\
		and\ \citenamefont {Blaak}}]{cap12}%
	\BibitemOpen
	\bibfield  {author} {\bibinfo {author} {\bibfnamefont {B.}~\bibnamefont
			{Capone}}, \bibinfo {author} {\bibfnamefont {I.}~\bibnamefont {Coluzza}},
		\bibinfo {author} {\bibfnamefont {F.}~\bibnamefont {LoVerso}}, \bibinfo
		{author} {\bibfnamefont {C.~N.}\ \bibnamefont {Likos}}, \ and\ \bibinfo
		{author} {\bibfnamefont {R.}~\bibnamefont {Blaak}},\ }\href@noop {}
	{\bibfield  {journal} {\bibinfo  {journal} {Phys. Rev. Lett.}\ }\textbf
		{\bibinfo {volume} {109}},\ \bibinfo {pages} {238301} (\bibinfo {year}
		{2012})}\BibitemShut {NoStop}%
	\bibitem [{\citenamefont {Capone}\ \emph {et~al.}(2013)\citenamefont {Capone},
		\citenamefont {Coluzza}, \citenamefont {Blaak}, \citenamefont {Lo~Verso},\
		and\ \citenamefont {Likos}}]{cap13}%
	\BibitemOpen
	\bibfield  {author} {\bibinfo {author} {\bibfnamefont {B.}~\bibnamefont
			{Capone}}, \bibinfo {author} {\bibfnamefont {I.}~\bibnamefont {Coluzza}},
		\bibinfo {author} {\bibfnamefont {R.}~\bibnamefont {Blaak}}, \bibinfo
		{author} {\bibfnamefont {F.}~\bibnamefont {Lo~Verso}}, \ and\ \bibinfo
		{author} {\bibfnamefont {C.~N.}\ \bibnamefont {Likos}},\ }\href@noop {}
	{\bibfield  {journal} {\bibinfo  {journal} {New Journal of Physics}\ }\textbf
		{\bibinfo {volume} {15}},\ \bibinfo {pages} {095002} (\bibinfo {year}
		{2013})}\BibitemShut {NoStop}%
	\bibitem [{\citenamefont {Yan}\ and\ \citenamefont {Xie}(2013)}]{yan13}%
	\BibitemOpen
	\bibfield  {author} {\bibinfo {author} {\bibfnamefont {L.-T.}\ \bibnamefont
			{Yan}}\ and\ \bibinfo {author} {\bibfnamefont {X.-M.}\ \bibnamefont {Xie}},\
	}\href@noop {} {\bibfield  {journal} {\bibinfo  {journal} {Progress in
				Polymer Science}\ }\textbf {\bibinfo {volume} {38}},\ \bibinfo {pages}
		{369–405} (\bibinfo {year} {2013})}\BibitemShut {NoStop}%
	\bibitem [{\citenamefont {Bernabei}\ \emph {et~al.}(2013)\citenamefont
		{Bernabei}, \citenamefont {Bacova}, \citenamefont {Moreno}, \citenamefont
		{Narros},\ and\ \citenamefont {Likos}}]{ber13}%
	\BibitemOpen
	\bibfield  {author} {\bibinfo {author} {\bibfnamefont {M.}~\bibnamefont
			{Bernabei}}, \bibinfo {author} {\bibfnamefont {P.}~\bibnamefont {Bacova}},
		\bibinfo {author} {\bibfnamefont {A.~J.}\ \bibnamefont {Moreno}}, \bibinfo
		{author} {\bibfnamefont {A.}~\bibnamefont {Narros}}, \ and\ \bibinfo {author}
		{\bibfnamefont {C.~N.}\ \bibnamefont {Likos}},\ }\href@noop {} {\bibfield
		{journal} {\bibinfo  {journal} {Soft Matter}\ }\textbf {\bibinfo {volume}
			{9}},\ \bibinfo {pages} {1287} (\bibinfo {year} {2013})}\BibitemShut
	{NoStop}%
	\bibitem [{\citenamefont {Han}\ \emph {et~al.}(2015)\citenamefont {Han},
		\citenamefont {Wang}, \citenamefont {Wang},\ and\ \citenamefont
		{Tang}}]{han15}%
	\BibitemOpen
	\bibfield  {author} {\bibinfo {author} {\bibfnamefont {T.}~\bibnamefont
			{Han}}, \bibinfo {author} {\bibfnamefont {X.}~\bibnamefont {Wang}}, \bibinfo
		{author} {\bibfnamefont {D.}~\bibnamefont {Wang}}, \ and\ \bibinfo {author}
		{\bibfnamefont {B.~Z.}\ \bibnamefont {Tang}},\ }\href@noop {} {\bibfield
		{journal} {\bibinfo  {journal} {Topics in Current Chemistry}\ }\textbf
		{\bibinfo {volume} {379}},\ \bibinfo {pages} {7} (\bibinfo {year}
		{2015})}\BibitemShut {NoStop}%
	\bibitem [{\citenamefont {Likos}\ \emph {et~al.}(2008)\citenamefont {Likos},
		\citenamefont {Mladek}, \citenamefont {Moreno}, \citenamefont {Gottwald},\
		and\ \citenamefont {Kahl}}]{lik08}%
	\BibitemOpen
	\bibfield  {author} {\bibinfo {author} {\bibfnamefont {C.~N.}\ \bibnamefont
			{Likos}}, \bibinfo {author} {\bibfnamefont {B.~M.}\ \bibnamefont {Mladek}},
		\bibinfo {author} {\bibfnamefont {A.~J.}\ \bibnamefont {Moreno}}, \bibinfo
		{author} {\bibfnamefont {D.}~\bibnamefont {Gottwald}}, \ and\ \bibinfo
		{author} {\bibfnamefont {G.}~\bibnamefont {Kahl}},\ }\href@noop {} {\bibfield
		{journal} {\bibinfo  {journal} {Comput. Phys. Commun.}\ }\textbf {\bibinfo
			{volume} {179}},\ \bibinfo {pages} {71} (\bibinfo {year} {2008})}\BibitemShut
	{NoStop}%
	\bibitem [{\citenamefont {Mohanty}\ \emph {et~al.}(2017)\citenamefont
		{Mohanty}, \citenamefont {N{\"o}jd}, \citenamefont {van Gruijthuijsen},
		\citenamefont {Crassous}, \citenamefont {Obiols-Rabasa}, \citenamefont
		{Schweins}, \citenamefont {Stradner},\ and\ \citenamefont
		{Schurtenberger}}]{moh17}%
	\BibitemOpen
	\bibfield  {author} {\bibinfo {author} {\bibfnamefont {P.~S.}\ \bibnamefont
			{Mohanty}}, \bibinfo {author} {\bibfnamefont {S.}~\bibnamefont {N{\"o}jd}},
		\bibinfo {author} {\bibfnamefont {K.}~\bibnamefont {van Gruijthuijsen}},
		\bibinfo {author} {\bibfnamefont {J.~J.}\ \bibnamefont {Crassous}}, \bibinfo
		{author} {\bibfnamefont {M.}~\bibnamefont {Obiols-Rabasa}}, \bibinfo {author}
		{\bibfnamefont {R.}~\bibnamefont {Schweins}}, \bibinfo {author}
		{\bibfnamefont {A.}~\bibnamefont {Stradner}}, \ and\ \bibinfo {author}
		{\bibfnamefont {P.}~\bibnamefont {Schurtenberger}},\ }\href@noop {}
	{\bibfield  {journal} {\bibinfo  {journal} {Scientific Reports}\ }\textbf
		{\bibinfo {volume} {7}},\ \bibinfo {pages} {1487} (\bibinfo {year}
		{2017})}\BibitemShut {NoStop}%
	\bibitem [{\citenamefont {Conley}\ \emph {et~al.}(2017)\citenamefont {Conley},
		\citenamefont {Aebischer}, \citenamefont {Nöjd}, \citenamefont
		{Schurtenberger},\ and\ \citenamefont {Scheffold}}]{gau17}%
	\BibitemOpen
	\bibfield  {author} {\bibinfo {author} {\bibfnamefont {G.~M.}\ \bibnamefont
			{Conley}}, \bibinfo {author} {\bibfnamefont {P.}~\bibnamefont {Aebischer}},
		\bibinfo {author} {\bibfnamefont {S.}~\bibnamefont {Nöjd}}, \bibinfo
		{author} {\bibfnamefont {P.}~\bibnamefont {Schurtenberger}}, \ and\ \bibinfo
		{author} {\bibfnamefont {F.}~\bibnamefont {Scheffold}},\ }\href@noop {}
	{\bibfield  {journal} {\bibinfo  {journal} {Science Advances}\ }\textbf
		{\bibinfo {volume} {3}},\ \bibinfo {pages} {e1700969} (\bibinfo {year}
		{2017})}\BibitemShut {NoStop}%
	\bibitem [{\citenamefont {Colla}, \citenamefont {Blaak},\ and\ \citenamefont
		{Likos}(2018)}]{col18}%
	\BibitemOpen
	\bibfield  {author} {\bibinfo {author} {\bibfnamefont {T.}~\bibnamefont
			{Colla}}, \bibinfo {author} {\bibfnamefont {R.}~\bibnamefont {Blaak}}, \ and\
		\bibinfo {author} {\bibfnamefont {C.~N.}\ \bibnamefont {Likos}},\ }\href@noop
	{} {\bibfield  {journal} {\bibinfo  {journal} {Soft Matter}\ }\textbf
		{\bibinfo {volume} {14}},\ \bibinfo {pages} {5106} (\bibinfo {year}
		{2018})}\BibitemShut {NoStop}%
	\bibitem [{\citenamefont {Doi}\ and\ \citenamefont {Edwards}(1986)}]{doi86}%
	\BibitemOpen
	\bibfield  {author} {\bibinfo {author} {\bibfnamefont {M.}~\bibnamefont
			{Doi}}\ and\ \bibinfo {author} {\bibfnamefont {S.~F.}\ \bibnamefont
			{Edwards}},\ }\href@noop {} {\emph {\bibinfo {title} {The Theory of Polymer
				Dynamics}}}\ (\bibinfo  {publisher} {Oxford University Press},\ \bibinfo
	{year} {1986})\BibitemShut {NoStop}%
	\bibitem [{\citenamefont {Karg}\ \emph {et~al.}(2019)\citenamefont {Karg},
		\citenamefont {Pich}, \citenamefont {Hellweg}, \citenamefont {Hoare},
		\citenamefont {Lyon}, \citenamefont {Crassous}, \citenamefont {Suzuki},
		\citenamefont {Gumerov}, \citenamefont {Schneider}, \citenamefont
		{Potemkin},\ and\ \citenamefont {Richtering}}]{kar19}%
	\BibitemOpen
	\bibfield  {author} {\bibinfo {author} {\bibfnamefont {M.}~\bibnamefont
			{Karg}}, \bibinfo {author} {\bibfnamefont {A.}~\bibnamefont {Pich}}, \bibinfo
		{author} {\bibfnamefont {T.}~\bibnamefont {Hellweg}}, \bibinfo {author}
		{\bibfnamefont {T.}~\bibnamefont {Hoare}}, \bibinfo {author} {\bibfnamefont
			{L.~A.}\ \bibnamefont {Lyon}}, \bibinfo {author} {\bibfnamefont {J.~J.}\
			\bibnamefont {Crassous}}, \bibinfo {author} {\bibfnamefont {D.}~\bibnamefont
			{Suzuki}}, \bibinfo {author} {\bibfnamefont {R.~A.}\ \bibnamefont {Gumerov}},
		\bibinfo {author} {\bibfnamefont {S.}~\bibnamefont {Schneider}}, \bibinfo
		{author} {\bibfnamefont {I.~I.}\ \bibnamefont {Potemkin}}, \ and\ \bibinfo
		{author} {\bibfnamefont {W.}~\bibnamefont {Richtering}},\ }\href@noop {}
	{\bibfield  {journal} {\bibinfo  {journal} {Langmuir}\ }\textbf {\bibinfo
			{volume} {35}},\ \bibinfo {pages} {6231} (\bibinfo {year}
		{2019})}\BibitemShut {NoStop}%
	\bibitem [{\citenamefont {Bosman}, \citenamefont {Janssen},\ and\ \citenamefont
		{Meijer}(1999)}]{bos99}%
	\BibitemOpen
	\bibfield  {author} {\bibinfo {author} {\bibfnamefont {A.~W.}\ \bibnamefont
			{Bosman}}, \bibinfo {author} {\bibfnamefont {H.~M.}\ \bibnamefont {Janssen}},
		\ and\ \bibinfo {author} {\bibfnamefont {E.~W.}\ \bibnamefont {Meijer}},\
	}\href@noop {} {\bibfield  {journal} {\bibinfo  {journal} {Chemical Reviews}\
		}\textbf {\bibinfo {volume} {99}},\ \bibinfo {pages} {1665} (\bibinfo {year}
		{1999})}\BibitemShut {NoStop}%
	\bibitem [{\citenamefont {Ballauff}\ and\ \citenamefont {Likos}(2004)}]{bal04}%
	\BibitemOpen
	\bibfield  {author} {\bibinfo {author} {\bibfnamefont {M.}~\bibnamefont
			{Ballauff}}\ and\ \bibinfo {author} {\bibfnamefont {C.~N.}\ \bibnamefont
			{Likos}},\ }\href@noop {} {\bibfield  {journal} {\bibinfo  {journal}
			{Angewandte Chemie International Edition}\ }\textbf {\bibinfo {volume}
			{43}},\ \bibinfo {pages} {2998} (\bibinfo {year} {2004})}\BibitemShut
	{NoStop}%
	\bibitem [{\citenamefont {Xu}, \citenamefont {Cao},\ and\ \citenamefont
		{Wu}(2010)}]{xu10}%
	\BibitemOpen
	\bibfield  {author} {\bibinfo {author} {\bibfnamefont {X.}~\bibnamefont
			{Xu}}, \bibinfo {author} {\bibfnamefont {D.}~\bibnamefont {Cao}}, \ and\
		\bibinfo {author} {\bibfnamefont {J.}~\bibnamefont {Wu}},\ }\href@noop {}
	{\bibfield  {journal} {\bibinfo  {journal} {Soft Matter}\ }\textbf {\bibinfo
			{volume} {6}},\ \bibinfo {pages} {4631} (\bibinfo {year} {2010})}\BibitemShut
	{NoStop}%
	\bibitem [{\citenamefont {Brittain}\ and\ \citenamefont
		{Minko}(2007)}]{brittain2007}%
	\BibitemOpen
	\bibfield  {author} {\bibinfo {author} {\bibfnamefont {W.~J.}\ \bibnamefont
			{Brittain}}\ and\ \bibinfo {author} {\bibfnamefont {S.}~\bibnamefont
			{Minko}},\ }\href@noop {} {\bibfield  {journal} {\bibinfo  {journal} {Journal
				of Polymer Science Part A: Polymer Chemistry}\ }\textbf {\bibinfo {volume}
			{45}},\ \bibinfo {pages} {3505} (\bibinfo {year} {2007})}\BibitemShut
	{NoStop}%
	\bibitem [{\citenamefont {Vincent}(1974)}]{vincent1974effect}%
	\BibitemOpen
	\bibfield  {author} {\bibinfo {author} {\bibfnamefont {B.}~\bibnamefont
			{Vincent}},\ }\href@noop {} {\bibfield  {journal} {\bibinfo  {journal}
			{Advances in Colloid and Interface Science}\ }\textbf {\bibinfo {volume}
			{4}},\ \bibinfo {pages} {193} (\bibinfo {year} {1974})}\BibitemShut {NoStop}%
	\bibitem [{\citenamefont {Van~der Waarden}(1951)}]{van1951adsorption}%
	\BibitemOpen
	\bibfield  {author} {\bibinfo {author} {\bibfnamefont {M.}~\bibnamefont
			{Van~der Waarden}},\ }\href@noop {} {\bibfield  {journal} {\bibinfo
			{journal} {Journal of Colloid Science}\ }\textbf {\bibinfo {volume} {6}},\
		\bibinfo {pages} {443} (\bibinfo {year} {1951})}\BibitemShut {NoStop}%
	\bibitem [{\citenamefont {Mackor}(1951)}]{mackor1951theoretical}%
	\BibitemOpen
	\bibfield  {author} {\bibinfo {author} {\bibfnamefont {E.}~\bibnamefont
			{Mackor}},\ }\href@noop {} {\bibfield  {journal} {\bibinfo  {journal}
			{Journal of Colloid Science}\ }\textbf {\bibinfo {volume} {6}},\ \bibinfo
		{pages} {492} (\bibinfo {year} {1951})}\BibitemShut {NoStop}%
	\bibitem [{\citenamefont {Mackor}\ and\ \citenamefont {Van~der
			Waals}(1952)}]{mackor1952statistics}%
	\BibitemOpen
	\bibfield  {author} {\bibinfo {author} {\bibfnamefont {E.}~\bibnamefont
			{Mackor}}\ and\ \bibinfo {author} {\bibfnamefont {J.}~\bibnamefont {Van~der
				Waals}},\ }\href@noop {} {\bibfield  {journal} {\bibinfo  {journal} {Journal
				of Colloid Science}\ }\textbf {\bibinfo {volume} {7}},\ \bibinfo {pages}
		{535} (\bibinfo {year} {1952})}\BibitemShut {NoStop}%
	\bibitem [{\citenamefont {Clayfield}\ and\ \citenamefont
		{Lumb}(1966)}]{clayfield1966theoretical}%
	\BibitemOpen
	\bibfield  {author} {\bibinfo {author} {\bibfnamefont {E.-J.}\ \bibnamefont
			{Clayfield}}\ and\ \bibinfo {author} {\bibfnamefont {E.-C.}\ \bibnamefont
			{Lumb}},\ }\href@noop {} {\bibfield  {journal} {\bibinfo  {journal} {Journal
				of Colloid and Interface Science}\ }\textbf {\bibinfo {volume} {22}},\
		\bibinfo {pages} {269} (\bibinfo {year} {1966})}\BibitemShut {NoStop}%
	\bibitem [{\citenamefont {Azzaroni}(2012)}]{azza12}%
	\BibitemOpen
	\bibfield  {author} {\bibinfo {author} {\bibfnamefont {O.}~\bibnamefont
			{Azzaroni}},\ }\href@noop {} {\bibfield  {journal} {\bibinfo  {journal}
			{Journal of Polymer Science Part A: Polymer Chemistry}\ }\textbf {\bibinfo
			{volume} {50}},\ \bibinfo {pages} {3225} (\bibinfo {year}
		{2012})}\BibitemShut {NoStop}%
	\bibitem [{\citenamefont {Wang}\ \emph {et~al.}(2020)\citenamefont {Wang},
		\citenamefont {Wang}, \citenamefont {Li}, \citenamefont {Li},\ and\
		\citenamefont {Hu}}]{wang20}%
	\BibitemOpen
	\bibfield  {author} {\bibinfo {author} {\bibfnamefont {S.}~\bibnamefont
			{Wang}}, \bibinfo {author} {\bibfnamefont {Z.}~\bibnamefont {Wang}}, \bibinfo
		{author} {\bibfnamefont {J.}~\bibnamefont {Li}}, \bibinfo {author}
		{\bibfnamefont {L.}~\bibnamefont {Li}}, \ and\ \bibinfo {author}
		{\bibfnamefont {W.}~\bibnamefont {Hu}},\ }\href@noop {} {\bibfield  {journal}
		{\bibinfo  {journal} {Mater. Chem. Front.}\ }\textbf {\bibinfo {volume}
			{4}},\ \bibinfo {pages} {692} (\bibinfo {year} {2020})}\BibitemShut {NoStop}%
	\bibitem [{\citenamefont {Krishnan}, \citenamefont {Weinman},\ and\
		\citenamefont {Ober}(2008)}]{kri08}%
	\BibitemOpen
	\bibfield  {author} {\bibinfo {author} {\bibfnamefont {S.}~\bibnamefont
			{Krishnan}}, \bibinfo {author} {\bibfnamefont {C.~J.}\ \bibnamefont
			{Weinman}}, \ and\ \bibinfo {author} {\bibfnamefont {C.~K.}\ \bibnamefont
			{Ober}},\ }\href@noop {} {\bibfield  {journal} {\bibinfo  {journal} {J.
				Mater. Chem.}\ }\textbf {\bibinfo {volume} {18}},\ \bibinfo {pages} {3405}
		(\bibinfo {year} {2008})}\BibitemShut {NoStop}%
	\bibitem [{\citenamefont {Grozea}\ and\ \citenamefont {Walker}(2009)}]{gro09}%
	\BibitemOpen
	\bibfield  {author} {\bibinfo {author} {\bibfnamefont {C.~M.}\ \bibnamefont
			{Grozea}}\ and\ \bibinfo {author} {\bibfnamefont {G.~C.}\ \bibnamefont
			{Walker}},\ }\href@noop {} {\bibfield  {journal} {\bibinfo  {journal} {Soft
				Matter}\ }\textbf {\bibinfo {volume} {5}},\ \bibinfo {pages} {4088} (\bibinfo
		{year} {2009})}\BibitemShut {NoStop}%
	\bibitem [{\citenamefont {Likos}(2001)}]{lik01}%
	\BibitemOpen
	\bibfield  {author} {\bibinfo {author} {\bibfnamefont {C.~N.}\ \bibnamefont
			{Likos}},\ }\href@noop {} {\bibfield  {journal} {\bibinfo  {journal} {Physics
				Reports}\ }\textbf {\bibinfo {volume} {348}},\ \bibinfo {pages} {267}
		(\bibinfo {year} {2001})}\BibitemShut {NoStop}%
	\bibitem [{\citenamefont {Ramakrishna}\ \emph {et~al.}(2001)\citenamefont
		{Ramakrishna}, \citenamefont {Mayer}, \citenamefont {Wintermantel},\ and\
		\citenamefont {Leong}}]{ram01}%
	\BibitemOpen
	\bibfield  {author} {\bibinfo {author} {\bibfnamefont {S.}~\bibnamefont
			{Ramakrishna}}, \bibinfo {author} {\bibfnamefont {J.}~\bibnamefont {Mayer}},
		\bibinfo {author} {\bibfnamefont {E.}~\bibnamefont {Wintermantel}}, \ and\
		\bibinfo {author} {\bibfnamefont {K.~W.}\ \bibnamefont {Leong}},\ }\href@noop
	{} {\bibfield  {journal} {\bibinfo  {journal} {Composites Science and
				Technology}\ }\textbf {\bibinfo {volume} {61}},\ \bibinfo {pages} {1189}
		(\bibinfo {year} {2001})}\BibitemShut {NoStop}%
	\bibitem [{\citenamefont {Geng}\ \emph {et~al.}(2007)\citenamefont {Geng},
		\citenamefont {Dalhaimer}, \citenamefont {Cai}, \citenamefont {Tsai},
		\citenamefont {Tewari}, \citenamefont {Minko},\ and\ \citenamefont
		{Discher}}]{gen07}%
	\BibitemOpen
	\bibfield  {author} {\bibinfo {author} {\bibfnamefont {Y.}~\bibnamefont
			{Geng}}, \bibinfo {author} {\bibfnamefont {P.}~\bibnamefont {Dalhaimer}},
		\bibinfo {author} {\bibfnamefont {S.}~\bibnamefont {Cai}}, \bibinfo {author}
		{\bibfnamefont {R.}~\bibnamefont {Tsai}}, \bibinfo {author} {\bibfnamefont
			{M.}~\bibnamefont {Tewari}}, \bibinfo {author} {\bibfnamefont
			{T.}~\bibnamefont {Minko}}, \ and\ \bibinfo {author} {\bibfnamefont {D.~E.}\
			\bibnamefont {Discher}},\ }\href@noop {} {\bibfield  {journal} {\bibinfo
			{journal} {Nature Nanotechnology}\ }\textbf {\bibinfo {volume} {61}},\
		\bibinfo {pages} {249} (\bibinfo {year} {2007})}\BibitemShut {NoStop}%
	\bibitem [{\citenamefont {Milner}(1991)}]{milner1991}%
	\BibitemOpen
	\bibfield  {author} {\bibinfo {author} {\bibfnamefont {S.~T.}\ \bibnamefont
			{Milner}},\ }\href@noop {} {\bibfield  {journal} {\bibinfo  {journal}
			{Science}\ }\textbf {\bibinfo {volume} {251}},\ \bibinfo {pages} {905}
		(\bibinfo {year} {1991})}\BibitemShut {NoStop}%
	\bibitem [{\citenamefont {Wijmans}\ and\ \citenamefont
		{Zhulina}(1993)}]{wijmans1993}%
	\BibitemOpen
	\bibfield  {author} {\bibinfo {author} {\bibfnamefont {C.}~\bibnamefont
			{Wijmans}}\ and\ \bibinfo {author} {\bibfnamefont {E.~B.}\ \bibnamefont
			{Zhulina}},\ }\href@noop {} {\bibfield  {journal} {\bibinfo  {journal}
			{Macromolecules}\ }\textbf {\bibinfo {volume} {26}},\ \bibinfo {pages} {7214}
		(\bibinfo {year} {1993})}\BibitemShut {NoStop}%
	\bibitem [{\citenamefont {Ballauff}\ and\ \citenamefont
		{Borisov}(2006)}]{ballauff2006}%
	\BibitemOpen
	\bibfield  {author} {\bibinfo {author} {\bibfnamefont {M.}~\bibnamefont
			{Ballauff}}\ and\ \bibinfo {author} {\bibfnamefont {O.}~\bibnamefont
			{Borisov}},\ }\href@noop {} {\bibfield  {journal} {\bibinfo  {journal}
			{Current Opinion in Colloid \& Interface Science}\ }\textbf {\bibinfo
			{volume} {11}},\ \bibinfo {pages} {316} (\bibinfo {year} {2006})}\BibitemShut
	{NoStop}%
	\bibitem [{\citenamefont {Zhang}\ and\ \citenamefont
		{M{\"u}ller}(2005)}]{zhang2005cylindrical}%
	\BibitemOpen
	\bibfield  {author} {\bibinfo {author} {\bibfnamefont {M.}~\bibnamefont
			{Zhang}}\ and\ \bibinfo {author} {\bibfnamefont {A.~H.}\ \bibnamefont
			{M{\"u}ller}},\ }\href@noop {} {\bibfield  {journal} {\bibinfo  {journal}
			{Journal of Polymer Science Part A: Polymer Chemistry}\ }\textbf {\bibinfo
			{volume} {43}},\ \bibinfo {pages} {3461} (\bibinfo {year}
		{2005})}\BibitemShut {NoStop}%
	\bibitem [{\citenamefont {Levin}(2002)}]{levin2002electrostatic}%
	\BibitemOpen
	\bibfield  {author} {\bibinfo {author} {\bibfnamefont {Y.}~\bibnamefont
			{Levin}},\ }\href@noop {} {\bibfield  {journal} {\bibinfo  {journal} {Reports
				on progress in physics}\ }\textbf {\bibinfo {volume} {65}},\ \bibinfo {pages}
		{1577} (\bibinfo {year} {2002})}\BibitemShut {NoStop}%
	\bibitem [{\citenamefont {Bakhshandeh}(2018)}]{bakhshandeh2018theoretical}%
	\BibitemOpen
	\bibfield  {author} {\bibinfo {author} {\bibfnamefont {A.}~\bibnamefont
			{Bakhshandeh}},\ }\href@noop {} {\bibfield  {journal} {\bibinfo  {journal}
			{Chemical Physics}\ }\textbf {\bibinfo {volume} {513}},\ \bibinfo {pages}
		{195} (\bibinfo {year} {2018})}\BibitemShut {NoStop}%
	\bibitem [{\citenamefont {Fisher}\ and\ \citenamefont
		{Levin}(1993)}]{fisher1993criticality}%
	\BibitemOpen
	\bibfield  {author} {\bibinfo {author} {\bibfnamefont {M.~E.}\ \bibnamefont
			{Fisher}}\ and\ \bibinfo {author} {\bibfnamefont {Y.}~\bibnamefont {Levin}},\
	}\href@noop {} {\bibfield  {journal} {\bibinfo  {journal} {Physical review
				letters}\ }\textbf {\bibinfo {volume} {71}},\ \bibinfo {pages} {3826}
		(\bibinfo {year} {1993})}\BibitemShut {NoStop}%
	\bibitem [{\citenamefont {Levin}\ and\ \citenamefont
		{Fisher}(1996)}]{levin1996criticality}%
	\BibitemOpen
	\bibfield  {author} {\bibinfo {author} {\bibfnamefont {Y.}~\bibnamefont
			{Levin}}\ and\ \bibinfo {author} {\bibfnamefont {M.~E.}\ \bibnamefont
			{Fisher}},\ }\href@noop {} {\bibfield  {journal} {\bibinfo  {journal}
			{Physica A: Statistical Mechanics and its Applications}\ }\textbf {\bibinfo
			{volume} {225}},\ \bibinfo {pages} {164} (\bibinfo {year}
		{1996})}\BibitemShut {NoStop}%
	\bibitem [{\citenamefont {Bakhshandeh}, \citenamefont {Dos~Santos},\ and\
		\citenamefont {Levin}(2011)}]{bakhshandeh2011weak}%
	\BibitemOpen
	\bibfield  {author} {\bibinfo {author} {\bibfnamefont {A.}~\bibnamefont
			{Bakhshandeh}}, \bibinfo {author} {\bibfnamefont {A.~P.}\ \bibnamefont
			{Dos~Santos}}, \ and\ \bibinfo {author} {\bibfnamefont {Y.}~\bibnamefont
			{Levin}},\ }\href@noop {} {\bibfield  {journal} {\bibinfo  {journal}
			{Physical review letters}\ }\textbf {\bibinfo {volume} {107}},\ \bibinfo
		{pages} {107801} (\bibinfo {year} {2011})}\BibitemShut {NoStop}%
	\bibitem [{\citenamefont {Bakhshandeh}\ \emph
		{et~al.}(2019{\natexlab{a}})\citenamefont {Bakhshandeh}, \citenamefont
		{Dos~Santos}, \citenamefont {Diehl},\ and\ \citenamefont
		{Levin}}]{bakhshandeh2019isothermal}%
	\BibitemOpen
	\bibfield  {author} {\bibinfo {author} {\bibfnamefont {A.}~\bibnamefont
			{Bakhshandeh}}, \bibinfo {author} {\bibfnamefont {A.~P.}\ \bibnamefont
			{Dos~Santos}}, \bibinfo {author} {\bibfnamefont {A.}~\bibnamefont {Diehl}}, \
		and\ \bibinfo {author} {\bibfnamefont {Y.}~\bibnamefont {Levin}},\
	}\href@noop {} {\bibfield  {journal} {\bibinfo  {journal} {The Journal of
				chemical physics}\ }\textbf {\bibinfo {volume} {151}},\ \bibinfo {pages}
		{084101} (\bibinfo {year} {2019}{\natexlab{a}})}\BibitemShut {NoStop}%
	\bibitem [{\citenamefont {Bakhshandeh}\ and\ \citenamefont
		{Segala}(2019)}]{bakhshandeh2019adsorption}%
	\BibitemOpen
	\bibfield  {author} {\bibinfo {author} {\bibfnamefont {A.}~\bibnamefont
			{Bakhshandeh}}\ and\ \bibinfo {author} {\bibfnamefont {M.}~\bibnamefont
			{Segala}},\ }\href@noop {} {\bibfield  {journal} {\bibinfo  {journal}
			{Journal of Molecular Liquids}\ }\textbf {\bibinfo {volume} {294}},\ \bibinfo
		{pages} {111673} (\bibinfo {year} {2019})}\BibitemShut {NoStop}%
	\bibitem [{\citenamefont {Lund}\ and\ \citenamefont
		{J{\"o}nsson}(2005)}]{lund2005charge}%
	\BibitemOpen
	\bibfield  {author} {\bibinfo {author} {\bibfnamefont {M.}~\bibnamefont
			{Lund}}\ and\ \bibinfo {author} {\bibfnamefont {B.}~\bibnamefont
			{J{\"o}nsson}},\ }\href@noop {} {\bibfield  {journal} {\bibinfo  {journal}
			{Biochemistry}\ }\textbf {\bibinfo {volume} {44}},\ \bibinfo {pages} {5722}
		(\bibinfo {year} {2005})}\BibitemShut {NoStop}%
	\bibitem [{\citenamefont {Lund}\ and\ \citenamefont
		{J{\"o}nsson}(2013)}]{lund2013charge}%
	\BibitemOpen
	\bibfield  {author} {\bibinfo {author} {\bibfnamefont {M.}~\bibnamefont
			{Lund}}\ and\ \bibinfo {author} {\bibfnamefont {B.}~\bibnamefont
			{J{\"o}nsson}},\ }\href@noop {} {\bibfield  {journal} {\bibinfo  {journal}
			{Q. Rev. Biophys}\ }\textbf {\bibinfo {volume} {46}},\ \bibinfo {pages} {265}
		(\bibinfo {year} {2013})}\BibitemShut {NoStop}%
	\bibitem [{\citenamefont {Trefalt}, \citenamefont {Behrens},\ and\
		\citenamefont {Borkovec}(2016)}]{trefalt2016charge}%
	\BibitemOpen
	\bibfield  {author} {\bibinfo {author} {\bibfnamefont {G.}~\bibnamefont
			{Trefalt}}, \bibinfo {author} {\bibfnamefont {S.~H.}\ \bibnamefont
			{Behrens}}, \ and\ \bibinfo {author} {\bibfnamefont {M.}~\bibnamefont
			{Borkovec}},\ }\href@noop {} {\bibfield  {journal} {\bibinfo  {journal}
			{Langmuir}\ }\textbf {\bibinfo {volume} {32}},\ \bibinfo {pages} {380}
		(\bibinfo {year} {2016})}\BibitemShut {NoStop}%
	\bibitem [{\citenamefont {da~Silva}\ and\ \citenamefont
		{J{\"o}nsson}(2009)}]{da2009polyelectrolyte}%
	\BibitemOpen
	\bibfield  {author} {\bibinfo {author} {\bibfnamefont {F.~L.~B.}\
			\bibnamefont {da~Silva}}\ and\ \bibinfo {author} {\bibfnamefont
			{B.}~\bibnamefont {J{\"o}nsson}},\ }\href@noop {} {\bibfield  {journal}
		{\bibinfo  {journal} {Soft Matter}\ }\textbf {\bibinfo {volume} {5}},\
		\bibinfo {pages} {2862} (\bibinfo {year} {2009})}\BibitemShut {NoStop}%
	\bibitem [{\citenamefont {Bakhshandeh}, \citenamefont {Frydel},\ and\
		\citenamefont {Levin}(2020)}]{D0CP03633A}%
	\BibitemOpen
	\bibfield  {author} {\bibinfo {author} {\bibfnamefont {A.}~\bibnamefont
			{Bakhshandeh}}, \bibinfo {author} {\bibfnamefont {D.}~\bibnamefont {Frydel}},
		\ and\ \bibinfo {author} {\bibfnamefont {Y.}~\bibnamefont {Levin}},\
	}\href@noop {} {\bibfield  {journal} {\bibinfo  {journal} {Phys. Chem. Chem.
				Phys.}\ }\textbf {\bibinfo {volume} {22}},\ \bibinfo {pages} {24712}
		(\bibinfo {year} {2020})}\BibitemShut {NoStop}%
	\bibitem [{\citenamefont {Bakhshandeh}\ \emph
		{et~al.}(2019{\natexlab{b}})\citenamefont {Bakhshandeh}, \citenamefont
		{Frydel}, \citenamefont {Diehl},\ and\ \citenamefont {Levin}}]{bk2019}%
	\BibitemOpen
	\bibfield  {author} {\bibinfo {author} {\bibfnamefont {A.}~\bibnamefont
			{Bakhshandeh}}, \bibinfo {author} {\bibfnamefont {D.}~\bibnamefont {Frydel}},
		\bibinfo {author} {\bibfnamefont {A.}~\bibnamefont {Diehl}}, \ and\ \bibinfo
		{author} {\bibfnamefont {Y.}~\bibnamefont {Levin}},\ }\href@noop {}
	{\bibfield  {journal} {\bibinfo  {journal} {Phys. Rev. Lett.}\ }\textbf
		{\bibinfo {volume} {123}},\ \bibinfo {pages} {208004} (\bibinfo {year}
		{2019}{\natexlab{b}})}\BibitemShut {NoStop}%
	\bibitem [{\citenamefont {Frydel}(2019)}]{derek2019}%
	\BibitemOpen
	\bibfield  {author} {\bibinfo {author} {\bibfnamefont {D.}~\bibnamefont
			{Frydel}},\ }\href@noop {} {\bibfield  {journal} {\bibinfo  {journal} {The
				Journal of Chemical Physics}\ }\textbf {\bibinfo {volume} {150}},\ \bibinfo
		{pages} {194901} (\bibinfo {year} {2019})}\BibitemShut {NoStop}%
	\bibitem [{\citenamefont {Podgornik}(2018)}]{pod2019}%
	\BibitemOpen
	\bibfield  {author} {\bibinfo {author} {\bibfnamefont {R.}~\bibnamefont
			{Podgornik}},\ }\href@noop {} {\bibfield  {journal} {\bibinfo  {journal} {The
				Journal of Chemical Physics}\ }\textbf {\bibinfo {volume} {149}},\ \bibinfo
		{pages} {104701} (\bibinfo {year} {2018})}\BibitemShut {NoStop}%
	\bibitem [{\citenamefont {Bakhshandeh}, \citenamefont {Dos~Santos},\ and\
		\citenamefont {Levin}(2020)}]{bakhshandeh2020interaction}%
	\BibitemOpen
	\bibfield  {author} {\bibinfo {author} {\bibfnamefont {A.}~\bibnamefont
			{Bakhshandeh}}, \bibinfo {author} {\bibfnamefont {A.~P.}\ \bibnamefont
			{Dos~Santos}}, \ and\ \bibinfo {author} {\bibfnamefont {Y.}~\bibnamefont
			{Levin}},\ }\href@noop {} {\bibfield  {journal} {\bibinfo  {journal} {The
				Journal of Physical Chemistry B}\ }\textbf {\bibinfo {volume} {124}},\
		\bibinfo {pages} {11762} (\bibinfo {year} {2020})}\BibitemShut {NoStop}%
	\bibitem [{NIN(1971)}]{NINHAM1971405}%
	\BibitemOpen
	\href@noop {} {\bibfield  {journal} {\bibinfo  {journal} {Journal of
				Theoretical Biology}\ }\textbf {\bibinfo {volume} {31}},\ \bibinfo {pages}
		{405} (\bibinfo {year} {1971})}\BibitemShut {NoStop}%
	\bibitem [{\citenamefont {Pelton}(2000)}]{pel00}%
	\BibitemOpen
	\bibfield  {author} {\bibinfo {author} {\bibfnamefont {R.}~\bibnamefont
			{Pelton}},\ }\href@noop {} {\bibfield  {journal} {\bibinfo  {journal} {R. .
				Adv. Colloid Interfac.}\ }\textbf {\bibinfo {volume} {85}},\ \bibinfo {pages}
		{1} (\bibinfo {year} {2000})}\BibitemShut {NoStop}%
	\bibitem [{\citenamefont {Jusufi}, \citenamefont {Likos},\ and\ \citenamefont
		{Löwen}(2002)}]{jus02}%
	\BibitemOpen
	\bibfield  {author} {\bibinfo {author} {\bibfnamefont {A.}~\bibnamefont
			{Jusufi}}, \bibinfo {author} {\bibfnamefont {C.~N.}\ \bibnamefont {Likos}}, \
		and\ \bibinfo {author} {\bibfnamefont {H.}~\bibnamefont {Löwen}},\
	}\href@noop {} {\bibfield  {journal} {\bibinfo  {journal} {The Journal of
				Chemical Physics}\ }\textbf {\bibinfo {volume} {116}},\ \bibinfo {pages}
		{11011} (\bibinfo {year} {2002})}\BibitemShut {NoStop}%
	\bibitem [{\citenamefont {Jusufi}, \citenamefont {Likos},\ and\ \citenamefont
		{Ballauff}(2004)}]{jus04}%
	\BibitemOpen
	\bibfield  {author} {\bibinfo {author} {\bibfnamefont {A.}~\bibnamefont
			{Jusufi}}, \bibinfo {author} {\bibfnamefont {C.~N.}\ \bibnamefont {Likos}}, \
		and\ \bibinfo {author} {\bibfnamefont {M.}~\bibnamefont {Ballauff}},\
	}\href@noop {} {\bibfield  {journal} {\bibinfo  {journal} {Colloid and
				Polymer Science}\ }\textbf {\bibinfo {volume} {282}},\ \bibinfo {pages} {910}
		(\bibinfo {year} {2004})}\BibitemShut {NoStop}%
	\bibitem [{\citenamefont {Jiang}, \citenamefont {Li},\ and\ \citenamefont
		{Wu}(2007)}]{jian07}%
	\BibitemOpen
	\bibfield  {author} {\bibinfo {author} {\bibfnamefont {T.}~\bibnamefont
			{Jiang}}, \bibinfo {author} {\bibfnamefont {Z.}~\bibnamefont {Li}}, \ and\
		\bibinfo {author} {\bibfnamefont {J.}~\bibnamefont {Wu}},\ }\href@noop {}
	{\bibfield  {journal} {\bibinfo  {journal} {Macromolecules}\ }\textbf
		{\bibinfo {volume} {40}},\ \bibinfo {pages} {334} (\bibinfo {year}
		{2007})}\BibitemShut {NoStop}%
	\bibitem [{\citenamefont {Jusufi}\ and\ \citenamefont {Likos}(2009)}]{jusu09}%
	\BibitemOpen
	\bibfield  {author} {\bibinfo {author} {\bibfnamefont {A.}~\bibnamefont
			{Jusufi}}\ and\ \bibinfo {author} {\bibfnamefont {C.~N.}\ \bibnamefont
			{Likos}},\ }\href@noop {} {\bibfield  {journal} {\bibinfo  {journal} {Rev.
				Mod. Phys.}\ }\textbf {\bibinfo {volume} {81}},\ \bibinfo {pages} {1753}
		(\bibinfo {year} {2009})}\BibitemShut {NoStop}%
	\bibitem [{\citenamefont {Colla}, \citenamefont {Likos},\ and\ \citenamefont
		{Levin}(2014)}]{col14}%
	\BibitemOpen
	\bibfield  {author} {\bibinfo {author} {\bibfnamefont {T.}~\bibnamefont
			{Colla}}, \bibinfo {author} {\bibfnamefont {C.~N.}\ \bibnamefont {Likos}}, \
		and\ \bibinfo {author} {\bibfnamefont {Y.}~\bibnamefont {Levin}},\
	}\href@noop {} {\bibfield  {journal} {\bibinfo  {journal} {The Journal of
				Chemical Physics}\ }\textbf {\bibinfo {volume} {141}},\ \bibinfo {pages}
		{234902} (\bibinfo {year} {2014})}\BibitemShut {NoStop}%
	\bibitem [{\citenamefont {P{\'e}rez-Ch{\'a}vez}, \citenamefont {Albesa},\ and\
		\citenamefont {Longo}(2021)}]{per21}%
	\BibitemOpen
	\bibfield  {author} {\bibinfo {author} {\bibfnamefont {N.~A.}\ \bibnamefont
			{P{\'e}rez-Ch{\'a}vez}}, \bibinfo {author} {\bibfnamefont {A.~G.}\
			\bibnamefont {Albesa}}, \ and\ \bibinfo {author} {\bibfnamefont {G.~S.}\
			\bibnamefont {Longo}},\ }\href@noop {} {\bibfield  {journal} {\bibinfo
			{journal} {Macromolecules}\ }\textbf {\bibinfo {volume} {54}},\ \bibinfo
		{pages} {2936} (\bibinfo {year} {2021})}\BibitemShut {NoStop}%
	\bibitem [{\citenamefont {Hansen}\ and\ \citenamefont
		{McDonald}(2013)}]{hansen}%
	\BibitemOpen
	\bibfield  {author} {\bibinfo {author} {\bibfnamefont {J.-P.}\ \bibnamefont
			{Hansen}}\ and\ \bibinfo {author} {\bibfnamefont {I.~R.}\ \bibnamefont
			{McDonald}},\ }\href@noop {} {\emph {\bibinfo {title} {Theory of simple
				liquids: with applications to soft matter}}},\ \bibinfo {edition} {4th}\ ed.\
	(\bibinfo  {publisher} {Elsevier Acad Press},\ \bibinfo {address} {Amsterdam
		u.a.},\ \bibinfo {year} {2013})\BibitemShut {NoStop}%
	\bibitem [{\citenamefont {de~Gennes}(1979)}]{gennes}%
	\BibitemOpen
	\bibfield  {author} {\bibinfo {author} {\bibfnamefont {P.~G.}\ \bibnamefont
			{de~Gennes}},\ }\href@noop {} {\emph {\bibinfo {title} {Scaling Concepts in
				Polymer Physics}}}\ (\bibinfo  {publisher} {Cornell University Press},\
	\bibinfo {address} {Ithaca},\ \bibinfo {year} {1979})\BibitemShut {NoStop}%
	\bibitem [{\citenamefont {Locatelli}, \citenamefont {Capone},\ and\
		\citenamefont {Likos}(2016)}]{loc16}%
	\BibitemOpen
	\bibfield  {author} {\bibinfo {author} {\bibfnamefont {E.}~\bibnamefont
			{Locatelli}}, \bibinfo {author} {\bibfnamefont {B.}~\bibnamefont {Capone}}, \
		and\ \bibinfo {author} {\bibfnamefont {C.~N.}\ \bibnamefont {Likos}},\
	}\href@noop {} {\bibfield  {journal} {\bibinfo  {journal} {The Journal of
				Chemical Physics}\ }\textbf {\bibinfo {volume} {145}},\ \bibinfo {pages}
		{174901} (\bibinfo {year} {2016})}\BibitemShut {NoStop}%
	\bibitem [{\citenamefont {Pincus}(1991)}]{pin91}%
	\BibitemOpen
	\bibfield  {author} {\bibinfo {author} {\bibfnamefont {P.}~\bibnamefont
			{Pincus}},\ }\href@noop {} {\bibfield  {journal} {\bibinfo  {journal}
			{Macromolecules}\ }\textbf {\bibinfo {volume} {24}},\ \bibinfo {pages} {2912}
		(\bibinfo {year} {1991})}\BibitemShut {NoStop}%
	\bibitem [{\citenamefont {Pincus}(1976)}]{pin76}%
	\BibitemOpen
	\bibfield  {author} {\bibinfo {author} {\bibfnamefont {P.}~\bibnamefont
			{Pincus}},\ }\href@noop {} {\bibfield  {journal} {\bibinfo  {journal}
			{Macromolecules}\ }\textbf {\bibinfo {volume} {9}},\ \bibinfo {pages} {386}
		(\bibinfo {year} {1976})}\BibitemShut {NoStop}%
	\bibitem [{\citenamefont {Wu}\ and\ \citenamefont {Li}(2007)}]{wu07}%
	\BibitemOpen
	\bibfield  {author} {\bibinfo {author} {\bibfnamefont {J.}~\bibnamefont
			{Wu}}\ and\ \bibinfo {author} {\bibfnamefont {Z.}~\bibnamefont {Li}},\
	}\href@noop {} {\bibfield  {journal} {\bibinfo  {journal} {Annual Review of
				Physical Chemistry}\ }\textbf {\bibinfo {volume} {58}},\ \bibinfo {pages}
		{85} (\bibinfo {year} {2007})}\BibitemShut {NoStop}%
	\bibitem [{Note1()}]{Note1}%
	\BibitemOpen
	\bibinfo {note} {In a planar case, this so-called ``box model'' corresponds
		to uniform density profiles for the polymer beads.}\BibitemShut {Stop}%
	\bibitem [{\citenamefont {Tergolina}\ and\ \citenamefont {dos
			Santos}(2017)}]{Vinicius}%
	\BibitemOpen
	\bibfield  {author} {\bibinfo {author} {\bibfnamefont {V.~B.}\ \bibnamefont
			{Tergolina}}\ and\ \bibinfo {author} {\bibfnamefont {A.~P.}\ \bibnamefont
			{dos Santos}},\ }\href@noop {} {\bibfield  {journal} {\bibinfo  {journal}
			{The Journal of Chemical Physics}\ }\textbf {\bibinfo {volume} {147}},\
		\bibinfo {pages} {114103} (\bibinfo {year} {2017})}\BibitemShut {NoStop}%
	\bibitem [{\citenamefont {Akinchina}, \citenamefont {Shusharina},\ and\
		\citenamefont {Linse}(2004)}]{akinchina2004diblock}%
	\BibitemOpen
	\bibfield  {author} {\bibinfo {author} {\bibfnamefont {A.}~\bibnamefont
			{Akinchina}}, \bibinfo {author} {\bibfnamefont {N.~P.}\ \bibnamefont
			{Shusharina}}, \ and\ \bibinfo {author} {\bibfnamefont {P.}~\bibnamefont
			{Linse}},\ }\href@noop {} {\bibfield  {journal} {\bibinfo  {journal}
			{Langmuir}\ }\textbf {\bibinfo {volume} {20}},\ \bibinfo {pages} {10351}
		(\bibinfo {year} {2004})}\BibitemShut {NoStop}%
	\bibitem [{\citenamefont {Wernersson}\ and\ \citenamefont
		{Linse}(2013)}]{wernersson2013spreading}%
	\BibitemOpen
	\bibfield  {author} {\bibinfo {author} {\bibfnamefont {E.}~\bibnamefont
			{Wernersson}}\ and\ \bibinfo {author} {\bibfnamefont {P.}~\bibnamefont
			{Linse}},\ }\href@noop {} {\bibfield  {journal} {\bibinfo  {journal}
			{Langmuir}\ }\textbf {\bibinfo {volume} {29}},\ \bibinfo {pages} {10455}
		(\bibinfo {year} {2013})}\BibitemShut {NoStop}%
	\bibitem [{\citenamefont {Angelescu}\ and\ \citenamefont
		{Linse}(2014)}]{angelescu2014monte}%
	\BibitemOpen
	\bibfield  {author} {\bibinfo {author} {\bibfnamefont {D.~G.}\ \bibnamefont
			{Angelescu}}\ and\ \bibinfo {author} {\bibfnamefont {P.}~\bibnamefont
			{Linse}},\ }\href@noop {} {\bibfield  {journal} {\bibinfo  {journal}
			{Macromolecules}\ }\textbf {\bibinfo {volume} {47}},\ \bibinfo {pages} {415}
		(\bibinfo {year} {2014})}\BibitemShut {NoStop}%
	\bibitem [{\citenamefont {Frenkel}\ and\ \citenamefont
		{Smit}(2001)}]{frenkel2001understanding}%
	\BibitemOpen
	\bibfield  {author} {\bibinfo {author} {\bibfnamefont {D.}~\bibnamefont
			{Frenkel}}\ and\ \bibinfo {author} {\bibfnamefont {B.}~\bibnamefont {Smit}},\
	}\href@noop {} {\emph {\bibinfo {title} {Understanding molecular simulation:
				from algorithms to applications}}},\ Vol.~\bibinfo {volume} {1}\ (\bibinfo
	{publisher} {Elsevier},\ \bibinfo {year} {2001})\BibitemShut {NoStop}%
	\bibitem [{\citenamefont {Metropolis}\ \emph {et~al.}(1953)\citenamefont
		{Metropolis}, \citenamefont {Rosenbluth}, \citenamefont {Rosenbluth},
		\citenamefont {Teller},\ and\ \citenamefont
		{Teller}}]{metropolis1953equation}%
	\BibitemOpen
	\bibfield  {author} {\bibinfo {author} {\bibfnamefont {N.}~\bibnamefont
			{Metropolis}}, \bibinfo {author} {\bibfnamefont {A.~W.}\ \bibnamefont
			{Rosenbluth}}, \bibinfo {author} {\bibfnamefont {M.~N.}\ \bibnamefont
			{Rosenbluth}}, \bibinfo {author} {\bibfnamefont {A.~H.}\ \bibnamefont
			{Teller}}, \ and\ \bibinfo {author} {\bibfnamefont {E.}~\bibnamefont
			{Teller}},\ }\href@noop {} {\bibfield  {journal} {\bibinfo  {journal} {The
				journal of chemical physics}\ }\textbf {\bibinfo {volume} {21}},\ \bibinfo
		{pages} {1087} (\bibinfo {year} {1953})}\BibitemShut {NoStop}%
	\bibitem [{\citenamefont {Li}\ and\ \citenamefont
		{Wu}(2006{\natexlab{a}})}]{li06}%
	\BibitemOpen
	\bibfield  {author} {\bibinfo {author} {\bibfnamefont {Z.}~\bibnamefont
			{Li}}\ and\ \bibinfo {author} {\bibfnamefont {J.}~\bibnamefont {Wu}},\
	}\href@noop {} {\bibfield  {journal} {\bibinfo  {journal} {Phys. Rev. Lett.}\
		}\textbf {\bibinfo {volume} {96}},\ \bibinfo {pages} {048302} (\bibinfo
		{year} {2006}{\natexlab{a}})}\BibitemShut {NoStop}%
	\bibitem [{\citenamefont {Li}\ and\ \citenamefont
		{Wu}(2006{\natexlab{b}})}]{li06_2}%
	\BibitemOpen
	\bibfield  {author} {\bibinfo {author} {\bibfnamefont {Z.}~\bibnamefont
			{Li}}\ and\ \bibinfo {author} {\bibfnamefont {J.}~\bibnamefont {Wu}},\
	}\href@noop {} {\bibfield  {journal} {\bibinfo  {journal} {The Journal of
				Physical Chemistry B}\ }\textbf {\bibinfo {volume} {110}},\ \bibinfo {pages}
		{7473} (\bibinfo {year} {2006}{\natexlab{b}})}\BibitemShut {NoStop}%
	\bibitem [{\citenamefont {Rosenfeld}(1989)}]{ros89}%
	\BibitemOpen
	\bibfield  {author} {\bibinfo {author} {\bibfnamefont {Y.}~\bibnamefont
			{Rosenfeld}},\ }\href@noop {} {\bibfield  {journal} {\bibinfo  {journal}
			{Phys. Rev. Lett.}\ }\textbf {\bibinfo {volume} {63}},\ \bibinfo {pages}
		{980} (\bibinfo {year} {1989})}\BibitemShut {NoStop}%
	\bibitem [{\citenamefont {Rosenfeld}(2002)}]{ros02}%
	\BibitemOpen
	\bibfield  {author} {\bibinfo {author} {\bibfnamefont {Y.}~\bibnamefont
			{Rosenfeld}},\ }\href@noop {} {\bibfield  {journal} {\bibinfo  {journal}
			{Journal of Physics: Condensed Matter}\ }\textbf {\bibinfo {volume} {14}},\
		\bibinfo {pages} {9141} (\bibinfo {year} {2002})}\BibitemShut {NoStop}%
	\bibitem [{\citenamefont {Roth}\ \emph {et~al.}(2002)\citenamefont {Roth},
		\citenamefont {Evans}, \citenamefont {Lang},\ and\ \citenamefont
		{Kahl}}]{roth02}%
	\BibitemOpen
	\bibfield  {author} {\bibinfo {author} {\bibfnamefont {R.}~\bibnamefont
			{Roth}}, \bibinfo {author} {\bibfnamefont {R.}~\bibnamefont {Evans}},
		\bibinfo {author} {\bibfnamefont {A.}~\bibnamefont {Lang}}, \ and\ \bibinfo
		{author} {\bibfnamefont {G.}~\bibnamefont {Kahl}},\ }\href@noop {} {\bibfield
		{journal} {\bibinfo  {journal} {Journal of Physics: Condensed Matter}\
		}\textbf {\bibinfo {volume} {14}},\ \bibinfo {pages} {12063} (\bibinfo {year}
		{2002})}\BibitemShut {NoStop}%
	\bibitem [{\citenamefont {Yu}\ and\ \citenamefont {Wu}(2002)}]{yu02}%
	\BibitemOpen
	\bibfield  {author} {\bibinfo {author} {\bibfnamefont {Y.-X.}\ \bibnamefont
			{Yu}}\ and\ \bibinfo {author} {\bibfnamefont {J.}~\bibnamefont {Wu}},\
	}\href@noop {} {\bibfield  {journal} {\bibinfo  {journal} {The Journal of
				Chemical Physics}\ }\textbf {\bibinfo {volume} {117}},\ \bibinfo {pages}
		{10156} (\bibinfo {year} {2002})}\BibitemShut {NoStop}%
	\bibitem [{\citenamefont {Roth}(2010)}]{roth10}%
	\BibitemOpen
	\bibfield  {author} {\bibinfo {author} {\bibfnamefont {R.}~\bibnamefont
			{Roth}},\ }\href@noop {} {\bibfield  {journal} {\bibinfo  {journal} {Journal
				of Physics: Condensed Matter}\ }\textbf {\bibinfo {volume} {22}},\ \bibinfo
		{pages} {063102} (\bibinfo {year} {2010})}\BibitemShut {NoStop}%
	\bibitem [{\citenamefont {Mansoori}\ \emph {et~al.}(1971)\citenamefont
		{Mansoori}, \citenamefont {Carnahan}, \citenamefont {Starling},\ and\
		\citenamefont {Leland}}]{man71}%
	\BibitemOpen
	\bibfield  {author} {\bibinfo {author} {\bibfnamefont {G.~A.}\ \bibnamefont
			{Mansoori}}, \bibinfo {author} {\bibfnamefont {N.~F.}\ \bibnamefont
			{Carnahan}}, \bibinfo {author} {\bibfnamefont {K.~E.}\ \bibnamefont
			{Starling}}, \ and\ \bibinfo {author} {\bibfnamefont {T.~W.}\ \bibnamefont
			{Leland}},\ }\href@noop {} {\bibfield  {journal} {\bibinfo  {journal} {The
				Journal of Chemical Physics}\ }\textbf {\bibinfo {volume} {54}},\ \bibinfo
		{pages} {1523} (\bibinfo {year} {1971})}\BibitemShut {NoStop}%
	\bibitem [{\citenamefont {Blum}(1975)}]{blum75}%
	\BibitemOpen
	\bibfield  {author} {\bibinfo {author} {\bibfnamefont {L.}~\bibnamefont
			{Blum}},\ }\href@noop {} {\bibfield  {journal} {\bibinfo  {journal}
			{Molecular Physics}\ }\textbf {\bibinfo {volume} {30}},\ \bibinfo {pages}
		{1529} (\bibinfo {year} {1975})}\BibitemShut {NoStop}%
	\bibitem [{\citenamefont {Li}\ and\ \citenamefont {Wu}(2004)}]{li04}%
	\BibitemOpen
	\bibfield  {author} {\bibinfo {author} {\bibfnamefont {Z.}~\bibnamefont
			{Li}}\ and\ \bibinfo {author} {\bibfnamefont {J.}~\bibnamefont {Wu}},\
	}\href@noop {} {\bibfield  {journal} {\bibinfo  {journal} {Phys. Rev. E}\
		}\textbf {\bibinfo {volume} {70}},\ \bibinfo {pages} {031109} (\bibinfo
		{year} {2004})}\BibitemShut {NoStop}%
	\bibitem [{\citenamefont {Patra}(2014)}]{pat14}%
	\BibitemOpen
	\bibfield  {author} {\bibinfo {author} {\bibfnamefont {C.~N.}\ \bibnamefont
			{Patra}},\ }\href@noop {} {\bibfield  {journal} {\bibinfo  {journal} {The
				Journal of Chemical Physics}\ }\textbf {\bibinfo {volume} {141}},\ \bibinfo
		{pages} {184702} (\bibinfo {year} {2014})}\BibitemShut {NoStop}%
	\bibitem [{\citenamefont {Colla}\ \emph {et~al.}(2016)\citenamefont {Colla},
		\citenamefont {Girotto}, \citenamefont {dos Santos},\ and\ \citenamefont
		{Levin}}]{col16}%
	\BibitemOpen
	\bibfield  {author} {\bibinfo {author} {\bibfnamefont {T.}~\bibnamefont
			{Colla}}, \bibinfo {author} {\bibfnamefont {M.}~\bibnamefont {Girotto}},
		\bibinfo {author} {\bibfnamefont {A.~P.}\ \bibnamefont {dos Santos}}, \ and\
		\bibinfo {author} {\bibfnamefont {Y.}~\bibnamefont {Levin}},\ }\href@noop {}
	{\bibfield  {journal} {\bibinfo  {journal} {The Journal of Chemical Physics}\
		}\textbf {\bibinfo {volume} {145}},\ \bibinfo {pages} {094704} (\bibinfo
		{year} {2016})}\BibitemShut {NoStop}%
	\bibitem [{\citenamefont {Daoud}\ and\ \citenamefont {Cotton}(1982)}]{dao82}%
	\BibitemOpen
	\bibfield  {author} {\bibinfo {author} {\bibfnamefont {M.}~\bibnamefont
			{Daoud}}\ and\ \bibinfo {author} {\bibfnamefont {J.~P.}\ \bibnamefont
			{Cotton}},\ }\href@noop {} {\bibfield  {journal} {\bibinfo  {journal} {J.
				Phys. Paris}\ }\textbf {\bibinfo {volume} {43}},\ \bibinfo {pages} {531}
		(\bibinfo {year} {1982})}\BibitemShut {NoStop}%
	\bibitem [{\citenamefont {Jiang}, \citenamefont {Ginzburg},\ and\ \citenamefont
		{Wang}(2018)}]{jiang18}%
	\BibitemOpen
	\bibfield  {author} {\bibinfo {author} {\bibfnamefont {J.}~\bibnamefont
			{Jiang}}, \bibinfo {author} {\bibfnamefont {V.~V.}\ \bibnamefont {Ginzburg}},
		\ and\ \bibinfo {author} {\bibfnamefont {Z.-G.}\ \bibnamefont {Wang}},\
	}\href@noop {} {\bibfield  {journal} {\bibinfo  {journal} {Soft Matter}\
		}\textbf {\bibinfo {volume} {14}},\ \bibinfo {pages} {5878} (\bibinfo {year}
		{2018})}\BibitemShut {NoStop}%
	\bibitem [{\citenamefont {Gilles}\ \emph {et~al.}(2016)\citenamefont {Gilles},
		\citenamefont {Tagliazucchi}, \citenamefont {Azzaroni},\ and\ \citenamefont
		{Szleifer}}]{Szleifer1}%
	\BibitemOpen
	\bibfield  {author} {\bibinfo {author} {\bibfnamefont {F.~M.}\ \bibnamefont
			{Gilles}}, \bibinfo {author} {\bibfnamefont {M.}~\bibnamefont
			{Tagliazucchi}}, \bibinfo {author} {\bibfnamefont {O.}~\bibnamefont
			{Azzaroni}}, \ and\ \bibinfo {author} {\bibfnamefont {I.}~\bibnamefont
			{Szleifer}},\ }\href@noop {} {\bibfield  {journal} {\bibinfo  {journal} {The
				Journal of Physical Chemistry C}\ }\textbf {\bibinfo {volume} {120}},\
		\bibinfo {pages} {4789} (\bibinfo {year} {2016})}\BibitemShut {NoStop}%
	\bibitem [{\citenamefont {Prusty}\ \emph {et~al.}(2020)\citenamefont {Prusty},
		\citenamefont {Nap}, \citenamefont {Szleifer},\ and\ \citenamefont
		{De~La~Cruz}}]{Szleifer2}%
	\BibitemOpen
	\bibfield  {author} {\bibinfo {author} {\bibfnamefont {D.}~\bibnamefont
			{Prusty}}, \bibinfo {author} {\bibfnamefont {R.}~\bibnamefont {Nap}},
		\bibinfo {author} {\bibfnamefont {I.}~\bibnamefont {Szleifer}}, \ and\
		\bibinfo {author} {\bibfnamefont {M.~O.}\ \bibnamefont {De~La~Cruz}},\
	}\href@noop {} {\bibfield  {journal} {\bibinfo  {journal} {Soft Matter}\
		}\textbf {\bibinfo {volume} {16}},\ \bibinfo {pages} {8832} (\bibinfo {year}
		{2020})}\BibitemShut {NoStop}%
\end{thebibliography}
\end{document}